\documentclass[a4paper, 11pt]{article}
\usepackage[top=2cm,bottom=2cm,left=2.5cm,right=2.5cm]{geometry}
\usepackage[utf8]{inputenc}
\usepackage{amsmath,amssymb}
\usepackage[colorlinks=true]{hyperref}
\usepackage{graphicx}
\usepackage{sidecap}
\sidecaptionvpos{figure}{c}
\usepackage{subcaption}
\usepackage[dvipsnames]{xcolor}
\usepackage{mathrsfs}
\usepackage{pgfplots}
\usepgfplotslibrary{external}
\numberwithin{equation}{section}

\newcommand{\sgn}{\mathrm{sgn}}
\newcommand{\eps}{\varepsilon}

\newcommand{\bs}{\boldsymbol}
\newcommand{\rmd}{\mathrm{d}}
\newcommand{\la}{\lambda}
\newcommand{\ud}{\mathrm{d}}

\newcommand{\eg}{\emph{e.g.}~}

\renewcommand{\o}{\omega}
\newcommand{\g}{\gamma}
\def\m{\mu}\newcommand{\G}{\Gamma}
\def\C{{\mathbb C}}
\renewcommand\L{\Lambda}
\def\R{{\mathbb R}}
\def\N{{\mathbb N}}
\def\hf{\frac{1}{2}}
\def\T{{\mathbb T}}
\def\Rscr{\mathcal R}
\def\be{\begin{equation}}
\def\ee{\end{equation}}
\def\bea{\begin{eqnarray}}
\def\eea{\end{eqnarray}}

\title{Soliton gas in integrable dispersive hydrodynamics}

\author{Gennady A. El }
\date{Department of Mathematics, Physics and Electrical
Engineering, \\ Northumbria University, Newcastle upon Tyne, United
Kingdom}

\begin{document}

\title{Soliton gas in integrable dispersive hydrodynamics}

\author{Gennady  El }
\date{Department of Mathematics, Physics and Electrical
Engineering, \\ Northumbria University, Newcastle upon Tyne, United
Kingdom}

\maketitle

\begin{abstract}
We review spectral theory of soliton gases in integrable dispersive hydrodynamic systems. We first present a phenomenological approach based on the consideration of phase shifts in pairwise soliton collisions and leading to the  kinetic equation for a non-equilibrium soliton gas. Then a more detailed theory is presented in which soliton gas dynamics are modelled by  a thermodynamic type limit of modulated finite-gap spectral solutions of the Korteweg-de Vries and the focusing nonlinear Schr\"odinger   equations.  For the focusing   nonlinear Schr\"odinger equation the notions of  soliton condensate and  breather gas are introduced that are related to the phenomena of spontaneous modulational instability and the rogue wave formation. Integrability properties of the kinetic equation for soliton gas are discussed and some  physically relevant solutions are presented and compared with direct numerical simulations of dispersive hydrodynamic systems.
\end{abstract}

\tableofcontents\vspace{1cm}
\section{Introduction} 
\label{sec:intro}

\subsection{Integrable turbulence and soliton gas}
Random  nonlinear dispersive waves have been the subject of an 
active research in nonlinear physics for more than  five  decades, most notably
in the contexts of water wave dynamics.
A significant portion of the work in this direction has
been centred around weak wave turbulence 
\cite{nazarenko_wave_2011}. 
The wave turbulence theory deals with out of equilibrium statistics of incoherent, weakly nonlinear dispersive
waves in non-integrable systems. One of the early and most significant results of the wave turbulence theory was the analytical determination by Zakharov \cite{zakharov_weak_1965} of the analogs of the Kolmogorov spectra  describing energy flux  through scales in  dissipative hydrodynamic turbulence. 
These spectra, called Kolmogorov-Zakharov spectra, were  obtained  as solutions of the kinetic equations for the evolution of the Fourier spectra of random weakly nonlinear dispersive waves in multidimensional non-integrable systems.  

More recently, a new theme in turbulence theory has emerged in
connection with the dynamics of strongly noninear random waves described by
one-dimensional integrable systems such as  the Korteweg-de Vries (KdV) and 1D nonlinear Schr\"odinger (NLS) equations.  This kind of random wave motion in nonlinear conservative systems,  dubbed   `integrable turbulence' \cite{zakharov_turbulence_2009},  has attracted significant attention  from both  fundamental and applied perspectives.
 The interest in integrable turbulence is motivated by the complexity of  
many natural or experimentally observed nonlinear wave phenomena often requiring a statistical description even though the underlying physical model is, in principle, amenable to the well-established mathematical techniques of integrable systems theory such as the inverse scattering transform (IST) or finite-gap theory  \cite{novikov_theory_1984}, \cite{belokolos_algebro-geometric_1994}, \cite{osborne_nonlinear_2010}. Indeed, integrable systems are known to capture  essential properties of many wave processes occurring in
real-world systems \cite{calogero_what_1991}. The integrable turbulence framework is particularly pertinent to the description of modulationally unstable systems whose solutions, under the effect of random noise, can exhibit highly complex nonlinear behaviour  that can be adequately described  in terms of the turbulence theory concepts, such as  probability distribution functions, ensemble averages, power spectra etc. \cite{Agafontsev:15}, \cite{Agafontsev:16}, \cite{kraych_statistical_2019}. We stress that the term `turbulence' in this context is understood as complex spatiotemporal  dynamics that requires probabilistic description and is not related to the energy cascades through scales, the prime feature of  hydrodynamic and weak turbulence. 

Localised nonlinear solitary waves are a ubiquitous  feature of nonlinear dispersive wave propagation whose discovery dates back to shallow water wave observations by
John Scott Russell in 1845 \cite{russell_report_1845}.  If the wave dynamics are described by one of the completely integrable equations the solitary waves  exhibit particle-like properties such as elastic, pairwise
interactions accompanied by certain phase/position shifts. Such solitary waves are called solitons
\cite{zabusky_interaction_1965} and have been extensively studied both
theoretically \cite{ablowitz_inverse_1974}, \cite{novikov_theory_1984}, \cite{newell_solitons_1985} and experimentally
\cite{remoissenet_waves_2013}. 
The main tool for the analysis of integrable nonlinear dispersive PDEs is the IST \cite{gardner_method_1967} based on the reformulation of a nonlinear PDE as a compatibility condition of two {\it linear} problems: a stationary spectral (scattering) problem and the evolution problem for the same auxiliary function. Within the scattering  problem solitons are associated with discrete values of the spectrum, while the integrable evolution preserves these spectral values in time.

Solitons can form ordered macroscopic coherent structures such as modulated soliton trains  and  dispersive shock waves \cite{maiden_solitonic_2018}, \cite{el_dispersive_2016}.  Furthermore, solitons  can form {\it irregular}, statistical ensembles that can be interpreted as soliton gases. The nonlinear wave field in such gases represents a particular case of integrable turbulence, often called soliton turbulence (with the caveat that the latter term has also been  used  in the context of nonintegrable wave dynamics,  see e.g. \cite{zakharov_soliton_1988}, \cite{kachulin_soliton_2020}). 
 Generally, soliton gas and soliton turbulence represent two complementary aspects of the same physical object, the natural counterparts of the wave-particle duality of a single soliton. In the soliton-gas description the focus is on the collective dynamics/kinetics of solitons as interacting (quasi)particles  characterised by  certain amplitude (or velocity) distribution function, while the soliton turbulence description emphasises the characterics  of the random nonlinear wave field associated with the soliton gas, such as probability density function, power spectrum etc.  
 The observations and analysis of irregular  soliton  complexes in the ocean  have been reported in \cite{costa_soliton_2014},   \cite{slunyaev_persistence_2021}. Recent laboratory experiments on the generation of shallow-water and deep water soliton gases were reported in \cite{redor_experimental_2019} and \cite{suret_nonlinear_2020} respectively. It has also been demonstrated  that  soliton gas dynamics in the focusing NLS equation enables a remarkably accurate description of the statistical properties of the nonlinear stage of noise-induced  modulational instability \cite{gelash_bound_2019} as well as provides important insights into the dynamical and statistical mechanisms of the spontaneous formation of rogue waves  \cite{gelash_strongly_2018}, \cite{agafontsev_rogue_2021}. 

\begin{figure}[h]
\begin{center}
  \includegraphics[scale=0.4]{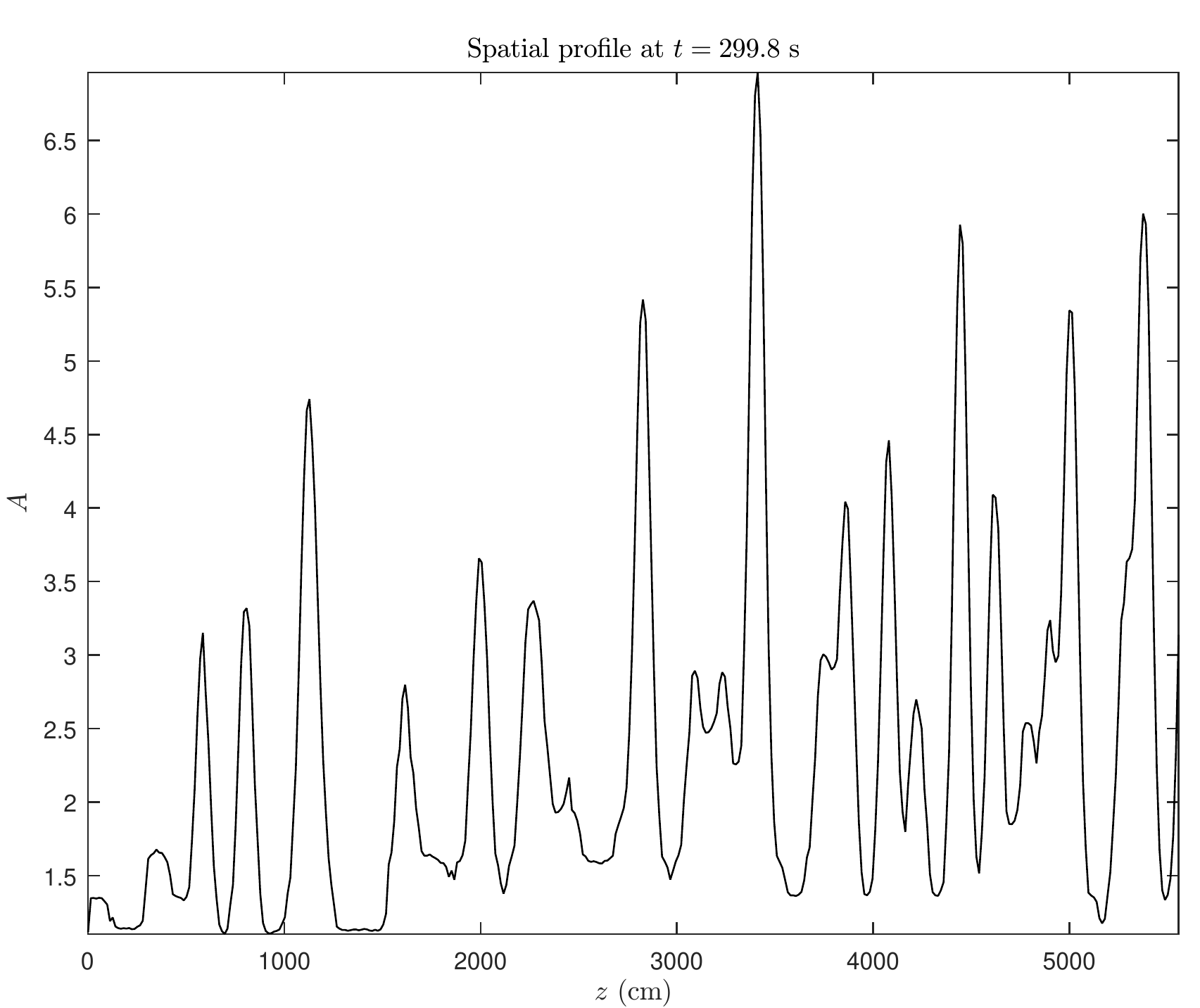}  \quad  \includegraphics[scale=0.38]{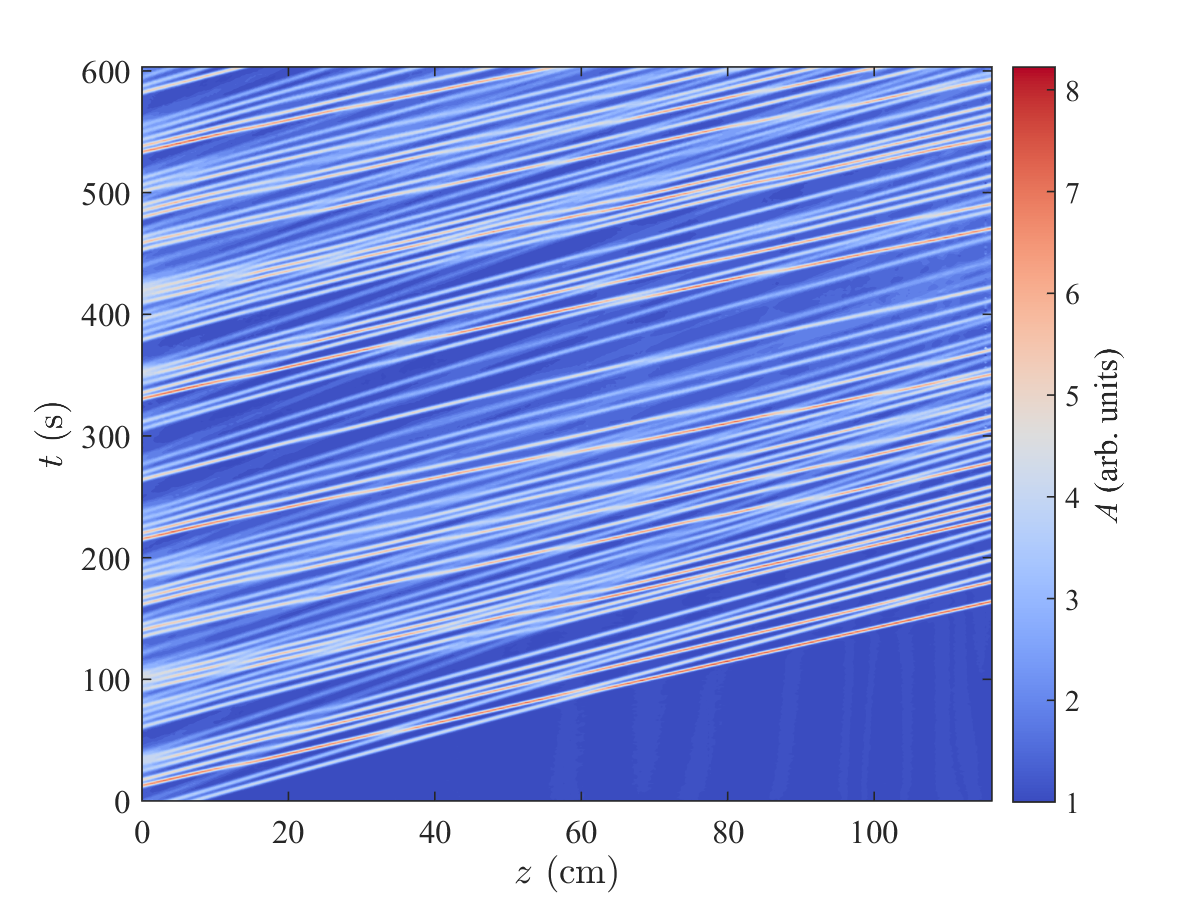}   \end{center} 
  \caption{Soliton gas in  dispersive hydrodynamics. Left: experimentally measured profile of the nonlinear wave field of a dense soliton  gas in a viscous fluid conduit; Right: $x,t$-contour image. The images are courtesy of the Dispersive Hydrodynamics Laboratory, Boulder, Colorado}
  \label{fig:conduit}
\end{figure}

Analytical description of  soliton gases in nonlinear dispersive wave systems was initiated in  the Zakharov's 1971 paper  \cite{zakharov_kinetic_1971}, where a spectral kinetic equation for KdV solitons was introduced using an IST-based phenomenological `flea gas' reasoning enabling the evaluation of an effective adjustment to the soliton's  velocity in a {\it rarefied gas} due to the interactions (collisions) between individual  solitons, accompanied by the well-defined phase-shifts. The  kinetic equation for a rarefied soliton gas describes the evolution of the distribution function of the solitons with respect to the spectral parameter and the positions of soliton centres, i.e. the {\it density of states} (DOS).  
 In a {\it dense gas}, however, the solitons  exhibit significant overlap and, as a result, are continuously involved in a strong nonlinear interaction with each other. This can be seen in Fig.~\ref{fig:conduit} displaying a laboratory realisation of a dense soliton gas in a viscous fluid conduit \cite{lowman_dispersive_2013}, a versatile fluid dynamics platform enabling high precision experiments on the generation and interaction of solitary waves that exhibit nearly elastic collisions \cite{lowman_interactions_2014}, \cite{maiden_solitary_2020}. One can appreciate that, for a dense gas the particle interpretation of individual solitons  becomes less  transparent and the wave aspect of the collective soliton  dynamics comes to the foreground. Indeed, a consistent generalisation of Zakharov's kinetic equation for  KdV solitons  to the case of  a dense soliton gas  has been achieved in \cite{el_thermodynamic_2003} in the framework of the nonlinear wave modulation (Whitham) theory \cite{whitham_linear_1999}.  It was proposed in \cite{el_soliton_2001}, \cite{el_thermodynamic_2003}  that the soliton gas can be modelled by the  thermodynamic type solitonic limit of the spectral finite-gap KdV solutions and their modulations \cite{flaschka_multiphase_1980}. The resulting kinetic equation has the form of a nonlinear integro-differential equation for the DOS in the IST spectral phase space. The structure of the kinetic equation obtained in \cite{el_thermodynamic_2003} suggested 
 that, remarkably,  in a dense gas the net effect of soliton interactions can be formally evaluated using the same phase-shift argument  that was used in  the rarefied gas theory \cite{zakharov_kinetic_1971}.  This observation, termed the collision rate assumption, has enabled an effective phenomenological theory  of  a dense soliton gas  for the focusing NLS  equation \cite{el_kinetic_2005} and more recently, for the defocusing and resonant NLS equations  \cite{congy_soliton_2021}.  The phenomenological  theory of soliton gas for the focusing NLS equation proposed in \cite{el_kinetic_2005} has been confirmed and substantially extended in \cite{el_spectral_2020} within the framework of the thermodynamic limit of spectral finite-gap  solutions of the focusing NLS equation and their modulations. This latter work has revealed a number of  new soliton gas phenomena due to a  very different structure of the spectral phase space of the focusing NLS equation compared to the KdV equation. In particular, the generalisation of soliton gas, termed {\it breather gas}, was introduced  by considering a special family of focusing NLS solitonic solutions called breathers. 
  Such a breather gas
 represents an intriguing type of integrable turbulence observed in the ocean \cite{osborne_highly_2019}  and recently realised numerically \cite{roberti_numerical_2021}.
   
Mathematical properties of  the  kinetic equation for soliton gas were studied in \cite{el_kinetic_2011}, \cite{pavlov_generalized_2012} where it was proved that it admits an infinite series of integrable linearly degenerate hydrodynamic reductions obtained  by a multi-component delta-function ansatz for the DOS. A further study of the classical integrability properties of the soliton gas kinetic equation was undertaken in \cite{bulchandani_classical_2017}. A very recent paper \cite{kuijlaars_minimal_2021} presents rigorous results for the uniqueness, existence and non-negativity of the solutions to the integral equations for the DOS  in the spectral kinetic  theory for KdV and focusing NLS soliton gases \cite{el_thermodynamic_2003}, \cite{el_spectral_2020}. 
 
Apart from the above line of  research  on  soliton gases  inspired by the Zakharov 1971 work  there have been many other developments---analytical, numerical and experimental ---exploring  various aspects of  soliton gas/soliton turbulence  dynamics in both integrable and nonintegrable classical wave systems (see \eg  \cite{meiss_drift-wave_1982}, \cite{schwache_properties_1997}, \cite{fratalocchi_time-reversal_2011},  \cite{schmidt_non-thermal_2012}, \cite{turitsyna_laminarturbulent_2013}, \cite{dutykh_numerical_2014},  \cite{soto-crespo_integrable_2016}, \cite{giovanangeli_soliton_2018},  \cite{girotti_rigorous_2018}, \cite{marcucci_topological_2019-1}).  
Additionally, there has been a recent surge of  related activity in  generalised hydrodynamics (see \cite{castro-alvaredo_emergent_2016}, \cite{bertini_transport_2016}, \cite{doyon_lecture_2020} and references therein), where the equations analogous to those arising in the soliton gas theory became pivotal for the understanding of large-scale, hydrodynamic properties of integrable quantum many-body systems.

\subsection{Dispersive hydrodynamics}
This review considers classical soliton gases from the perspective of  dispersive hydrodynamics modelled by hyperbolic or elliptic conservation laws
regularized by small conservative, dispersive corrections \cite{biondini_dispersive_2016}, the KdV equation $u_t+6uu_x +  u_{xxx}=0$ being the simplest example. 
Smallness of the dispersive
term is understood in the sense that the typical coherence length
 of the medium---determined by a balance
between nonlinear and dispersive effects---is much less than the scale
of initial, boundary, or intermediate flow conditions where long
wavelength, nonlinear, hydrodynamic effects dominate. This
distinguishes dispersive hydrodynamic problems from typical
formulations in classical soliton theory which usually involve variations of the wave field on the scale of the coherence length.  The dispersive
hydrodynamic framework arises naturally in the description of problems
related to the large-scale dynamics of shallow water or internal
waves, but also proves to be extremely useful in the modelling of
various phenomena in nonlinear optics including the ``atom optics'' of
quantum fluids such as Bose-Einstein condensates.

In the scalar case general dispersive hydrodynamics are described by the equations of the form
\begin{equation}
\label{eq:scalar_dh}
u_t + F(u)_x =  (D[u])_x , 
\end{equation}
where $F(u)$ is the nonlinear hyperbolic flux and $D[u]$ is a
differential (generally integro-differential) operator, possibly
nonlinear, that gives rise to a real-valued  dispersion
relation 
$\omega=\omega_0({k})$ for linearised waves. 

Scalar integrable dispersive hydrodynamics of the form
\eqref{eq:scalar_dh}, such as the KdV, modified KdV, Camassa-Holm or
Benjamin-Ono equations often arise as small-amplitude,
`unidirectional' approximations of more general Eulerian
bidirectional systems (see \cite{whitham_linear_1999})
\begin{equation}
\label{eq:disp_Euler}
\begin{split}
\rho_t + (\rho u)_x &= (D_1[\rho,u])_x\, , \\
(\rho u)_t + \left ( \rho u^2 + \sigma P(\rho) \right )_x &=
(D_2[\rho,u])_x \,, \quad \sigma=\pm 1,
\end{split}
\end{equation}
where $D_{1,2}[\rho,u]$ are conservative, dispersive operators,
 $\rho$ and
$u$ are interpreted as a mass density and fluid velocity,
respectively and  $\sigma P(\rho)$, where $P(\rho)>0$ is the pressure law.  For $\sigma=+1$ this class of equations generalise the shallow water
and isentropic gas dynamics equations while encompassing many of the
integrable dispersive hydrodynamic models such as the Kaup-Boussinesq
system~\cite{kaup_higher_1975}, the hydrodynamic form of the
defocusing NLS equation~\cite{kamchatnov_nonlinear_2000}, the
Calogero-Sutherland system describing the dispersive hydrodynamics of
quantum many-body systems \cite{abanov_integrable_2009} and many others.  
For $\sigma = -1$ equations \eqref{eq:disp_Euler} describe `elliptic' dispersive hydrodynamics, the most prominent example being the focusing NLS equation describing, in particular, the evolution of weakly nonlinear narrow-band wave packets on deep water  and the propagation of light in optical media in the self-focusing, cubic nonlinearity regime. Integrable bidirectional dispersive hydrodynamics admit Lax representation similar to \eqref{lax_gen}  with $\psi(x,t,\lambda)$ being a vector function.  

Integrability of dispersive hydrodynamics \eqref{eq:scalar_dh} is understood as the existence of the Lax pair,
a system of linear differential equations for an auxiliary (generally vector) function $\psi(x,t; \lambda)$:
\be\label{lax_gen}
\mathcal{L} \psi = \lambda \psi, \qquad
\psi_t = \mathcal{A} \psi, 
\ee
where $\mathcal{L}$ and $\mathcal{A}$ are linear differential operators depending on $u(x,t)$ and its derivatives, and $\lambda$ is a spectral parameter
so that the equation \eqref{eq:scalar_dh} can be written in the operator form $\mathcal{L}_t+(\mathcal{L}\mathcal{A} - \mathcal{A}\mathcal{L})=0$ under the additional requirement of {\it isospectrality}, $\lambda_t=0$. For sufficiently rapidly decaying fields $u(x,t) \to 0$ as $|x| \to \pm \infty$ the spectrum of the operator $\mathcal{L}$, often called the Lax spectrum, consists of two components: discrete and continuous. The points of discrete spectrum correspond to solitons in the asymptotic solution at $t \gg 1$, while the continuous spectrum corresponds to dispersive radiation.  The solutions whose spectrum consists of $N$ discrete points  only are called $N$-soliton solutions.

A prominent example of  a multiscale nonlinear wave structure supported by dispersive hydrodynamics is  dispersive shock wave exhibiting coherence at both microscopic (soliton) and macroscopic (wave modulation) scales \cite{el_dispersive_2016}. Contrastingly, soliton gas  as a dispersive hydrodynamic structure  
exhibits  coherence at the microscopic
scale while being macroscopically incoherent, in the sense that the
values of the wave field at two points separated by a distance much
larger than the intrinsic dispersive length of the system (the soliton
width), are not dynamically related.  The source of randomness in dispersive hydrodynamics is
typically related to some sort of stochastic large-scale initial or boundary
conditions although one can envisage dynamical mechanisms of the
effective randomization of the wave field and soliton gas generation
\cite{gurevich_development_1999}, \cite{el_dam_2016}.   In particular, statistical soliton ensembles can be naturally generated from both non-vanishing
deterministic (e.g.  quasiperiodic) and random initial
conditions via the processes of soliton fission
\cite{trillo_experimental_2016}, \cite{trillo_observation_2016} or modulation
instability \cite{gelash_bound_2019}. 

The inherent scale separation in dispersive hydrodynamic flows suggests the use of asymptotic methods for their description.   One such method, the Whitham modulation theory  \cite{whitham_linear_1999} proved  particularly effective and has been extensively developed in the contexts of both integrable and nonintegrable systems.
Within the Whitham theory the dispersive hydrodynamic system is asymptotically reduced to a system of quasilinear equations,  which describe large-scale variations of the wave's modulation parameters such as the amplitude, the wavenumber, the mean etc. The Whitham hydrodynamic equations can be derived by applying a multiple-scale singular perturbation theory or, equivalently, by averaging the dispersive hydrodynamic conservation laws  over rapid oscillations.  Although the Whitham method does not rely on integrability, the presence of integrable structure in the original dispersive hydrodynamic system greatly enhances  the structure of the modulation  system as well. The inherited integrability of the Whitham modulation equations  is realised via the generalised hodograph transform \cite{tsarev_poisson_1985}, \cite{dubrovin_hydrodynamics_1989}. 

Importantly,  the Whitham theory  for integrable dispersive PDEs  admits spectral formulation within the extension of IST called the finite-gap theory  \cite{novikov_theory_1984}. The finite-gap theory describes an important class of periodic or quasiperiodic (multiphase) solutions to integrable systems, whose Lax spectrum (the set of admissible values of the spectral parameter $\lambda$ in  \eqref{lax_gen} corresponding to $L^2$  eigenfunctions $\psi$)  lies in the {\it finite} union of disjoint  bands. 
In 1980 Flaschka, Forest and McLaughlin  showed that the Whitham equations obtained via multiphase averaging of the KdV equation  describe slow evolution of the 
band spectrum of the finite-gap Lax operator $\mathcal{L}$ \cite{flaschka_multiphase_1980}. It has then been shown by Lax and Levermore \cite{lax_small_1983} that the spectral Whitham equations derived in \cite{flaschka_multiphase_1980}  also describe the weak, distribution limits of dispersive conservation laws obtained in the limit of small dispersion.

It has been noted by P. Lax in \cite{lax_zero_1991} that the Whitham modulation equations can be viewed  as certain analogs of the moment equations  in classical statistical fluid mechanics \cite{monin_statistical_2011}.
This observation, combined with the spectral formulation of the Whitham theory  \cite{flaschka_multiphase_1980} and the ideas of Venakides on the continuum limit of theta-functions \cite{venakides_continuum_1989} has enabled the development of the spectral theory of soliton gas  \cite{el_soliton_2001}, \cite{el_thermodynamic_2003}, \cite{el_spectral_2020} providing the justification of the phenomenological approach of \cite{el_kinetic_2005} and a general mathematical  framework for the study of soliton gases/soliton turbulence  in classical integrable dispersive hydrodynamic  systems.

\bigskip
The structure of the review is as follows. In Section~\ref{sec:kin_phenomen} we present a phenomenological theory of soliton gas in unidirectional and bidirectional systems of integrable dispersive hydrodynamics using the KdV and NLS equations as prototype examples. The main result is the construction of the kinetic equation describing the evolution of the density of states in a soliton gas. In the bi-directional case we identify two different types of soliton gases: isotropic and anisotropic, differing in the properties of overtaking and head-on solitons. 
In Section~\ref{sec:spectral_theory} we develop a detailed spectral theory of soliton gas for the KdV equation and of soliton and breather gases  for the focusing NLS equations. This is done by applying a special thermodynamic type limit to the spectral finite-gap solutions and their modulations described by the appropriate Whitham equations. For the focusing NLS equation we introduce the notion of soliton condensate and also consider three types of `rogue wave' breather  gases.
In Section~\ref{sec:hyd_red} hydrodynamic reductions of the spectral kinetic equation are derived  by employing a multi-component delta-function ansatz for the density of states and their integrability is investigated. In Section~\ref{sec:riemann} we apply the hydrodynamic reductions to consider a canonical Riemann problem describing collision of two multi-component soliton gases.  Finally, the Conclusion  provides a brief summary and outlines some directions of future research.

\section{Kinetic equation for soliton gas: phenomenological construction}
\label{sec:kin_phenomen}
We first introduce soliton gas phenomenologically, as an infinite  ensemble of interacting solitons randomly distributed on the line with non-zero density. This intuitive definition lacks precision but it is sufficient for the purposes of this section. A more elaborate mathematical model of a soliton gas will be described in Section \ref{sec:spectral_theory}. 

\subsection{Unidirectional soliton gas}
\label{sec:unidir_sol_gas}
It is convenient to introduce the basic ideas of soliton gas theory using the KdV equation as a prototype model of nonlinear dispersive wave propagation in a broad variety of physical systems.
We consider the KdV equation in the  `physical' form
\begin{equation}\label{KdV}
 u_t\ +\ 6uu_{x}\ +\ u_{xxx}\ =\ 0 \, .
\end{equation}
The KdV equation \eqref{KdV} belongs to the family of  completely  integrable equations. For a  broad class of initial conditions its integrability  is realised via the 
IST method \cite{novikov_theory_1984}.
The inverse scattering theory  associates   soliton  of the KdV equation \eqref{KdV} with a point of discrete spectrum $\la_n$,  of the Schr\"odinger operator, which is the Lax operator for the KdV equation (cf. \eqref{lax_gen}),
\be\label{schr}
\mathcal{L}= -\partial_{xx}^2  - u(x,t) \,  .
\ee
Assuming $u \to 0$ as $x \to \pm \infty$ the KdV soliton solution corresponds to $\la_i = -\eta^2_i$, $\eta_i>0$, is given by
\be\label{kdv1sol}
u_s(x,t; \eta_i)=2\eta_i^2 \hbox{sech}^2 [\eta_i(x-4\eta_i^2 t - x_i^{0})],
\ee
where the soliton amplitude $a_i= 2\eta_i^2$,  the  speed $s_i=4\eta_i^2$ and $x_i^0$ is the initial position or  `phase'. Along with the simplest single-soliton solution the KdV equation supports $N$-soliton solutions $u_N(x,t)$ characterised by $N$ discrete spectral parameters $\eta_1 < \eta_2 < \dots <\eta_N$ and the set of initial positions  $\{x_i^0 | i=1, \dots, N\}$. Since the discrete spectrum of the (self-adjoint) Schr\"odinger operator \eqref{schr} is non-degenerate (i.e. $\eta_i \ne \eta_j \iff i \ne j$) all solitons in the $N$-soliton solution have different velocities and the long-time asymptotic solution  assumes the form of a rank-ordered soliton train,
\be\label{train}
t  \to \infty: \quad u_N(x,t) \sim  \sum \limits_{j=1}^N 2 \eta_i^2 \hbox{sech}^2 [\eta_i(x-4\eta_i^2 t - x_i^0 + \Delta_i)],
\ee
where the phase (position) shifts $\Delta_i$ occur due to the interaction of individual solitons at the initial stage of the evolution \cite{drazin_solitons_1987}, \cite{novikov_theory_1984}. These phase shifts play important role in our consideration as detailed below.

The integrable structure of the KdV equation has profound implications for the soliton interaction dynamics:

\medskip
(i)  the KdV evolution preserves the IST spectrum,  $\partial_t \eta_j=0$, implying that solitons retain their `identity' (amplitude, speed) upon  interactions;

\medskip
(ii) the  collision of two solitons with spectral parameters $\eta_i$ and $\eta_j$, $i \ne j$ results in their phase  (position) shifts given by
\be \label{shift_kdv}
\Delta_{ij} \equiv \Delta (\eta_i, \eta_j)= \frac{\sigma_{ij}}{\eta_i}   \ln\left|\frac{\eta_i + \eta_j}{\eta_i-\eta_j} \right|, \quad
\sigma_{ij}= \sgn(\eta_i-\eta_j),
\ee
so that the taller soliton acquires shift forward and the smaller one -- shift backwards; 

\medskip
(iii) solitons interact pairwise, i.e. the resulting, accumulated phase shift $\Delta_i$ of a given  soliton with spectral parameter $\eta_i$  after its interaction with  $M$ solitons with parameters $\eta_j$, $j \ne i$, is equal to the sum of the individual phase shifts,
\be\label{delta_total}
\Delta_i= \sum \limits_{j =1, j\ne i}^M \Delta_{ij}.
\ee
It is important to stress that  the collision phase shifts are the far-field effects and mathematically, the artefacts of the asymptotic representation of the exact two-soliton solution of the KdV equation  in the form of a sum of two individual solitons: $u_2(x,t; \eta_1, \eta_2) \simeq u_s(x+\Delta_{12},t; \eta_1)+u_s(x+\Delta_{21},t; \eta_2)$, which
is only valid if solitons are sufficiently separated (the long-time asymptotics). The interaction of solitons is a complex nonlinear process  \cite{lax_integrals_1968} and the resulting wave field in the interaction region  cannot be represented as a superposition of the phase-shifted one-soliton solutions.

Motivated by the above properties of $N$-soliton solutions we now introduce a rarefied soliton gas by generalising the  asymptotic soliton train solution \eqref{train} in the following way. We let $N \to \infty$ in \eqref{train} and introduce the  
 {\it density of states} (DOS) $ f(\eta, x,t)$ such that $f(\eta_0, x_0, t_0)\rmd\eta \rmd x$ is the number of solitons found at $t=t_0$   
 in the element  $ [\eta_0, \eta_0+ \rmd\eta] \times [x_0, x_0 + \rmd x]$ of the  phase space $\mathfrak{S}=\Gamma \times \mathbb{R}$, where  $\Gamma = [\eta_{\min}, \eta_{\max}] \subset \mathbb{R}^+$ is the spectral support of DOS (it is assumed in the above definition of DOS that the  interval $[x_0, x_0 + \rmd x]$ contains a sufficiently large number of solitons).  Also, without loss of generality one can assume $\Gamma =[0,1]$.  The parameter controlling the total spatial density of the soliton gas is then
\begin{equation}\label{kappa}
\beta = \int_0^1 f(\eta) \rmd \eta \, .
\end{equation}
For a rarefied gas $\beta \ll 1$  (this criterion is understood in the asymptotic sense since the actual, numerical,  value of $\beta$ depends on the definition of the unit interval of $x$. 

We further assume (to be validated later) the Poisson distribution with  density  $\beta$ for the soliton centres $x_j = x_j^0 + 4 \eta^2 t  \in \mathbb{R}$  so that the  distances $d_k=x_{k}-x_{k-1}$  between solitons are independent random values distributed
with probability density $\mathcal{P}(d)=\beta \exp {(-\beta d)}$.  We thus arrive at a `stochastic soliton lattice' 
\be\label{rar_u}
u_{\infty} (x,t) := \sum \limits_{i = 1} ^{\infty} 2\eta_i^2 \hbox{sech}^2 [\eta_i(x - 4\eta_i^2 t  - x_i^0)] , 
\quad \eta_i \in \Gamma , \ \ x_i^0 \in \mathbb{R}, \ \ \beta \ll 1\, ,
\ee
that can be viewed as an approximate model of a rarefied soliton gas.

 Each realisation of the random process $u_\infty (x,t)$ \eqref{rar_u}  satisfies the KdV equation \eqref{KdV} almost everywhere  in $x \in \mathbb{R}$.  Due to the small spatial density $\beta$, the individual solitons in a soliton gas overlap only in the regions of their exponential tails, except for the rare events of their collisions where a complex nonlinear interaction occurs \cite{lax_integrals_1968} affecting  the statistical characteristics of the random field \eqref{rar_u} \cite{pelinovsky_two-soliton_2013}.

We first look at the equilibrium, or uniform,  soliton gas for which $\partial f/\partial x=0$. In view of isospectrality of the KdV evolution, the spatially independent density of states $f(\eta)$ at $t=0$ also implies that  $\partial f / \partial t=\partial f/\partial x=0$ $\forall t>0$. 

Consider  propagation of a  `tracer' soliton with spectral parameter $\eta=\eta_1 \in [0,1]$ in a uniform soliton gas with a given DOS $f(\eta)$, $\eta \in [0,1]$ and assume that the gas is rarefied, $\int_0^1 f(\eta) \rmd \eta \ll 1$.  
Due to  the collisions of the tracer `$\eta_1$-soliton'  with the `$\mu$-solitons' ($\mu \ne \eta_1$) in the gas, each collision leading to the phase shift \eqref{shift_kdv}, the effective (mean) velocity of the trial soliton is approximately evaluated as 
\begin{equation}\label{s1}
  s(\eta_1)\ \approx \ 4\eta_1^2 + \frac{1}{\eta_1}\int \limits_0^1 \ln\left|\frac{\eta_1 + \mu}{\eta_1-\mu}\right|f(\mu)[4\eta_1^2-4\mu^2]\,\ud\mu.
\end{equation}
The integral correction term in \eqref{s1} follows from the continuum limit of eq. \eqref{delta_total} assuming  that in a rarefied gas $s(\eta_1)= 4 \eta_1^2$ to leading order. Then the total spatial shift of $\eta_1$-soliton over the time interval  $\rmd t$ due to the interactions with $\mu$-solitons, $\mu \in [0,1]$ is given by $\Delta_1 = [\int_0^1 \Delta(\eta_1, \mu) \Xi(\eta_1, \mu) \rmd \mu ] \rmd t$, where $\Xi(\eta_1, \mu_0) \rmd \mu \approx |4\eta_1^2 - 4 \mu_0^2| f(\mu_0) \rmd \mu$ is the average collision rate of $\eta_1$-soliton with $\mu$-solitons having their spectral parameter 
$\mu \in [\mu_0, \mu_0+\rmd\mu]$.  
Then using the expression \eqref{shift_kdv} for the KdV phase shift we obtain \eqref{s1}. The effective velocity $s(\eta_1)$ has a natural interpretation of the transport velocity in soliton gas.

\begin{figure}
\begin{center}
  \includegraphics[scale=0.4]{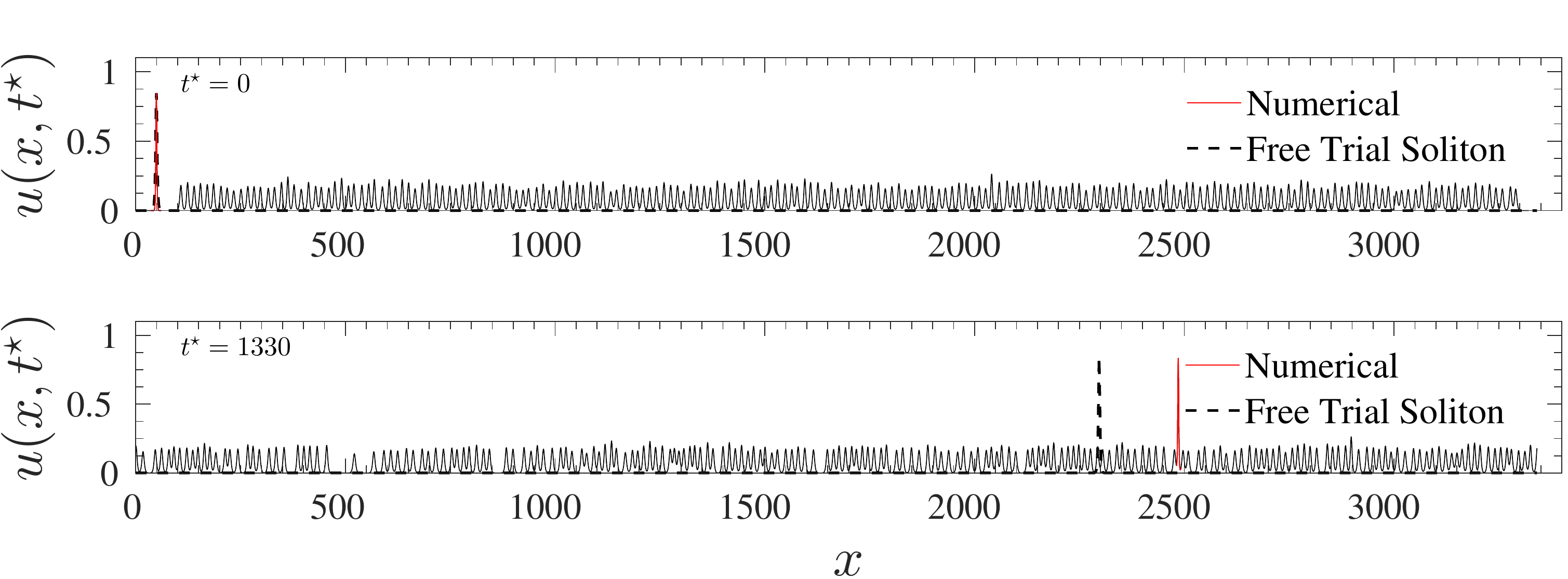}    
  \end{center}
    \caption{(adapted from \cite{carbone_macroscopic_2016}) Comparison for the propagation of a free soliton with the spectral parameter $\eta=\eta_1$ in a void (black dashed line) with the propagation of the trial soliton with the same spectral parameter $\eta_1$ (red solid line) through a one-component rarefied soliton gas with the DOS $f_0 \delta(\eta - \eta_0)$. Parameters used in the numerical simulations are: $\eta_1=0.65$,   
  $\eta_0 = 0.3$ and density $f_0 = 0.048$. One can see that the trial soliton  propagates faster in the gas due to the interactions with smaller solitons.}
    \label{fig:trial_sol1}  
  \end{figure}

  We note that the spectral parameter $\eta_1$ of the `special' soliton should not necessarily belong to the support $\Gamma=[0,1]$ of $f(\eta)$.  If $\eta_1 \notin \Gamma$ we shall call such a soliton  a `trial' soliton.
  The important distinction between tracer and trial solitons will become more transparent later. 
  The modification of the `free' velocity of a  trial soliton due its  propagation through a uniform soliton gas  is illustrated in Fig.~\ref{fig:trial_sol1}.
  
We now consider a non-equilibrium (non-uniform) soliton gas with $f(\eta) \equiv f(\eta, x,t)$, $s(\eta) \equiv s(\eta, x, t)$ so  that spatiotemporal variations of $f$ and $s$ occur on the scales much larger than those associated with variations of the nonlinear wave field $u(x,t)$ in  \eqref{rar_u}.
Isospectrality of the KdV dynamics implies the conservation equation for the density in the phase space 
$\mathfrak{S}$ (i.e. DOS), 
\be \label{kin_eq0}
f_t+(sf)_x = 0,
\ee
which, together with the expression for the effective transport velocity $s(\eta)$ given by  \eqref{s1}, can be viewed as kinetic equation for rarefied soliton gas first introduced by Zakharov in 1971 \cite{zakharov_kinetic_1971}. 
As we mentioned,   $x$ and $t$  in (\ref{kin_eq0}) are `slow' variables  compared to $x$ and $t$ in the KdV equation \eqref{KdV}.  
Thus the kinetic equation \eqref{kin_eq0}, \eqref{s1}  can be viewed as a modulation equation for the soliton gas \eqref{rar_u}. We shall see later that this analogy has a more precise meaning, with a hierarchy of spatiotemporal scales involved.

Zakharov's approximate equation \eqref{s1} for the effective transport velocity in a soliton gas was generalised in  \cite{el_thermodynamic_2003}  to the case of dense ($\beta= \mathcal{O}(1)$) gas.  This was done by evaluating the thermodynamic limit  of the  nonlinear dispersion relations associated with the spectral finite-gap solutions of the KdV equation \eqref{KdV}, see Section~\ref{sec:spectral_theory_kdv} below. The result has the form of a linear integral equation
\begin{equation}\label{eq_state_kdv} 
s(\eta)=4\eta^2+\frac{1}{\eta}\int \limits_{0}^1 \ln
\left|\frac{\eta + \mu}{\eta-\mu}\right|f(\mu)[s(\eta)-s(\mu)]\rmd \mu\,   ,
\end{equation}
which essentially provides the relation between the spectral flux density $v=fs$ and the DOS $f$ in the transport equation \eqref{kin_eq0} and so can be viewed as the equation of state of the soliton gas.
The expression \eqref{s1} for the effective soliton velocity in a rarefied soliton gas  represents an approximate first order solution of  the equation of state \eqref{eq_state_kdv}  obtained by assuming that the integral term is a small correction to the free soliton velocity $4\eta^2$. 
 
One can notice that  equation \eqref{eq_state_kdv} can be formally written down by utilising the same  phase-shift argument used in the derivation of the approximate expression \eqref{s1}, i.e. by a formal replacement in  \eqref{s1} of the leading order value $4 \eta^2$ for the soliton velocity with its effective value $s(\eta)$. The validity  of this replacement  in the KdV equation case suggests the general {\it collision rate assumption}  whereby the total position shift of the soliton with $\eta=\eta_1$ due to soliton collisions in a gas with DOS $f(\mu)$ over the time interval $\rmd t$  is given by 
\be \label{collision_rate}
\Delta_1 = [\int_\Gamma \Delta(\eta_1, \mu)|s(\eta_1) - s(\mu)| f(\mu)\rmd \mu ] \rmd t,
\ee 
where $\Gamma$ is a spectral support of the DOS $f(\eta)$.
This   assumption  was used in \cite{el_kinetic_2005}  for the   construction of the kinetic equation for the dense soliton gas  for the focusing NLS equation, see Section~\ref{sec:kin_nls} below. We note that in a different context,  the collision rate assumption is at heart of the generalised hydrodynamics (GHD),  the theory  for the large-scale dynamics of quantum many-body integrable systems. In GHD the pointlike quasiparticles are subject to the instantaneous velocity-dependent spatial shifts upon colliding (see \cite{castro-alvaredo_emergent_2016}, \cite{bertini_transport_2016}, \cite{doyon_geometric_2018}, \cite{doyon_soliton_2018}, \cite{doyon_lecture_2020}, \cite{spohn_hydrodynamic_2021} and references therein). 
It is important to stress, however,  that  the validity of \eqref{collision_rate} in the context of classical soliton gases, although intuitively suggestive, is far from being obvious.  Indeed, as we have already mentioned,  the very notion of the phase shift in soliton theory is only applicable in the context of the long-time asymptotics, i.e. when solitons have sufficient time to separate from each other  after the interaction, which can only happen in a rarefied gas. The fact that, in a dense gas, where solitons experience significant overlap and continual interaction, the net effect of soliton collisions on the mean velocity is expressed in the same way as in the rarefied gas is  quite remarkable. 
 
Along  with the notion of the phase shift, the phenomenological definition of the DOS $f(\eta)$ introduced above for a rarefied gas also requires a more careful treatment as the procedure of identifying individual solitons within a dense soliton gas is not obvious (as a matter of fact, the dense soliton gas is no longer described by an approximate `soliton lattice' expression \eqref{rar_u}).  This issue can be partially  addressed  by assuming that for any sufficiently broad interval $x \in [x_0, x_0+L]$, $L \gg 1$ the soliton gas can be approximated by an exact $N$-soliton solution and then using the `windowing'  procedure introduced in \cite{congy_soliton_2021} to `release' the solitons contained within this interval and count them when they get sufficiently separated. This procedure will be described  in Section~\ref{sec:ensemble_aver}.
Later, in Section \ref{sec:spectral_theory}  the DOS for dense soliton gases for the KdV and FNLS equations will be introduced in a more mathematically satisfactory way via the thermodynamic limit of the finite-gap spectra of nonlinear multiphase solutions.

\medskip

With all the above caveats, the 
equation of state for a general unidirectional soliton gas can be formulated.  Let the soliton solution of integrable dispersive hydrodynamics be characterised by  the value $\lambda=\lambda_i$ of the discrete spectrum of the Lax operator, and the position shift in the overtaking 
two-soliton collisions  be $\Delta(\la, \mu) = \sgn(\la-\mu)G(\la, \mu)$, where $G(\la, \mu)>0$. Then  we obtain upon using  \eqref{collision_rate},
 \begin{equation}\label{eq:state}
s(\la, x,t)= s_0(\la) + \int_{\Gamma}
G(\la, \mu) f(\mu, x,t)[s(\la,x,t) - s(\mu, x,t)] \rmd \mu
\end{equation}
under the additional assumption $s'(\la)>0$ (to be verified  in concrete cases). Here $s_0(\la)$ is the velocity of a free soliton and  $\Gamma$ is the support of the DOS $f(\la)$. We use the general notation $\lambda$ for the spectral parameter in \eqref{eq:state} with the understanding that it could be some function of the eigenvalue of the Lax operator (as is the case for the KdV equation, cf. \eqref{schr}, \eqref{kdv1sol}).

By re-arranging integral equation \eqref{eq:state} a useful representation  for $s(\la)$ can be obtained  
\begin{equation}\label{speed_trial}
s(\la)=\frac{s_0(\la) - \int_{\Gamma}G (\la, \mu)  f(\mu) s(\mu) \rmd \mu}{1- \int_{\Gamma } G (\la, \mu)  f(\mu) \rmd \mu}.
\end{equation}
Although the equation of state \eqref{eq:state}, as written,  implies that $s(\la)$ is defined  for  $\la \in \Gamma$, its consequence  \eqref{speed_trial} can be used to find the speed $s(\la_1)$ of the `trial' soliton with $\la= \la_1 \notin \Gamma$ propagating through the soliton gas with a given DOS $f(\la)$ and  transport velocity  $s(\la)$,  $\la \in \Gamma$.  This extension can be readily justified  by formally replacing $f(\la) \to f(\la) + w \delta{(\la - \la_1)}$ in \eqref{eq:state} and subsequently letting $w \to 0$.

It is suggested  by \eqref{speed_trial} that the case of soliton gas with $f(\lambda)$, $s(\lambda)$ satisfying 
 \be \label{cond_kdv}
  \int_{\Gamma } G (\la, \mu)  f(\mu) \rmd \mu = 1, \qquad    \int_{\Gamma}G (\la, \mu)  f(\mu) s(\mu) \rmd \mu = s_0(\la)
    \ee
is special.  Indeed, this case corresponds to the peculiar type of soliton gas termed {\it soliton condensate} \cite{el_spectral_2020}. Soliton condensate will be considered in Section~\ref{sec:sol_cond} in the context of the focusing NLS equation.

\subsection{Bidirectional soliton gas}
\label{sec:bidirectional}

In this section, following \cite{congy_soliton_2021}, we will derive the kinetic equation for integrable bidirectional Eulerian
dispersive hydrodynamics~\eqref{eq:disp_Euler} using the phenomenological construction outlined in the previous section.   For convenience, we reproduce system \eqref{eq:disp_Euler} here in a slightly simplified form covering  majority of dispersive hydrodynamic systems arising in applications:
\begin{equation}
\label{eq:disp_Euler1}
\begin{split}
\rho_t + (\rho u)_x &= 0\, , \\
(\rho u)_t + \left ( \rho u^2 + \sigma P(\rho) \right )_x &=
(D[\rho,u])_x \,, \quad \sigma=\pm 1,
\end{split}
\end{equation}
where $D[\rho,u])$ is dispersion operator that gives rise to a real-valued dispersion relation $\omega=\omega_0(k)$ for linearised solutions. Bidirectionality of \eqref{eq:disp_Euler1} implies that
the linear dispersion relation has two branches: $\omega_0^-(k)$ and $\omega_0^+(k)$--- 
corresponding to slow and fast waves respectively, i.e. $\omega_0^-(k)/k < \omega^+_0(k)/k$ in the
long wavelength limit $k \to 0$.  

Suppose that  system~\eqref{eq:disp_Euler1} supports a family of
bidirectional soliton solutions that bifurcate from the two branches
of the linear wave spectrum $\omega=\omega^{\pm}_0(k)$.  We denote the corresponding soliton
families $(\rho^-_{\rm s},u^-_{\rm s})$ and
$(\rho^+_{\rm s},u^+_{\rm s})$.  Let these soliton solutions be
parametrised by a real-valued spectral (IST) parameter $\lambda$ so
that $\lambda \in \Gamma_+ $ for the ``fast'' branch and
$\lambda \in \Gamma_-$ for the ``slow'' branch, where $\Gamma_\pm$ are
simply-connected subsets of $\mathbb{R}$ with one intersection point
at most.  Let the respective soliton velocities be
$c_\pm(\lambda)$. For convenience we assume that
$c_{\pm}'(\lambda)>0$, and $c_-(\lambda_1) \ < c_+(\lambda_2)$ if
$\lambda_1 \in \Gamma_-$ and $\lambda_2 \in \Gamma_+$,
$\lambda_1 \ne \lambda_2$.  If $\Gamma_- \cap \Gamma_+= \{\lambda_*\}$
we assume $c_-(\lambda_*)=c_+(\lambda_*)$. The generalisation to the case $c_{\pm}'(\lambda)<0$,  $c_-(\lambda_1) \ > c_+(\lambda_2)$ will be straightforward.

\begin{figure}[h]
\begin{center}
\includegraphics[scale= .9]{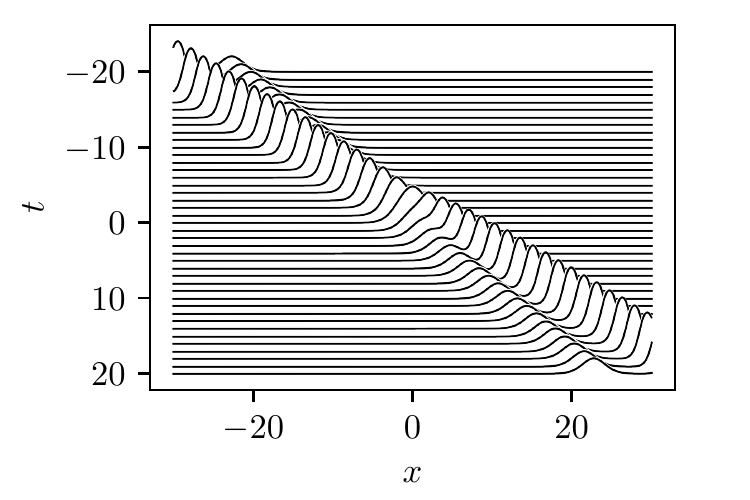} \qquad   \includegraphics[scale= .9]{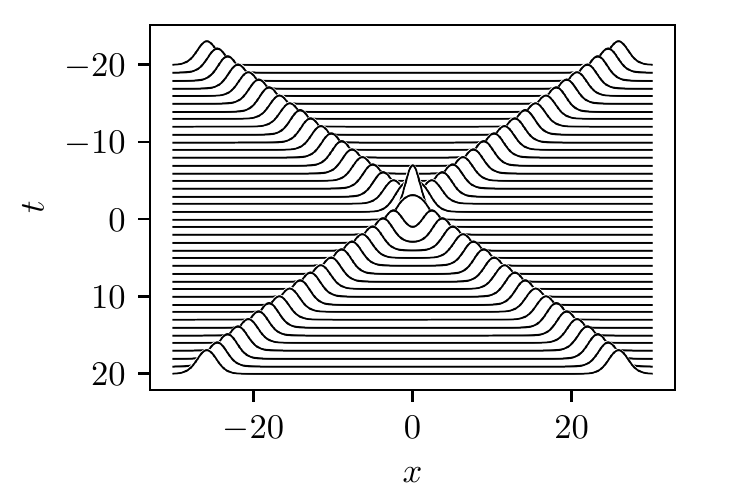}     
\end{center} 
  \caption{Soliton collisions in a bidirectional shallow-water soliton gas: a) overtaking; \ b) head-on. }
  \label{fig:headon_overt}
\end{figure}

One can distinguish between two types of  pairwise collisions in a
bidirectional soliton gas: the overtaking collisions between
solitons belonging to the same spectral branch and characterised by
the position shifts $\Delta_{++}$ and $\Delta_{--}$ respectively, and
the head-on collisions between solitons of different branches,
characterized by the position shifts $\Delta_{+ -}$ and $\Delta_{-+}$.
Let $\lambda \ne \mu$, and $\Delta_{\pm \pm} (\lambda,\mu)$ and
$\Delta_{\pm \mp} (\lambda,\mu)$ denote the position shifts of a
$\lambda$-soliton due to its collision with a $\mu$-soliton, with the
first and the second signs $\pm$ in the subscript indicating the
branch correspondence of the $\lambda$-soliton and the $\mu$-soliton
respectively, e.g. $\Delta_{-+}(\lambda, \mu)$ is the position shift
of a $\lambda$-soliton with $\lambda \in \Gamma_-$ in a collision with
a $\mu$-soliton with $\mu \in \Gamma_+$. The typical wave patterns in the overtaking and head-on
soliton interactions in a bidirectional shallow-water soliton gas are shown in Fig.~\ref{fig:headon_overt}

\subsubsection{Isotropic and anisotropic soliton gases}

We call the bidirectional soliton gas {\it isotropic} if the position
shifts for the overtaking and head-on collisions between $\lambda$-
and $\mu$- solitons satisfy the following sign conditions:
\begin{equation}\label{sign_cond}
\sgn [\Delta_{++}]= \sgn[\Delta_{+-}], \quad  \sgn[\Delta_{--}]=
\sgn[\Delta_{-+}],
\end{equation}
i.e.  the $\lambda$-soliton experiences a shift of a certain sign, say
shift forward (and the $\mu$-soliton--- the shift of an opposite sign)
irrespectively of the type of the collision---overtaking or head-on.
If conditions \eqref{sign_cond} are not satisfied, i.e. the sign of
the phase shift depends on the type of the collision, we shall call
the corresponding soliton gas {\it anisotropic}.  The difference
between the phase shifts in isotropic and anisotropic collisions is illustrated in
Fig.~\ref{fig:phase_shift} using concrete examples.

Following the construction for unidirectional soliton gas outlined in
Section~\ref{sec:unidir_sol_gas}, we now consider bidirectional soliton gases for
integrable Eulerian equations~\eqref{eq:disp_Euler}. We introduce two
separate DOS's $f_-(\lambda,x,t)$ and $f_+(\lambda,x,t)$ for the
populations of solitons whose spectral parameters belong to the
slow ($\Gamma_-$) and fast ($\Gamma_+$) branches of the
spectral set $\Gamma$ respectively.  The isospectrality of integrable
evolution implies now two separate conservation laws:
\begin{equation}
\label{eq:kin2}
(f_-)_t + (s_- f_-)_x = 0,\quad (f_+)_t + (s_+ f_+)_x = 0,
\end{equation}
where $s_-(\lambda,x,t)$ and $s_+(\lambda,x,t)$ are the  transport
velocities  associated with slow and  fast
spectral branches $\Gamma_-$ and $\Gamma_+$
respectively.  

We derive the equations of state for $s_\pm$ by extending  the
 phenomenological approach of  Section~\ref{sec:unidir_sol_gas} based on the collision rate assumption. 
Consider  a 
$\lambda$-soliton from the slow branch, $\lambda \in \Gamma_-$, and
compute its displacement in a gas over a `mesoscopic' time
interval $\rmd t$, sufficiently large to incorporate a large number of
collisions, but sufficiently small to ensure that the spatiotemporal
field $f_\pm(\lambda,x,t)$ is stationary over $\rmd t$ and
homogeneous on a typical spatial scale $c_\pm(\lambda) \rmd t$. Having this in
mind, we drop the $(x,t)$-dependence for convenience.  Each overtaking
collision with a $\mu$-soliton of the same branch, $\mu \in \Gamma_-$, shifts
the $\lambda$-soliton by the distance $\Delta_{--}(\lambda,\mu)$.
Then, invoking the collision rate assumption \eqref{collision_rate} the displacement of the $\lambda$-soliton over the time $\rmd t$ due
to the overtaking collisions with $\mu$-solitons, where  $\mu \in [\mu_0, \mu_0 + \rmd \mu]$,
is given by
$\int_{\Gamma_-} \Delta_{--}(\lambda,\mu) f_-(\mu)
|s_-(\lambda)-s_-(\mu)| \rmd t\, \rmd \mu$. 
Additionally, each head-on collision
with a fast soliton, $\mu \in \Gamma_+$, shifts the slow $\lambda$-soliton with $\lambda \in \Gamma_-$
by $\Delta_{-+}(\lambda,\mu)$, and the resulting displacement after a
time $\rmd t$ is given
$\int_{\Gamma_+}\Delta_{-+}(\lambda,\mu) f_+(\mu)
|s_+(\lambda)-s_-(\mu)| \rmd t\, \rmd\mu$. A similar consideration is applied
to the fast soliton branch, $\lambda \in \Gamma_+$, in the gas.
Equating the total displacements of the slow and fast
$\lambda$-solitons to $s_-(\lambda) \rmd t$ and $s_+(\lambda) \rmd t$
respectively, we obtain the equation of state of a bidirectional gas
in the form of two coupled linear integral equations:
\begin{equation}
\label{eq:s2}
\begin{split}
&s_-(\lambda) = c_-(\lambda) + \int_{\Gamma_-} 
\Delta_{--} (\lambda,\mu) f_-(\mu) |s_-(\lambda)-s_-(\mu)| \rmd \mu+
\int_{\Gamma_+}  \Delta_{-+} (\lambda,\mu) f_+(\mu)
|s_-(\lambda)-s_+(\mu)| \rmd\mu, \\
&s_+(\lambda) = c_+(\lambda) + \int_{\Gamma_+} 
\Delta_{++} (\lambda,\mu) f_+(\mu) |s_+(\lambda)-s_+(\mu)| \rmd \mu +
\int_{\Gamma_-}  \Delta_{+-} (\lambda,\mu) f_-(\mu)
|s_+(\lambda)-s_-(\mu)| \rmd\mu,
\end{split}
\end{equation}
where $\lambda \in \Gamma_-$ for the first equation and
$\lambda \in \Gamma_+$ for the second equation. 

 If the spectral
support $\Gamma = \Gamma_- \cup \Gamma_+ \subset \mathbb{R}$ is a
simply connected set and the gas is isotropic, the distinction between
the fast and slow branches becomes unnecessary and the kinetic
equation \eqref{eq:kin2},\eqref{eq:s2} for bidirectional soliton gas
is naturally reduced to the unidirectional gas equation
\eqref{kin_eq0},\eqref{eq:state} for a single DOS $f(\lambda)$ defined on the
entire set $\Gamma$.  It was shown in \cite{congy_soliton_2021} that the dynamics governed by the kinetic equations
\eqref{kin_eq0},\eqref{eq:state} and \eqref{eq:kin2},\eqref{eq:s2}
are in a very good agreement with the results of direct numerical
simulations of isotropic and anisotropic bidirectional soliton gases
respectively.

\subsubsection{Kinetic equations for  soliton gas in NLS dispersive hydrodynamics}
\label{sec:kin_nls}

\medskip
As a representative  example of bidirectional integrable dispersive hydrodynamics \eqref{eq:disp_Euler1}, we consider the
system
\begin{equation}
\label{eq:NLS_hyd}
\begin{split}
&\rho_t+(\rho u)_x =0,\\
&(\rho u)_t+\left(\rho u^2 + \sigma \frac{\rho^2}{2} \right)_x =
\frac{\mu}{4} \left[ \rho \left ( \ln{\rho} \right )_{xx}
\right]_x \, , \quad \sigma = \pm 1, \quad \mu = \pm 1.
\end{split}
\end{equation}
For $\mu =1$,  system \eqref{eq:NLS_hyd} is equivalent to the
NLS equation:
\begin{equation}
\label{eq:NLS}
i \psi_t + \tfrac12 \psi_{xx}-\sigma |\psi|^2 \psi=0, 
\end{equation}
with the mapping between the two representations being realised by the so-called Madelung transform:
\begin{equation}\label{madelung}
\psi=\sqrt{\rho}
\exp \left(i \phi \right), \quad \phi_x=u.
\end{equation}
The case $\sigma=+1$ in \eqref{eq:NLS} corresponds to the defocusing NLS  equation
describing the propagation of light beams through optical fibres in the
regime of normal dispersion, as well as nonlinear matter waves in
quasi-1D repulsive Bose-Einstein condensates, see for instance
\cite{yang_nonlinear_2010}.  
If $\sigma=-1$, equation \eqref{eq:NLS} is the focusing NLS  equation, which is a canonical model for the description of modulationally unstable wave systems
such as deep water waves or the propagation of light in optical fibres in the regime of anomalous dispersion.

For $\sigma=+1$, $\mu =-1$ system \eqref{eq:NLS} is equivalent to the NLS equation in a `quantum potential' also known as the resonant NLS  equation \cite{pashaev_resonance_2002}, \cite{lee_solitons_2007},
\begin{equation}
\label{eq:RNLS}
i \psi_t + \tfrac12 \psi_{xx}-|\psi|^2 \psi=|\psi|_{xx} \psi/|\psi|,
\end{equation}
This equation, in particular,
describes long magneto-acoustic waves in a cold plasma propagating
across the magnetic field \cite{gurevich_origin_1988}. It is also directly related  to the Kaup-Boussinesq  system, an integrable model for bidirectional shallow water waves \cite{kaup_higher_1975} (see \cite{congy_soliton_2021} for the transformation between the resonant NLS and the Kaup-Boussinesq system).

We now look at the soliton gas descriptions for each of the above NLS systems
\bigskip

{\it (i) Defocusing NLS equation}

\medskip
The inverse scattering theory of the defocusing NLS equation was constructed in ~\cite{zakharov_interaction_1973}.   It was shown that the defocusing NLS equation supports a family of dark
(or grey) soliton solutions  on the finite background, which, up to the initial position and phase, can be most conveniently represented in terms of the hydrodynamic variables $\rho, u$ as
\begin{equation}
\label{eq:dark}
\rho_{\rm s}^\pm = 1- (1-\lambda^2){\rm sech}^2
\big[\sqrt{1-\lambda^2}(x- c_\pm t) \big],\quad u_{\rm s}^\pm =
\lambda\left(1-\frac{1}{\rho_{\rm s}^\pm(x,t)} \right),\quad c_\pm =
\lambda \in \Gamma_\pm,
\end{equation}
where $\lambda \in \mathbb{R}$ is the discrete spectral parameter in the linear scattering problem associated with the defocusing NLS equation \cite{zakharov_interaction_1973} , $\Gamma_- = (-1,0]$ for the slow soliton branch and
$\Gamma_+=[0,+1)$ for the fast soliton branch; note that solutions
$(\rho_{\rm s}^+,u_{\rm s}^+)$ and $(\rho_{\rm s}^-,u_{\rm s}^-)$ have
the same analytical expression. Also, despite the same analytical expression for the soliton velocities $c_+$ and $c_-$ in the defocusing NLS case we keep the  the formal notational distinction to  retain the connection with general 
 equation of state \eqref{eq:s2} for bidirectional gas, where the expressions for $c_+(\lambda)$ and $c_-(\lambda)$ can be, in principle, different.  Additionally, without loss of generality we assumed in \eqref{eq:dark} the unit density background. 
 
\begin{figure}[h]
\begin{subfigure}{0.5\textwidth}
\centering
\hspace*{-10mm}
\includegraphics{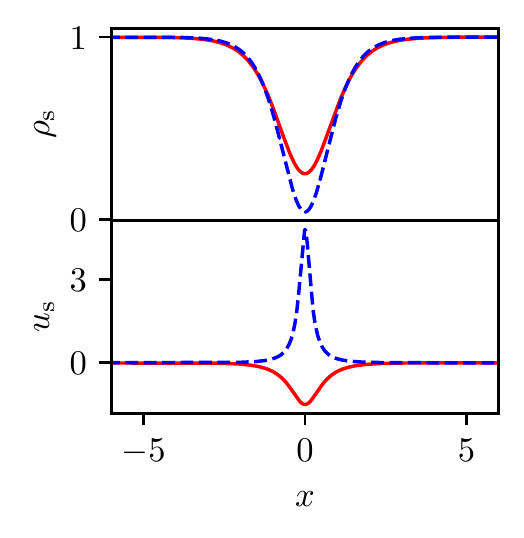}
{\caption{Dark soliton
solutions of the  defocusing NLS \\ equation~\eqref{eq:dark}: $\lambda =
+0.5,-0.2$.}}
\label{fig:solitona}
\end{subfigure}\hfill
\begin{subfigure}{0.5\textwidth}
\centering
\hspace*{-10mm}{
\includegraphics{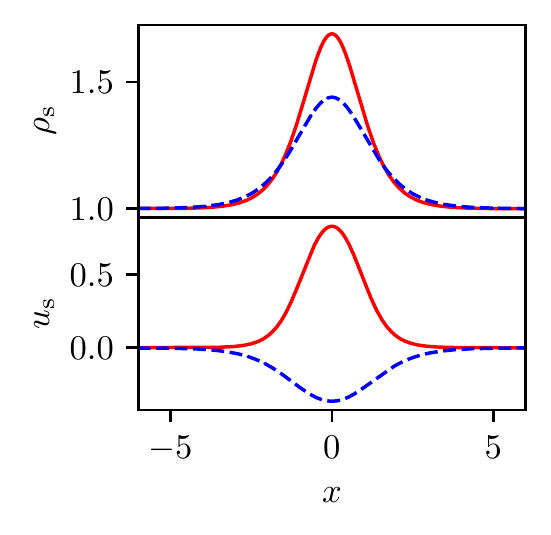}
\caption{Anti-dark soliton solutions of the resonant NLS \\
equation~\eqref{eq:sol_RNLS}: $\lambda = +1.3,-1.2$.}}
\label{fig:solitonb}
\end{subfigure}\hfill
\caption{Soliton solutions: solid lines correspond to the fast branch
solutions~$(\rho_{\rm s}^+, u_{\rm s}^+)$ and dashed lines to the slow
branch solutions~$(\rho_{\rm s}^-, u_{\rm s}^-)$.}
\label{fig:soliton}
\end{figure}

 Typical dark soliton solutions for $\rho$ and $u$ are displayed in
Fig.~\ref{fig:soliton}a.  The position shifts in the defocusing NLS overtaking
and head-on soliton collisions are given by the same analytical
expression
$\Delta_{\pm\pm} (\lambda, \mu)= \Delta_{\pm\mp} (\lambda, \mu) \equiv
\Delta(\lambda, \mu)$, where
\begin{equation}
\label{eq:Delta_DNLS}
\Delta(\lambda,\mu) = \frac{\sgn(\lambda-\mu)}{2 \sqrt{1-\lambda^2}}
\ln \frac{(\lambda-\mu)^2 + \big(\sqrt{1-\lambda^2}+\sqrt{1-\mu^2}
\big)^2} {(\lambda-\mu)^2 + \big(\sqrt{1-\lambda^2}-\sqrt{1-\mu^2}
\big)^2} \equiv \sgn(\lambda-\mu) G_1(\lambda, \mu),
\end{equation}
for all $\lambda, \mu \in (-1, 1)$. Expressions \eqref{eq:Delta_DNLS}  were obtained  in \cite{zakharov_interaction_1973}. Note that, unlike  in the KdV equation, one needs to distinguish between the position and phase shifts for soliton collisions in the NLS dispersive hydrodynamics. The expressions for the phase shifts can  also be found in \cite{zakharov_interaction_1973} but we do not consider them here. 

\begin{figure}[h]
\begin{subfigure}{0.4\textwidth}
\centering
\includegraphics{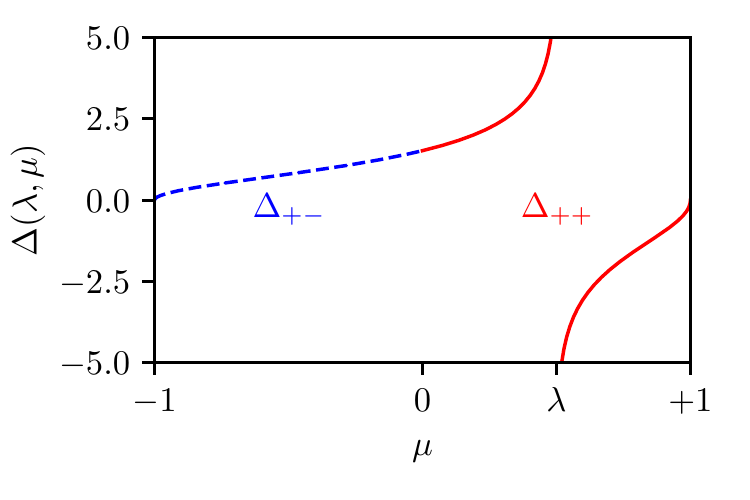}
\caption{Defocusing NLS ($\lambda, \mu$)-interaction with $\lambda=1/2$.}
\label{fig:phase_shifta}
\end{subfigure}\hfill
\begin{subfigure}{0.49\textwidth}
\centering
\vspace*{-1mm}
\includegraphics{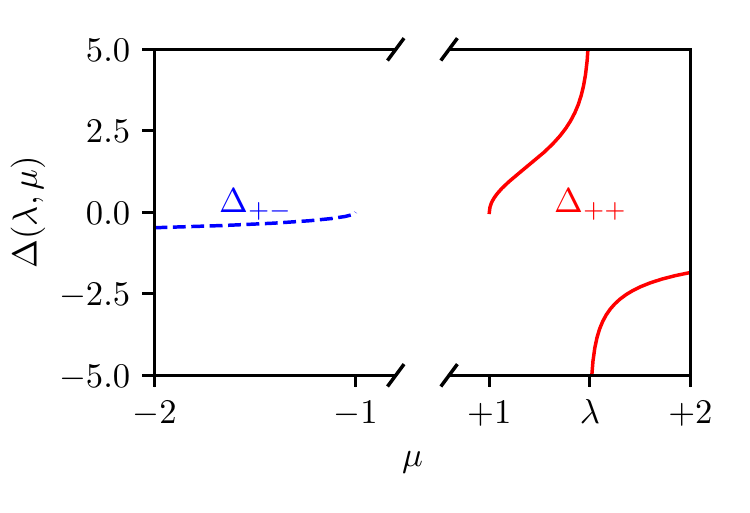}
\caption{Resonant NLS ($\lambda, \mu$)-interaction with $\lambda=3/2$.}
\label{fig:phase_shiftb}
\end{subfigure}
\caption{Variation of the phase shifts  in the isotropic (a) and anisotropic (b) interactions 
of solitons with spectral parameters $\lambda$ and $\mu$; $\lambda$ is fixed and  belongs to the ``fast'' spectral branch $\Gamma_+$.}
\label{fig:phase_shift}
\end{figure}

One can readily verify that the soliton
position shifts given by \eqref{eq:Delta_DNLS} satisfy the isotropy
conditions \eqref{sign_cond}. The variation of $\Delta(\lambda,\mu)$
with respect to $\mu$ for a fixed $\lambda$ is displayed in
Fig.~\ref{fig:phase_shifta}. One can see that the position shifts for
the head-on and overtaking collisions lie on the same curve with
$\Delta(\lambda, \mu)$ being continuous at $\lambda=0$, the point of
intersection of $\Gamma_-$ and $\Gamma_+$.  Due to the isotropic
nature of the defocusing NLS soliton interactions the coupled kinetic
equation~\eqref{eq:kin2},\eqref{eq:s2} for the bidirectional defocusing NLS gas
reduces to the single spectral transport equation~\eqref{kin_eq0} with the
equation of state 
\begin{equation}
\label{eq:s_DNLS}
s(\lambda,x,t)=\lambda + \int \limits_{-1}^{+1} G_1(\lambda,
\mu)f(\mu,x,t)[s(\lambda,x,t)- s(\mu,x,t)] \rmd \mu , \quad \lambda \in
(-1,1).
\end{equation}
One can verify by direct computation that the condition $s'(\lambda)>0$  necessary  for the validity of \eqref{eq:s_DNLS} is satisfied. 

\medskip
{\it (ii) Resonant NLS equation}

\medskip
The resonant NLS equation \eqref{eq:RNLS}  is reducible to the well-known integrable Kaup-Boussinesq system  for shallow water waves for which the inverse spectral theory  was constructed in \cite{kaup_higher_1975}.  It is not difficult to show  that the resonant NLS  equation has a family of spectral anti-dark soliton solutions given by~\cite{lee_solitons_2007} 
\begin{equation}
\label{eq:sol_RNLS}
\rho_{\rm s}^\pm = 1+ (\lambda^2-1){\rm sech}^2
\big[\sqrt{\lambda^2-1}(x-c_\pm
t) \big],\quad
u_{\rm s}^\pm = \lambda\left(1-\frac{1}{\rho_{\rm s}^\pm(x,t)} \right),\quad c_\pm =
\lambda \in \Gamma_\pm;
\end{equation}
 Typical anti-dark soliton
solutions of the resonant NLS are displayed in Fig.~\ref{fig:soliton}b.
In contrast with the defocusing NLS system, the admissible spectral set $\Gamma$ of the
resonant NLS solitons is spanned by two disconnected subsets: $\Gamma_- =
(-\infty,-1]$ for slow solitons and $\Gamma_+ = [+1,+\infty)$. The position shifts in the head-on and overtaking
collisions of the resonant NLS solitons are given by the same analytical expression $ \Delta_{\pm
\pm}(\lambda, \mu)= \Delta_{\pm \mp}(\lambda, \mu) \equiv
\Delta(\lambda, \mu)$, where
\begin{equation}
\label{eq:DeltaRNLS}
\Delta(\lambda, \mu)= \frac{\sgn(\lambda-\mu)}{2
\sqrt{\lambda^2-1}} \ln \frac{(\lambda-\mu)^2 -
\big(\sqrt{\lambda^2-1}+\sqrt{\mu^2-1} \big)^2} {(\lambda-\mu)^2 -
\big(\sqrt{\lambda^2-1}-\sqrt{\mu^2-1} \big)^2} \equiv
\sgn(\lambda-\mu)
G_2(\lambda, \mu).
\end{equation}
However, now one can verify that, unlike in the defocusing NLS case, the isotropy
condition \eqref{sign_cond} is not satisfied. Indeed, it follows from
\eqref{eq:DeltaRNLS} that
$\sgn[\Delta_{\pm \pm} (\lambda,\mu)] = \sgn(\lambda -\mu)$, whereas
$\sgn[\Delta_{\pm \mp} (\lambda,\mu)] = -\sgn(\lambda -\mu)$, that is
in a head-on collision between a $\lambda$-soliton and a $\mu$-soliton
with $\lambda>\mu$, the $\lambda$-soliton's position is now shifted
backwards.  The variation of $\Delta_{\pm \pm}(\lambda, \mu)$ for the
resonant NLS equation is shown in Fig.~\ref{fig:phase_shiftb}. One can see that it is
qualitatively different from the variation of
$\Delta_{\pm \mp}(\lambda, \mu)$ for the defocusing NLS equation
(Fig.~\ref{fig:phase_shifta}).

The kinetic equation for the anisotropic resonant NLS soliton gas then takes the
form of two continuity equations~\eqref{eq:kin2} complemented by the
coupled equations of state
\begin{equation}
\label{eq:s_RNLS}
\begin{split}
&s_-(\lambda) = \lambda + \int \limits_{-\infty}^{-1} G_2(\lambda,
\mu) f_-(\mu) (s_-(\lambda)-s_-(\mu)) \rmd \mu + \int
\limits_{+1}^{\infty} G_2(\lambda, \mu) f_+(\mu)
(s_-(\lambda)-s_+(\mu)) \rmd\mu,\\
&s_+(\lambda) = \lambda + \int \limits_{+1}^{+\infty} G_2(\lambda,
\mu) f_+(\mu) (s_+(\lambda)-s_+(\mu)) \rmd \mu + \int
\limits_{-\infty}^{-1} G_2(\lambda, \mu) f_-(\mu)
(s_+(\lambda)-s_-(\mu)) \rmd\mu,
\end{split}
\end{equation}
assuming that  $s_\pm'(\lambda)>0$ and $s_+>s_-$ (verified by direct computation).

\medskip
{\it (iii) Focusing NLS equation}

\medskip
The case of the focusing NLS equation (equation \eqref{eq:NLS} with $\sigma=-1$) is special as it represents an example of the {\it focusing} dispersive hydrodynamics, where the long-wave, `hydrodynamic' motion is described by an elliptic system.   The focusing NLS equation is a canonical model for the description of modulationally unstable systems in fluid dynamics and nonlinear optics.   Nevertheless, the focusing NLS equation supports stable soliton solutions that can propagate in both directions so the focusing NLS soliton gas  should be  classified as bidirectional.

We  consider the focusing NLS equation  in the following normalisation: 
\be\label{eq:fNLS}
i \psi_t + \psi_{xx} +2 |\psi|^2 \psi=0, 
\ee
which is standard in the context of the  IST analysis.  The IST for the focusing NLS equation was introduced in the celebrated paper of Zakharov and Shabat  \cite{Zakharov:72}, where it was shown that
the spectral problem associated with   \eqref{eq:fNLS} has the form
\begin{equation}\label{ZS}
\mathcal{L}Y = \lambda Y, \qquad  \mathcal{L} =      \begin{pmatrix}
-i \partial_x & \psi \\ \overline{\psi} & i \partial_x \\
\end{pmatrix},
   \end{equation}
where  $\lambda$ is the  spectral parameter and $Y=Y(x,t, \lambda)$ is a vector; $\overline \psi$ denotes complex conjugate.  The Lax operator $\mathcal{L}$ in \eqref{ZS} is often called the Zakharov-Shabat  operator. The main difference of the Zakharov-Shabat operator from the Lax operators for the KdV, defocusing and resonant NLS equations is that it is not self-adjoint, implying that the spectral parameter $\lambda$ is  complex, $\lambda \in \mathbb{C}$.

 A single-soliton solution of equation \eqref{eq:fNLS} is characterised by a discrete complex eigenvalue,  $\lambda_1 =a+ib$  and c.c., of the Zakharov-Shabat operator and is given by
\begin{equation}\label{fnls_soliton}
\psi_S (x,t)= 2ib \, \hbox{sech}[2b(x+4at-x_0)]e^{-2i(ax + 2(a^2-b^2)t)+i\phi_0},
   \end{equation}
where $x_0$ is the initial position of the soliton and $\phi_0$ is the initial phase.  One can see that  the focusing NLS soliton represents a localised wavepacket with the envelope propagating with the group velocity  $c_g=-4 a= -4 \hbox{Re} \lambda_1$ and the carrier wave having the phase velocity $c_p=(b^2-a^2)/a = -2\hbox{Re} (\lambda_1^2)/\hbox{Re} \lambda_1$. Also, similar to the defocusing and resonant NLS  equations, and in contrast with KdV equation, the amplitude and velocity of the focusing NLS soliton 
are two independent parameters.  We identify the two  families ($\pm$) of focusing NLS solitons according to the sign of their group velocity, $\hbox{sgn} ( c_g) = - \hbox{sgn}  (\hbox{Re} \lambda)$.

 Similar to other integrable NLS models, the solitons of the focusing NLS equation interact pairwise and experience both position and phase shifts upon the interaction.  The position shifts in the focusing NLS overtaking
and head-on soliton collisions  are given by the same expression, $\Delta_{\pm\pm} (\lambda, \mu)= \Delta_{\pm\mp} (\lambda, \mu) \equiv
\Delta(\lambda, \mu)$, where \cite{Zakharov:72}
\be \label{fnls_phase_shift}
\Delta(\lambda, \mu) = \frac{\hbox{sgn} [\hbox{Re} (\mu - \lambda)]}{ \hbox{Im} \lambda}\ln \left|\frac{\mu-\bar{\lambda}}{\mu-\lambda}\right|. 
\ee
One can see that the 
position shifts  \eqref{fnls_phase_shift} satisfy 
conditions \eqref{sign_cond} so we classify the focusing NLS soliton collisions as isotropic.

The DOS $f(\lambda)$ of the focusing NLS soliton gas is generally supported on some compact Schwarz symmetric 2D set $\Lambda \subset \mathbb{C}$ so it is sufficient to consider only the upper half plane part $\Lambda^+$. Here Schwarz symmetry means  that if $\lambda \in \mathbb{C}$
 is  a  point of the spectrum then so is the c.c. point $ \bar{\lambda}$. The kinetic equation for the focusing NLS soliton gas then assumes the form \cite{el_kinetic_2005}

\begin{equation}\label{FNLS_kin}
\begin{split}
&f_t + (fs)_x= 0, \\
&s(\lambda, x, t) = -4 \hbox{Re} \lambda + \frac{1}{ \hbox{Im} \lambda}   \iint \limits_{\Lambda^+}  \ln \left|\frac{\mu- \bar{\lambda}}{\mu-\lambda}\right| [s(\lambda, x, t) - s(\mu, x, t)] f(\mu, x, t) \rmd \xi \rmd\zeta,
\end{split}
\end{equation}
where $\mu = \xi +i \zeta$ and $\Lambda^+ \subset \mathbb{C}^+ \setminus i \mathbb{R}^+$.

The case $\Lambda^+ \subset i \mathbb{R}^+$ requires a separate consideration. Solitons of the focusing NLS equation can form stationary complexes described by  the special $N$-soliton solutions, called bound states, for which all discrete spectrum points are located on the imaginary axis \cite{Zakharov:72}. Since for the corresponding bound state soliton gas  $\hbox{Re} \lambda =0$ the equation of state in \eqref{FNLS_kin} immediately yields $s(\lambda) = 0$ resulting in the equilibrium DOS, $f_t=0$.   It was shown  in \cite{gelash_bound_2019} that  the  turbulent wave field $\psi(x,t)$ associated with the DOS $f(\lambda)$ in a dense bound state  soliton gas 
exhibits very peculiar  statistical properties, shedding light on the fundamental phenomenon of spontaneous modulational instability. This special soliton gas  will be considered in Section~\ref{sec:sol_cond}.

\subsection{Ensemble averages and modulation equations for soliton turbulence}
\label{sec:ensemble_aver}

Within the spectral kinetic description  of soliton gas the solitons are viewed as quasiparticles moving with the speeds determined by the nonlocal equation of state \eqref{eq:state}. The DOS $f(\la)$ in this description represents a comprehensive spectral characteristics of soliton gas. However, of ultimate interest in dispersive hydrodynamics is the description of the turbulent nonlinear wave field $u(x,t)$ associated with the underlying spectral soliton gas dynamics. 
A natural question arises then:  how to solve the inverse scattering problem i.e. how to `translate' the spectral characterisation of soliton gas (i.e. the DOS) into the statistical characterisation (ensemble averages, probability density function, correlations, etc.) of the associated nonlinear random wave field $u(x,t)$ satisfying an integrable  dispersive hydrodynamic equation, \eg the KdV equation?

Within the classical, deterministic IST setting the inverse spectral  problem is solved via the Gelfand-Levitan-Marchenko equation (see \eg \cite{drazin_solitons_1987}, \cite{ablowitz_solitons_1981}, \cite{novikov_theory_1984}) or utilising the Riemann-Hilbert problem approach (see \eg \cite{faddeev_hamiltonian_2007}). While the construction of a comprehensive  extension of the IST  for random potentials seems to be far away at present  (see \eg \cite{kotani_kdv_2008}, \cite{dyachenko_primitive_2016} for some of the important developments), some valuable quantitative results can be obtained by elementary means. We shall describe some of these results using the KdV equation as the simplest accessible example, the generalisation to other integrable unidirectional and bidirectional dispersive hydrodynamic equations being straightforward.

As is well known (see \eg \cite{monin_statistical_2011}, \cite{nazarenko_wave_2011}) a turbulent wave field is  characterised by the moments  $\langle u^n \rangle$ over the statistical ensemble, which, assuming ergodicity, can be computed as spatial averages $\overline{u^n} = \frac{1}{\ell}\int ^\ell_0 u^n(\tilde x, t)d \tilde x$, over a sufficiently large `mesoscopic' interval $l \ll \ell  \ll L$, where $l$ is the typical scale for spatial variations of $u(x,t)$ in a soliton gas, i.e. the typical soliton width, and $L$ is the typical scale for  spatial variations of the density of states $f(\la, x,t)$ in the non-uniform soliton gas. 
\begin{figure}[h]
\begin{subfigure}{0.3\textwidth}
\centering
\includegraphics[scale=0.7]{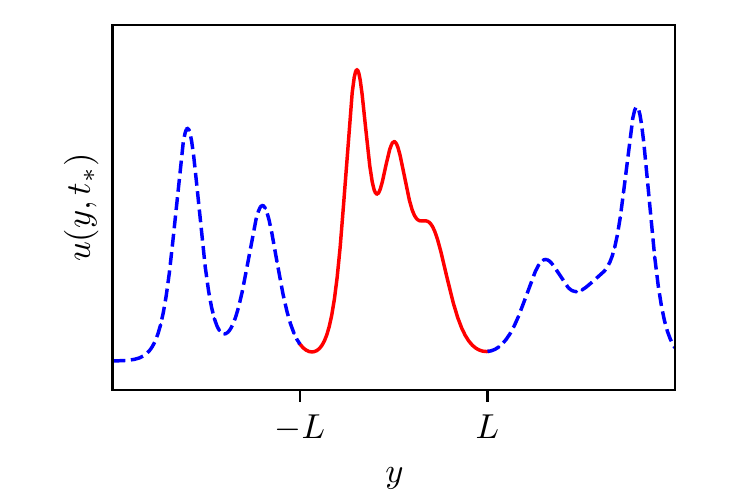}
\caption{Typical distribution $u(y,t_*)$ for a KdV
soliton gas.}
\label{fig:trunca}
\end{subfigure}\hfill
\begin{subfigure}{0.3\textwidth}
\centering
\includegraphics[scale=0.7]{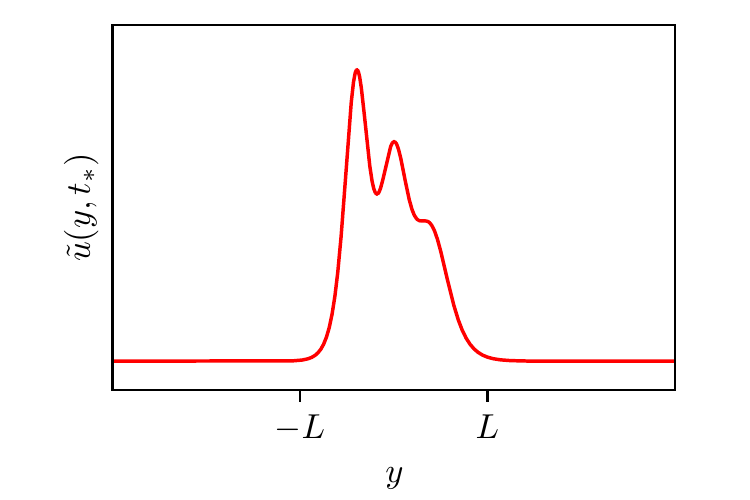}
\caption{Truncation of the distribution
$u(y,t_*)$ for $y \in (-L,L)$.}
\label{fig:truncb}
\end{subfigure}\hfill
\begin{subfigure}{0.3\textwidth}
\centering
\includegraphics[scale=0.7]{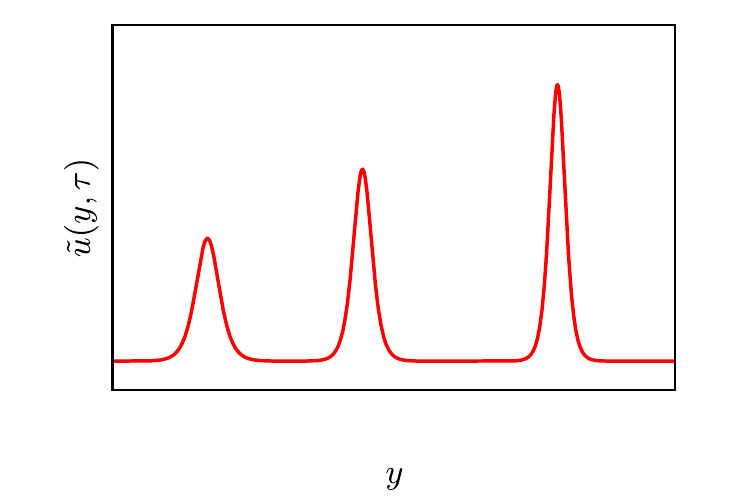}
\caption{Variation of the truncated distribution $\tilde u(y,\tau)$
at time $\tau \gg t_*$.}
\label{fig:truncc}
\end{subfigure}
\caption{Schematic for the evaluation of the integral~\eqref{cons_I}
in soliton gas using the `windowing' procedure.}
\label{fig:trunc}
\end{figure}
One of the fundamental properties of integrable dispersive hydrodynamics  is the availability of 
an infinite set of local conservation laws 
\be\label{cons_law}
(P_i)_t+(Q_i)_x=0, \ i=1, 2, \dots, 
\ee
where the  $P_i$ and $Q_i$ are functions of the field variable $u$ and its derivatives. For decaying initial data, the integrals $\int_{-\infty}^\infty P_i \rmd x$
are conserved in time.
For the KdV equation the existence of an infinite series of local polynomial conservation laws was established in \cite{miura_kortewegvries_1968}. Their conserved densities sometimes called  Kruskal integrals  can be deduced from the Lax pair via appropriate expansions for large values of the spectral parameter, see e.g.  \cite{novikov_theory_1984},  \cite{ablowitz_nonlinear_2011}. E.g. for KdV \eqref{KdV}  $P_1=u$, $P_2=u^2$, $P_3=\frac{u_x^2}{2} + u^3$.
Here we
describe a simple heuristic approach proposed in \cite{congy_soliton_2021} that enables one to link the
spectral DOS $f(\la)$ of a soliton gas with
the ensemble averages of $P_i$ of the integrable dispersive hydrodynamic 
system~\eqref{eq:scalar_dh}. We describe the general principle using the KdV equation as a prototype example.

Consider a uniform KdV soliton gas, i.e. a gas whose
statistical properties, particularly the DOS $f(\eta)$, do not depend on
$x,t$ (we return to the conventional use of the spectral variable $\eta = \sqrt{-\la}$  in the context of the KdV solitons, see  Section~\ref{sec:unidir_sol_gas}). We now make a natural assumption  that
the nonlinear wave field in a homogeneous soliton gas represents an
ergodic random process, both in $x$ and $t$ (we note in passing that
ergodicity is inherent in the spectral model of soliton gas that will be presented in Section~\ref{sec:spectral_theory}). The ergodicity
property implies that the ensemble-averages $\langle P_i[u] \rangle$
 in the
soliton gas can be replaced by the corresponding spatial
averages. Generally, for any functional $H[u(x,t)] $ we
have
\begin{equation}
\label{eq:psi2b_def}
\langle H[ u] \rangle =  \lim \limits_{L \to \infty} \frac{1}{2L}
\int\limits_{x-L}^{x+L} H[u(y,t)] \rmd y.
\end{equation}
for a single representative realization of soliton gas.  
We define
\begin{equation}\label{cons_I}
I_j = \int\limits_{x-L}^{x+L} P_j[u (y,t)] \rmd y, \quad j=1, 2, \dots,
\end{equation}
where $P_j[u]$ is the conserved density and $L \gg 1$.  Then
$\langle P_j[u] \rangle =  I_1/(2L) + \mathcal{O}(L^{-1})$.

Let $u(y,t)$ be a representative realization of a soliton gas and let
$\tilde u(y,t)$ be defined in such a way that for
some $t=t_*$ one has
$(\tilde u(y,t_*) = u(y,t_*)$ for
$y \in (x-L, x+L)$ and
$(\tilde u(y,t_*)= 0$ outside of this
interval so that the conserved densities $P_i[\tilde u(y,t_*)]$ and respective fluxes $Q_i[\tilde u(y,t_*)]$ also vanish outside $(x-L, x+L)$. To avoid complications we assume that the transition between
the two behaviors is smooth but sufficiently rapid so that such a
`windowed' portion of a soliton gas (see
Fig.~\ref{fig:trunca},\ref{fig:truncb}) can be approximated by the spectral
$N$-soliton solution  for some $N \gg 1$, with
the discrete IST spectrum points $\eta_i, i=1, \dots N$ being distributed on  $\Gamma=[0,1]$ with density
$\varphi(\eta) \approx 2Lf (\eta)$ (recall the definition of DOS in Sec.~\ref{sec:intro}).

The integrals \eqref{cons_I} then can be re-written as
\begin{equation}
\label{eq:IJ}
I_j = \int\limits_{-\infty}^{+\infty}  P_j[\tilde u(y,t)] 
\rmd y.
\end{equation}
Since the integrals \eqref{eq:IJ} do not depend on time they can be computed  at
 $t=\tau \gg t_*$ when the `windowed' solution $\tilde u(x,t)$ gets resolved into a  train of $N$ well-separated solitons $u_s(x,t; \eta_i), i=1, \dots, N$,  see Fig.~\ref{fig:truncc}. Assuming that the overlap between (exponentially decaying) solitons in the train becomes negligible as $t \to \infty$ the integral \eqref{eq:IJ} can be represented as a sum of integrals over the individual soliton solutions,
\begin{equation}
\label{eq:IJ1}
I_j = \sum \limits_{j=1}^N m_j(\eta), \quad m_j= \int\limits_{-\infty}^{+\infty}  P_j[u_s(y,t; \eta_i)] \rmd y .
\end{equation}
E.g. for the KdV equation  the two first integrals in \eqref{eq:IJ1} are readily evaluated 
using the one-soliton solution \eqref{kdv1sol} to give:
\begin{equation}
\label{eq:rhob_s}
m_1(\eta) = \int\limits_{-\infty}^{+\infty}
u_s(y,t; \eta) \rmd y = 4 \eta, \qquad  m_2(\eta)  = \int\limits_{-\infty}^{+\infty}
u_s^2(y,t; \eta) \rmd y = \frac{16}{3} \eta^3.
\end{equation}
These are  the `spectral mass' and `spectral momentum' of the $\eta$-soliton respectively.
Now, using \eqref{kdv1sol} and  taking the continuous limit as $N \to \infty$ in \eqref{eq:IJ1}  according to 
$\sum_i F(\eta_i)\to \int_{\Gamma} F(\eta) \varphi(\eta) \rmd \eta \, $,  we obtain the expressions for the first two statistical moments in the KdV soliton turbulence  defined according to \eqref{eq:psi2b_def}:
\begin{equation}\label{moments_KdV}
\langle {u} \rangle \ =\ 4\int_{0}^{1}\eta f(\eta )\,\ud\eta \,, \qquad
\langle {u^{2}} \rangle \ =\ \dfrac{16}{3}\int_{0}^{1}\eta ^{3}f(\eta)\,\ud\eta \, .
\end{equation}

One fundamental restriction imposed on the distribution function $f(\eta)$ follows from non-negativity of the variance
\begin{equation}\label{var}
  \mathcal{A}^2\ =\ \langle {u^2} \rangle \ -\ \langle {u} \rangle ^2\ \geqslant\ 0.
\end{equation}
Some consequences of this restriction have been explored in \cite{el_critical_2016}.  In particular, considering the `cold' soliton gas characterised by the Dirac delta-function DOS, $f(\eta)=w \delta(\eta - \eta_0)$ we obtain from \eqref{moments_KdV} $\langle {u} \rangle = 4 \eta_0 w$, 
$\langle {u^{2}} \rangle = \frac{16}{3} (\eta_0)^3 w$. Then the condition \eqref{var} yields the constraint
\be \label{crit_kdv}
w \le \frac{\eta_0}{3}.
\ee

We note that the method presented here only requires one to integrate the
single-soliton solution and thus can be readily applied to any
integrable dispersive hydrodynamic system supporting the soliton
resolution scenario.   In particular, the ensemble averages for the density $\langle \rho \rangle$, velocity 
$\langle u \rangle$ and momentum $\langle \rho u \rangle$ in the bidirectional soliton gases for the  defocusing and resonant NLS  equations were obtained in 
\cite{congy_soliton_2021}. Next, we note that the above simple derivation implies that 
the  expressions \eqref{moments_KdV} apply to both dense and rarefied gas. Indeed, the same expressions  were obtained in~\cite{dutykh_numerical_2014}
for a rarefied KdV gas (see also \cite{shurgalina_nonlinear_2016} for the
similar modified KdV equation results). In Section~\ref{sec:spectral_theory} we will show how these expressions follow from the general spectral construction of soliton gas, not invoking the heuristic `windowing' procedure.

In the above consideration of homogeneous soliton gases the ensemble
averages \eqref{eq:psi2b_def} are constant. For a non-uniform (non-equilibrium) gas
the DOS is a slowly varying function of $x,t$ and so are the ensemble
averages that now need to be interpreted as ``local averages'' in the
spirit of modulation theory \cite{whitham_linear_1999}.  Essentially,
one invokes  the ``mesoscopic'' scale $\ell$: $l \ll \ell \ll L$ ---
so that the DOS is approximately constant on any
interval $(x-\ell, x+\ell)$. Then the constant ensemble averages
\eqref{eq:psi2b_def} are replaced by the slowly varying quantities:
\begin{equation}
\label{eq:ell}
\langle H[u] \rangle_\ell (x,t)= \frac{1}{2\ell} \int
\limits_{x-\ell}^{x+\ell} H[u(y,t)] \rmd y.
\end{equation}
The `local' averages $\langle H[u] \rangle_\ell$ do not depend on
$\ell$ at leading order, and their spatiotemporal variations occur on
$x,t$-scales that correspond to the scales associated with variations
of $f(\eta)$ and are much larger than typical (fast) variation scales of the nonlinear wave field $u(x,t)$.  Using the scale separation and applying the ensemble averaging to the infinite set of conservation laws for integrable dispersive hydrodynamics we obtain the modulation system for soliton gas
\begin{equation}\label{stoch_Whitham}
\frac{\partial }{\partial t} \langle P_i[u ] \rangle + \frac{\partial }{\partial x} \langle Q_i[u]  \rangle = 0, \quad
i=1, 2, \dots
\end{equation}
Equations \eqref{stoch_Whitham} can be viewed as a stochastic version of the Whitham modulation equations \cite{whitham_linear_1999} first discussed in \cite{lax_zero_1991}. 
Equations \eqref{stoch_Whitham}  can also be viewed as the integrable turbulence counterpart of  the moment equations in the classical hydrodynamic  turbulence theory \cite{monin_statistical_2011}. As a matter of fact the infinite-component stochastic modulation system \eqref{stoch_Whitham} is consistent with the integro-differential kinetic equation \eqref{kin_eq0}, \eqref{eq_state_kdv}. We note that the local ensemble averages of the infinite set of conserved densities  $\langle P_i[u ] \rangle(x,t)$ satisfying system \eqref{stoch_Whitham} can be evaluated in terms of the spectral moments of the DOS: $\int_0^1 \eta^{2n+1} f(\eta, x, t) \rmd \eta$, $n=0,1,2, \dots$, where $f(\eta, x,t)$ is a relevant
solution  of the kinetic equation
\eqref{kin_eq0},\eqref{eq_state_kdv}. 

\section{Nonlinear spectral theory of soliton gas}
\label{sec:spectral_theory}

The phenomenological kinetic theory of soliton gas described  in the previous section is essentially based on the interpretation of  solitons as quasi-particles experiencing short-range pairwise interactions accompanied by the well-defined phase/position shifts. As  was already stressed, although this theoretical framework is justifiable in the case  of a rarefied gas, it is not quite satisfactory for a dense gas where solitons experience significant overlap and continual nonlinear interactions so that they could become indistinguishable as separate entities.
Clearly it would be desirable to have a  more mathematically robust approach  to the description of a general, dense soliton gas that would, in particular, provide a formal justification of the collision rate assumption \eqref{collision_rate} and, consequently, of the equation of state \eqref{eq:state}, at least in some concrete cases of integrable dispersive hydrodynamics. 

The above discussion strongly suggests that we need to look at the wave side of the soliton's `dual identity'.  As we already mentioned, there are insightful parallels between kinetic theory of soliton gas and the nonlinear wave modulation theory introduced by G.B. Whitham in 1965 \cite{whitham_non-linear_1965},
\cite{whitham_linear_1999}, just before the discovery of the IST method. The Whitham theory describes slow evolution (modulation) of the parameters characterising  nonlinear periodic and quasiperiodic waves, such as  amplitude, wavenumber,  frequency, mean etc. For integrable equations such as the KdV and NLS equations there is an elegant and powerful spectral approach to the derivation of the modulation equations first developed  by Flaschka, Forest and McLaughlin for the KdV equation \cite{flaschka_multiphase_1980} and then extended to other integrable equations such as the NLS, Benjamin-Ono, sine-Gordon equations etc. The spectral modulation theory is based on the extension of the IST method called the finite-gap theory \cite{novikov_theory_1984}, \cite{belokolos_algebro-geometric_1994}. 
In this section we will show how the finite-gap theory and modulation equations can be used to construct a mathematical model of soliton gas and justify the kinetic equation \eqref{kin_eq0}, \eqref{eq_state_kdv} for the soliton gas of the KdV equation and further, equation \eqref{FNLS_kin} for the soliton gas of the focusing NLS equation.  

\subsection{The Big Picture}
\label{sec:big_pic}

We start with  a high-level description of the spectral theory of soliton gas based on properties of the so-called finite-gap modulated solutions of integrable dispersive hydrodynamic equations. 
For simplicity we shall refer to scalar dispersive hydrodynamics \eqref{eq:scalar_dh}  although the general principles  are equally applicable to the vector, bidirectional case.
Our qualitative description will be substantiated in the following sections by considering  the examples of the KdV and the focusing NLS equations. As we have seen in Section \ref{sec:kin_phenomen}, the description of soliton gas includes two aspects:  (i) the `microscopic' structure of an equilibrium (uniform) gas, characterised by the equation of state \eqref{eq:state}; (ii) the slow evolution of the non-equilibrium (non-uniform)  soliton gas' parameters described by the spectral transport equation \eqref{kin_eq0}. This setting bears a strong resemblance to the Whitham modulation theory in which the local microscopic structure is given by an exact periodic or quasiperiodic solution of a dispersive hydrodynamic equation while the slow evolution is described by the modulation equations obtained via some period-averaging procedure or formal multiple scales analysis.

\subsubsection{Spectral modulation theory of multiphase waves}
The simplest yet instructive example of modulation theory  arises when considering the evolution of linear modulated waves  in dispersive media, the so-called kinematic wave theory \cite{whitham_linear_1999}. Dispersive hydrodynamics \eqref{eq:scalar_dh} upon linearisation about an equilibrium  $u=u_0$ admits harmonic multiperiodic  solutions 
in  the form of  a finite Fourier series 
\begin{equation}\label{fourier}
\tilde u =  \sum \limits_{j=1}^N a_je^{i(k_jx - \omega_j t + \theta_j^0)}, \ \ N \in \N, 
\end{equation}
where $\tilde u = u-u_0$, $a_j $ are  the  amplitudes of the Fourier modes, ${\bs k} = (k_1, k_2, \dots, k_N)$, ${\bs \omega} = (\omega_1, \omega_2, \dots , \omega_N)$ are the wavenumber and frequency vectors respectively, and $\bs{\theta}^0=(\theta_1^0, \dots, \theta_N^0)$ is the initial phase vector.  The frequency components in \eqref{fourier} satisfy the linear dispersion relation of \eqref{eq:scalar_dh}, i.e. $\omega_j=\omega_0(k_j)$. Assuming non-commensurability of the components $k_j$ and $\omega_j$ of the respective wavenumber and frequency vectors  in  \eqref{fourier}, this solution is defined on  $N$-dimensional $2 \pi$-torus, 
\be \label{torus}
 {\bs \theta} = {\bs k} x - {\bs \omega}t + {\bs \theta}^0 \in [0, 2\pi ) \times [0,2\pi) \times \dots \times  [0,2\pi) = \mathbb{T}^N.
 \ee
For a modulated linear wave all parameters in the Fourier series \eqref{fourier} become slow functions of $x,t$. To capture the modulations one introduces a small parameter $\eps$   and writes an  approximate  solution as $\tilde u({\bs \theta}, X,T)  \sim \sum_{j=1}^N a_j(X,T)e^{i \theta_j(x,t)}$, where $X=\eps x, T=\eps t$ are the slow $x$ and $t$ variables respectively, and  $\theta_j(x,t)=\eps^{-1}S_j(X,T)$ are the fast generalised phases.  We now define the local wavenumbers and frequencies by 
\be \label{local_kom}
k_j(X,T):=\partial_x\theta_j = \partial_X S, \quad \omega_j(X,T):= - \partial_t \theta_j = -\partial_T S(X,T),
\ee
 so that in the absence of modulations the approximate solution reduces to \eqref{fourier}. The phase consistency conditions $\partial_{xt}\theta_j= \partial_{tx}\theta_j$ then yield $N$ wave conservation laws, 
\begin{equation}\label{lin_kinem}
 {\bs k}_T + {\bs \omega}_X=0, \quad \omega_j = \omega_0(k_j), \quad j=1, \dots, N.
 \end{equation}
System \eqref{lin_kinem} represents the simplest example of modulation system. 
We note that the second part of   \eqref{lin_kinem} implies that, despite the slow $x,t$-variations,  the components of the frequency vector 
remain to be locally (on the typical scale of the fast oscillations in $\theta$) related to the wavenumber vector components by the same linear dispersion relations as in the non-modulated wave \eqref{fourier}. This result is non-trivial and is established by  a multiple scales analysis of the linearised dispersive hydrodynamics with the approximate solution $\tilde u({\bs \theta}, X,T)$ as the leading order term in the perturbation series in $\eps$. This analysis also yields the modulation equations for the amplitudes $a_j(X,T)$, which we do not consider here. We also note that the initial phases $\theta_j^0$ can be viewed as independent random values, each uniformly distributed on $[0, 2\pi)$. Then solution \eqref{fourier} becomes  a random process, whose slow modulations are described by eq. \eqref{lin_kinem} complemented by the appropriate amplitude modulation equations. 

We now turn to the nonlinear  analogue of the multiphase solution \eqref{fourier}. The one-phase nonlinear  periodic solutions are quite common in dispersive hydrodynamics and describe travelling waves. However, the existence of {\it nonlinear multiphase, quasiperiodic}  solutions is an exceptional feature  of integrable PDEs. 
Let the integrable dispersive hydrodynamics \eqref{eq:scalar_dh} admit multiphase solutions
\begin{equation} \label{nonlin_multiphase}
u(x,t) = F_N(\theta_1, \dots, \theta_N), \quad \theta_j=k_j x -\omega_j t + \theta_j^0,
\end{equation}
so that 
\be 
F_N(\theta_1, \dots, \theta_j + 2 \pi, \dots, \theta_N) = F_N(\theta_1, \dots, \theta_j,  \dots, \theta_N) , \quad  
j=1, 2, \dots, N,
\ee
i.e. ${\bs \theta} = (\theta_1, \theta_2, \dots, \theta_N)\in \mathbb{T}^N$.  
Quasiperiodic solution \eqref{nonlin_multiphase} can be viewed as a nonlinear analogue of the linearised Fourier solution \eqref{fourier}. 

The existence and properties of nonlinear multiphase  solutions to the KdV equation were established in 1970-s in the series of  pioneering works \cite{novikov_periodic_1974}, \cite{lax_periodic_1975}, \cite{its_periodic_1975}, \cite{dubrovin_periodic_1975} where the finite-gap theory, a nontrivial extension of the IST to periodic and quasiperiodic potentials has been developed  (see also the monographs \cite{novikov_theory_1984}, \cite{belokolos_algebro-geometric_1994} and a historical review \cite{matveev_30_2008}).  It  was shown that
the most natural  parametrisation of the multiphase solutions \eqref{nonlin_multiphase}  is achieved in terms of the spectrum  of the corresponding Lax operator.  For the preliminary discussion of this section it is convenient to assume that the Lax operator is self-adjoint so that its spectrum is real valued.  This is the case for the  (unidirectional) KdV  and (bidirectional) defocusing NLS equations. The case of complex band spectrum arises for the focusing NLS equation, and this case will be considered separately in Section \ref{sec:FNLS}. The fundamental result of the finite gap theory is that the Lax spectrum $\mathscr{S}_N$ of the $N$-phase solution \eqref{nonlin_multiphase}  lies in the union of $N+1$ disjoint  bands $\gamma_j =[\lambda_{2j-1}, \lambda_{2j}]$ (one of which could be semi-infinite, then $\gamma_{N+1}=[\lambda_{2N+1}, + \infty)$) separated by $N$ finite gaps $c_j=(\lambda_{2j}, \lambda_{2j+1})$, 
\be \label{lax_spectrum}
\lambda \in \mathscr{S}_N \equiv \cup _{i=1}^{N+1} \gamma_i \subset \mathbb{R}, \quad \gamma_i \cap \gamma_j= \emptyset, \ \ i \ne j.
\ee  
For that reason  multiphase solutions of integrable equations are often called finite-gap potentials (we recall that the Lax spectrum of a general periodic potential consists of an infinite number of bands and has even more complicated structure for almost periodic potentials \cite{pastur_spectra_1992}).  Thus the real spectrum of a finite-gap potential is fully parametrised by the state vector ${\bs \lambda}={(\lambda_1, \lambda_2, \dots, \lambda_\mathcal{D})}$, where   $\mathcal{D}=2N+1$ or $\mathcal{D}=2N+2$ depending on the presence or absence of a semi-infinite band.

One of the outcomes of the finite-gap theory are the nonlinear dispersion relations linking the physical parameters of the multiphase solution \eqref{nonlin_multiphase} such as the wavenumbers, frequencies, mean etc. with the components of the $\mathcal{D}$-dimensional spectral state vector ${\bs \lambda}$.
In particular, for the $N$-component wavenumber and  frequency vectors ${\bs k}$ and $\bs \omega$ in \eqref{nonlin_multiphase}  we  have
\be\label{n_gap_disp}
k_j = {K}_j({\bs \lambda}), \qquad \omega_j=  \Omega_j ({\bs \lambda}),   \quad j=1, \dots, N.
\ee
 E.g. for the KdV equation the nonlinear dispersion relations \eqref{n_gap_disp} have the form \eqref{kdv_nonlin_disp}.

 By manipulating the endpoints of spectral bands $\lambda_j$ one can modify the waveform of the solution \eqref{nonlin_multiphase}. Two limiting configurations are of particular interest.

\medskip
(i) {\it Harmonic (linear wave) limit}  is achieved by collapsing  spectral {\it gaps}, $|c_j|=\lambda_{2j+1} - \lambda_{2j} \to 0$, $j=1, \dots, N$. In this limit the $N$-gap solution converts into the linear quasiperiodic  solution \eqref{fourier}, and the nonlinear dispersion relations \eqref{n_gap_disp}  into the dispersion relation $\omega_j=\omega_0(k_j)$ for linearised wave modes. This is not surprising as the inverse scattering theory, including its finite-gap extension, essentially represents a nonlinear analogue of the Fourier method \cite{ablowitz_solitons_1981}.

\medskip
(ii) {\it Solitonic limit}.  In this limit the $N$-gap quasiperiodic solution \eqref{nonlin_multiphase} transforms into the  $N$-soliton solution exponentially decaying as $|x| \to \infty$ \cite{novikov_periodic_1974} (see also  monograph \cite{novikov_theory_1984}). 
 Spectrally, the solitonic limit is realised by collapsing all finite {\it bands} into double points, $|\gamma_j| =\lambda_{2j} - \lambda_{2j-1}\to 0$, while keeping the gaps open, i.e.  one  requires vanishing of the  band/gap ratio, $r_j=|\gamma_j|/|c_j| \to 0$.  The collapsed bands correspond to the discrete spectrum points in the traditional IST for decaying potentials.   
On the other hand, the spectral limit $r_j \to 0$  implies vanishing of the corresponding wavenumber, $k_j \to 0$, which is another, physically suggestive way to describe the solitonic limit of finite-gap potentials, particularly useful in the context of soliton gases.  
 
As a simple example  illustrating the harmonic and solitonic limits in finite-gap potentials we consider the  one-gap (single-phase) KdV solution. For $N=1$ the KdV Lax spectrum $\mathscr{S}_1=[\lambda_1, \lambda_2] \cup [\lambda_3, \infty)$ and the corresponding  KdV solution \eqref{nonlin_multiphase} assumes the form of a `cnoidal wave' 
(see \eg \cite{drazin_solitons_1987}); without loss of generality we set  $\lambda_3=0$,
\be \label{dn}
u(x,t)= F_1(\theta; {\bs \lambda})=\lambda_2-\lambda_1-2\lambda_2
  \mathrm{cn}^2\left(
    \frac{\mathrm{K}(m)}{\pi} \theta; m \right),   
    \ee
where $\mathrm{cn}[\cdot]$ is the Jacobi elliptic function, $m= \lambda_2 /  \lambda_1$ is the modulus, $\mathrm{K}(m)$ is the complete elliptic integral of the first kind, and $\theta=kx-\omega t + \theta^0$ is the phase with the wavenumber $k$ and the frequency $\omega$  given by the nonlinear dispersion relations 
(cf. \eqref{n_gap_disp})
\be \label{onegap_nldr}
 k = \frac{\pi \sqrt{- \lambda_1}}{\mathrm{K}(m)}, \quad \omega= -2k(\lambda_1+ \lambda_2)\, .
 \ee
The band/gap ratio in the one-gap solution is given by $r=1-m$.  The solitonic limit $r \to 0$ ($\lambda_2, \lambda_1 \to -\eta^2$) of \eqref{dn}, \eqref{onegap_nldr} is then evaluated using the asymptotic behaviour of elliptic functions \cite{abramowitz_handbook_1972},
\be \label{sol_lim_kdv}
r \to 0: \quad k \sim \frac{1}{\ln \frac{1}{r}} \to 0, \quad \frac{\omega}{k} \to 4\eta^2, \quad \mathrm{cn} \to \mathrm{sech},
\ee
so that  solution \eqref{dn} takes the form of a soliton \eqref{kdv1sol}, 
\be\label{kdv1sol1}
u(x,t)=u_s(x,t; \eta)=2\eta^2 \hbox{sech}^2 [\eta(x-4\eta^2 t - x^{0})].
\ee
The opposite, harmonic limit of the cnoidal wave solution \eqref{dn} is realised by closing the spectral gap (i.e. letting $m \to 0$) but we do not consider this limit here as it does not play a role in the soliton gas construction.

The above consideration provides a good intuition for what happens in general case $N>1$. Indeed, a $N$-gap KdV solution can be represented as a nonlinear superposition of $N$ cnoidal waves \cite{osborne_solitons_1995}, \cite{osborne_nonlinear_2010} with the interaction between the nonlinear modes described by the so-called Riemann period matrix $\mathrm{B}$ (see eq.~\eqref{B} below). In the $N$-soliton limit ($r_j \to 0 \ 
\forall j=1, 2, \dots, N$), one has (see \eg \cite{matveev_30_2008}, \cite{osborne_nonlinear_2010})
\be \label{nsol_lim}
 k_j \to 0,  \quad \mathrm{B}_{jj} \to \infty, \quad  \mathrm{B}_{ij} \to  \dfrac{i}{\pi}\ln \left| \dfrac{\eta_i+\eta_j}{\eta_i-\eta_j} \right| , \ \ i\ne j, \ \ i,j=1,2 \dots, N,
\ee   so that the off-diagonal elements of the interaction matrix transform into the normalised two-soliton phase shifts (cf.~\eqref{shift_kdv}). 

\bigskip
Modulation theory of nonlinear multiphase waves $F_N({\bs \theta}; {\bs \lambda})$ describes slow  evolution of the endpoints of the band spectrum,  ${\bs \lambda}={\bs \lambda}(X,T)$, where $X=\varepsilon x$, $T=\varepsilon t$, $\varepsilon \ll 1$ \cite{flaschka_multiphase_1980}, \cite{dubrovin_hydrodynamics_1989}. 
Similar to the linear modulation theory,  slow modulations  necessitate the introduction of the generalised phase vector ${\bs \theta}={\bs S}(X,T)/\eps$ so that the local wavenumber and local frequency vectors are defined by (cf.~\eqref{local_kom})
\be \label{kom_nonlin_multiphase}
 \quad {\bs k}(X,T) : = {\bs \theta}_x={\bs S}_X, \quad {\boldsymbol \omega}(X,T) : = - {\bs \theta}_t=-{\bs S}_T,  
\ee
The consistency condition ${\bs \theta}_{xt} = {\bs \theta}_{tx}$ then leads to the system of $N$ wave conservation equations, an analogue of the kinematic equation \eqref{lin_kinem} for linear waves with an important difference that, instead of the linear dispersion relation $\omega = \omega_0(k)$ linking the components of the frequency and wavenumber vectors it now includes the nonlinear dispersion relations \eqref{n_gap_disp}  for finite-gap potentials,
\begin{equation}\label{nlin_kinem}
 {\bs k}_T + {\bs \omega}_X=0,  \qquad {\bs k} = {\bs K}[\bs{\lambda}(X,T)],  \quad {\bs \omega}= {\bs \Omega} [\bs{\lambda}(X,T)] 
  \end{equation}
Importantly, the kinematic modulation system \eqref{nlin_kinem} for a general case of multiphase nonlinear waves  is not closed since $\dim ({\bs \lambda}) = \mathcal{D}> N$.
Indeed, the full modulation system  contains  $\mathcal{D}$ equations for  $\lambda_j(X,T)$, and the wave conservation equations are always consistent with (but not equivalent to) the full system (see \cite{flaschka_multiphase_1980} for the complete description of the KdV spectral modulation theory).
However, in the harmonic and soliton limits corresponding to the collapsed spectral gaps  or  bands  the dimension $\mathcal{D}$ of the state vector ${\bs \lambda}$  decreases,  enabling  the necessary closure  for the system of wave conservation laws in \eqref{nlin_kinem} under the additional constraint of constant background  
(see  \cite{maiden_solitonic_2018}, \cite{congy_interaction_2019} for the relevant theory of  the dynamic wave-mean flow interaction showing how the effects of nonconstant background can be included). In particular, in the harmonic limit system \eqref{nlin_kinem} transforms into the kinematic system \eqref{lin_kinem}. The solitonic limit is a singular one and requires a more delicate treatment.

\subsubsection{Thermodynamic limit of  finite-gap spectral solutions}
\label{sec:therm_lim_kdv}
 The main idea of the spectral construction of soliton gas is to take simultaneously  the solitonic limit  and the limit $N \to \infty$ of $N$-gap potential in such a way that 
 \be \label{therm_k}
 \forall k_j \to 0, \quad N \to \infty \quad \hbox{ but} \quad \lim \limits_{N \to \infty}\sum \limits_{j=1}^N k_j  = \beta = \mathcal{O}(1),
\ee
with a similar behaviour for the frequency components $\omega_j$.
The limit \eqref{therm_k} is the thermodynamic type limit for nonlinear multiphase waves (as we shall see, $\beta$ in \eqref{sec:therm_lim_kdv}  agrees with the  soliton gas density introduced earlier in \eqref{kappa}). This limit suggests the following scaling for the wavenumbers and frequencies:
\be \label{kom_scale}
k_j \sim \omega_j \sim \frac{1}{N},  \qquad N \gg 1.
\ee
Analysis of the nonlinear dispersion relations \eqref{n_gap_disp} for the KdV and NLS equations yields the asymptotic structure of the  spectral set $\mathscr{S}_N$ compatible with the thermodynamic scaling \eqref{kom_scale}. It turns out that the large $N$ behaviour \eqref{kom_scale} of $k_j$, $\omega_j$ can be achieved by 
introducing  the special distribution  of spectral bands $\gamma_j$ and gaps $c_j$ on a fixed spectral interval $\Gamma = [\lambda_1, \lambda_{\mathcal{D}}]$  (an arc in the complex plane for the focusing NLS) such that  
\be\label{therm_spectr}
 \qquad |c_j| \sim -\ln^{-1}|\gamma_j| \sim \frac{1}{N},    \qquad N \gg 1,
\ee
i.e. by making the bands  exponentially narrow compared to the gaps. 
We note that the band-gap scaling \eqref{therm_spectr} is inspired by the  spectrum of the periodic Lax operator \eqref{schr} in the semiclassical limit \cite{weinstein_asymptotic_1987}, \cite{venakides_continuum_1989}.

We shall call the band-gap scaling \eqref{therm_spectr} the  exponential {\it thermodynamic spectral scaling} and denote the limit as $N \to \infty$ of a function $F({\bs \lambda})$ considered on this scaling as $\mathfrak{T} \text{-}\lim F({\bs \lambda}) $.
For the thermodynamic scaling  \eqref{therm_spectr} we have $r_j = |\gamma_j|/|c_j| \to 0$ $\forall j$, yielding  essentially an infinite-soliton limit, in full agreement with our intention to describe soliton gas.   Other meaningful scalings (sub-exponential, super-exponential) compatible with \eqref{kom_scale} and leading to special cases of soliton gases are possible and will be discussed later.

We will show that the application of the thermodynamic limit  to the kinematic modulation system \eqref{nlin_kinem} results in the kinetic equation for soliton gas
\be \label{kin_eq_mod}
f_{\tilde T} + (sf)_{\tilde X}=0, \quad  s(\la) = \mathcal{S}[f(\la)],
\ee
where $\mathcal{S}[\dots]$ is a functional and
\be
f(\la)=\frac{\rmd}{\rmd \la} \{ \mathfrak{T}\text{-}\lim  \sum \limits_{j=1}^{M \le N} k_j\},  \quad s(\la) = \mathfrak{T} \text{-}\lim \frac{\omega_j}{k_j} \ 
 \ee
are the density of states and the transport velocity  respectively,  both depending on the continuous  parameter $\la \in \Gamma$ and on the `superslow' spacetime variables $\tilde X= \delta x$, $\tilde T = \delta t$, where $\delta \ll \varepsilon \ll 1$  is a small parameter that scales the typical spatiotemporal modulations of soliton gas, which are much slower than those associated with the typical Whitham modulations of finite-gap potentials. In practice we will not be introducing the small parameters $\varepsilon$ and $\delta$ explicitly, assuming that the typical $x,t$-variations of $u$ and $f(\la)$ occur on disparate micro- and macroscopic scales respectively.

The outlined construction was concerned with the spectral characterisation of soliton gas defined as a thermodynamic limit of finite-gap potentials and can be symbolically represented as $ \mathscr{S}_N \longrightarrow f(\la)$.  However, this description is incomplete as the  Lax spectrum $\mathscr{S}_N $ determines the finite-gap potential \eqref{nonlin_multiphase} only up to $N$ initial phases,  $\bs{\theta}^0 \in \mathbb{T}^N$.  For a general finite-gap potential the respective incommensurability of the components of the wavenumber and frequency vectors in \eqref{nonlin_multiphase} implies dense winding on the phase torus.
Then the natural assumption for the construction of soliton gas  would be to let the components of ${\bs \theta}^0$ be independent random values, each  distributed uniformly on $[0,  2\pi)$. This is the so-called Random Phase Approximation, the standard assumption in the wave turbulence theory \cite{nazarenko_wave_2011}, which was also used for the construction of random finite-gap solutions of the KdV equation  in \cite{osborne_behavior_1993} and of the focusing NLS equation in  \cite{bertola_rogue_2016}, \cite{osborne_breather_2019}.
Following the classical construction of the configuration space of the ideal one-dimensional gas of non-interacting particles (see \eg \cite{sinai_introduction_1977}) it can be shown  that, upon the thermodynamic limit the uniform distribution of the initial phase vector ${\bs \theta}^0$ over the invariant torus $\mathbb{T}^N$ transforms into the Poisson distribution on $\mathbb{R}$ with the density  $\beta=\int_0^1 f(\la) \rmd \la$ for the  position phases $x_j=\theta_j^0/k_j$ \cite{el_soliton_2001}, which is consistent with the distribution of the soliton centres in the phenomenologically introduced rarefied soliton gas \eqref{rar_u}. The derivation of the Poisson distribution for the position phases in the KdV soliton gas will be presented in Section~\ref{sec:poisson}.

\subsection{Soliton gas for the KdV equation}
\label{sec:spectral_theory_kdv}

\subsubsection{Thermodynamic limit and nonlinear dispersion relations for soliton gas}
We now realise  the  thermodynamic spectral limit construction of soliton gas  for the KdV equation \eqref{KdV} following \cite{el_soliton_2001}, \cite{el_thermodynamic_2003}. The Lax spectrum \eqref{lax_spectrum} of the $N$-phase KdV solution  \eqref{nonlin_multiphase} lies in the union
of bands,
\begin{equation}
  \label{eq:27}
\lambda \in  [\lambda_1,\lambda_2] \cup [\lambda_3,\lambda_4] \cup
  \cdots \cup [\lambda_{2N+1}, \infty) .
\end{equation}
The state vector  ${\bs \lambda}= (\lambda_1, \lambda_2, \dots, \lambda_{2N+1})$ parametrises the
$N$-gap KdV solution up to  $N$ initial
phases $\theta^0_j$, which we assume to be independent random values, each uniformly distributed on $[0, 2\pi)$.  In what follows, we take advantage of some known results from the
KdV finite-gap theory \cite{novikov_theory_1984}, \cite{belokolos_algebro-geometric_1994}, \cite{flaschka_multiphase_1980} and apply them to the description of the KdV soliton gas. 

\medskip
(i) {\it Nonlinear dispersion relations for finite-gap potentials}

\medskip

\begin{figure}[h]
\centerline{\includegraphics[scale=0.7]{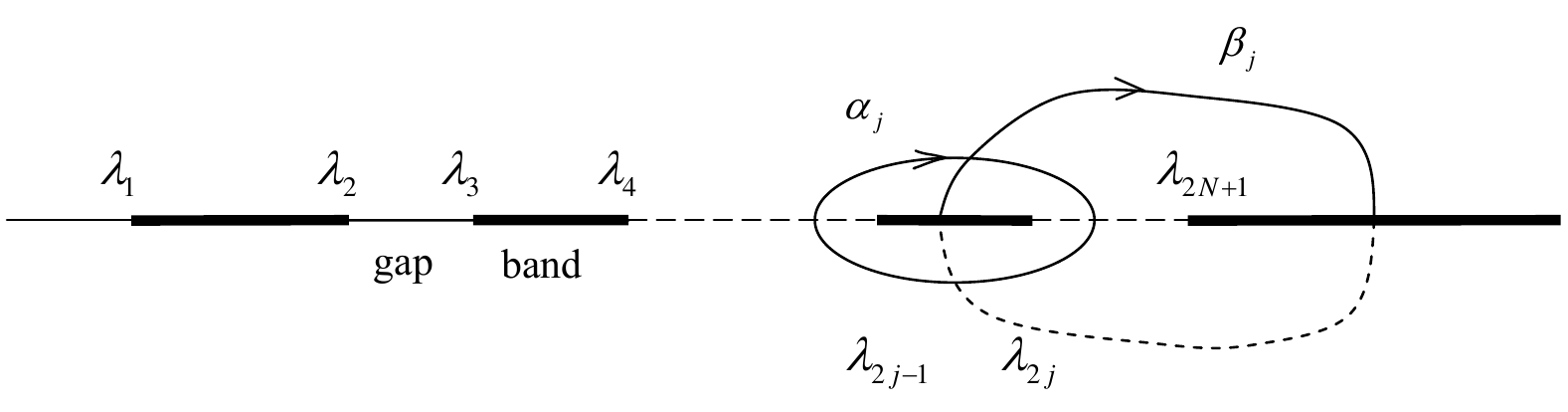}}
\caption{Spectrum of  $N$-gap solution of the KdV equation and the canonical homology basis on the Riemann surface $\mathcal{R}$ \eqref{RS}.}
\label{fig:kdv_cycles}
\end{figure}

To formulate the nonlinear dispersion relations \eqref{n_gap_disp} for the multiphase KdV solutions we need to introduce several fundamental  objects underlying the algebraic structure of finite-gap potentials expressible in terms of multidimensional Riemann theta-functions \cite{dubrovin_theta_1981}. In the finite-gap theory the vector ${\bs \lambda}=(\lambda_1, \lambda_2, \dots, \lambda_{2N+1})$ of the endpoints of spectral bands defines the 
two-sheeted hyperelliptic Riemann surface $\mathcal{R}$ of
genus $N$ via 
\begin{equation}  \label{RS}
 {R(z)} =\prod\limits_{j=1}^{2N+1}(z -\lambda
_{j})^{1/2}  \,,\qquad z \in {\mathbb{C}} \, ,
\end{equation}
with $R \sim z^{N+1/2}$ as $z \to \infty$. We make the branch cuts of $R(z)$ along the spectral bands $[\lambda_1, \lambda_2]$, \dots 
$[\lambda_{2j-1}, \lambda_{2j}]$, \dots, $[\lambda_{2N+1}, \infty)$ and
introduce a canonical homology basis on $\mathcal{R}$ as follows (see
Fig.~\ref{fig:kdv_cycles}): the $\alpha_j$-cycle surrounds the $j$-th band clockwise on the
upper sheet, and the $\beta_j$- cycle is canonically conjugated to $\alpha_j$
 such that the closed contour $\beta_j$ starts at $\lambda_{2j}$ , goes to
$+\infty$ on the upper sheet and returns to $\lambda_{2j}$ on the lower
sheet.

We introduce a basis of  holomorphic differentials on $\mathcal{R}$:
\begin{equation}\label{chi}
w_j = \sum \limits_{k=0}^{N-1}\kappa_{jk} \frac{z^{k}}{R(z)}\rmd z \,
, \qquad j=1, \dots, N,
\end{equation}
where the coefficients $\kappa_{jk}({\bs \lambda})$ are determined by the
normalisation over the $\alpha$-cycles
\begin{equation} \label{norm}
\oint \limits_{\beta_k}w_{j} = \delta _{jk}\, ,
\end{equation}
while the integrals over the $\beta $-cycles give the  entries of
the symmetric $N \times N$ Riemann period matrix $\mathrm{B}$,
\be \label{B} \mathrm{B}_{ij}= \oint \limits
_{\beta_j}w_i \, 
\ee
with positive definite imaginary part.

The components  $k_j$, $\omega_j$ of the wavenumber and frequency vectors are expressed
in terms of the branch points $\lambda_j$ of the spectral Riemann surface \eqref{RS} by the relations \cite{flaschka_multiphase_1980}
\begin{equation} \label{kdv_nonlin_disp}
\begin{split}
{\bs k} &= -4\pi i \mathrm{B}^{-1}{\bs \kappa}^{(N-1)}\, , \\
{\bs \omega}  &= 8\pi i \mathrm{B}^{-1} \left({\bs \kappa}^{(N-1)}\sum \limits_{j=1}^{2N+1}\lambda_j + 2{\bs \kappa}^{(N-2)}\right) \, ,
\end{split}
\end{equation}
where $[{\bs \kappa}^{(M)}]_i \equiv \kappa_{iM}$.
These are the nonlinear dispersion relations \eqref{n_gap_disp} for  finite-gap KdV solutions.

\bigskip
{\it (ii) Thermodynamic spectral scaling}

\medskip
We now introduce the exponential  thermodynamic scaling \eqref{therm_spectr} for the spectrum \eqref{eq:27} of the KdV finite-gap solutions. We fix $\la_1=-1$ and $\lambda_{2N+1}=0$ and, following \cite{venakides_continuum_1989}, consider the lattice of points $0 <  \eta _N<\eta _{N-1} <\ldots < \eta _1 <1$ ,
where
\be \label{lat}
-\eta _j^2=\frac 12\left( \lambda_{2j-1}+\lambda_{2j}\right)\,
\ee
are centres of spectral bands. We then define two positive smooth functions on $[0,1]$:

(a) \ The normalised density $\varphi(\eta)>0$  of the lattice points $\eta_j$,  introduced such that
\be\label{phieta}
\eta_j-\eta_{j+1}  \sim  \frac{1}{N \varphi(\eta_j) } ,  \quad \int_{0}^1 \varphi(\eta)\rmd \eta =1,
\ee
i.e. $\varphi(\eta) \rmd \eta$ is the probability measure on $[0,1]$.

(b) \ The normalised logarithmic band width distribution $\tau(\eta)>0$ defined by
\be\label{nueta}
\tau(\eta_j) \sim  -\frac{1}{N} \ln (\lambda_{2j}-\lambda_{2j-1}),  \quad j=1, \dots, N, \ N \gg 1.
\ee
The  functions $\varphi (\eta)$ and $\tau(\eta)$  asymptotically define the
 Riemann surface $\mathcal{R}$ \eqref{RS} for $N \gg 1$.  
The asymptotics \eqref{phieta}, \eqref{nueta} imply the exponential spectral scaling (cf. \eqref{therm_spectr})
\be \label{exp_scale}
 |c_j| \sim \frac{1}{\varphi(\eta_j) N}, \qquad |\gamma_j| =\delta_i \sim \exp\{-N \tau(\eta_j)\}, \quad j=1, \dots, N, \quad N \gg 1.
\ee
Other spectral scalings of interest are the sub-exponential  scaling:
$e^{- a N } \ll  |\gamma_j|  \ll \frac{1}{N}, \quad j= 1, \dots, N$ for any $a>0$, and super-exponential scaling:
$e^{-a N } \gg  |\gamma_j|,  \quad j= 1, \dots, N$ for any $a>0$. The latter  corresponds to the case of an `ideal' soliton gas consisting of noninteracting solitons, and the former to the special kind of soliton gas termed {\it soliton condensate}. These scalings were introduced in \cite{el_spectral_2020}  in the context of the focusing NLS soliton gas and will be considered in Section~\ref{sec:FNLS}.

\medskip
{\it (iii) \  Nonlinear dispersion relations for soliton gas} 

\medskip
We  re-write the nonlinear dispersion relations \eqref{kdv_nonlin_disp} as
\begin{equation} \label{kdv_nonlin_disp1}
\mathrm{B} {\bs k} = -4\pi i {\bs \kappa}^{(N-1)}\, , \quad
\mathrm{B} {\bs \omega}  = 8\pi i  \left({\bs \kappa}^{(N-1)}\sum \limits_{j=1}^{2N+1}\lambda_j + 2{\bs \kappa}^{(N-2)} \right) \, 
\end{equation}
and apply  the thermodynamic spectral scaling \eqref{exp_scale}. For that, we introduce the lattice \eqref{lat} and the large $N$ expansions  \eqref{phieta}, \eqref{nueta} in \eqref{chi}, \eqref{norm}, \eqref{B}, which  gives at leading order \cite{venakides_continuum_1989} (cf. \eqref{nsol_lim}):
\be\label{Bij}
\mathrm{B}_{jj} \sim  \frac{i}{\pi} N \tau({\eta_j}), \quad \mathrm{B}_{kj} \sim \dfrac{i}{\pi}\ln \left| \dfrac{\eta_k+\eta_j}{\eta_k-\eta_j}
\right| \ \ \text{when} \ \  k \ne j,
\ee
\be \label{ca}
{\kappa}_{j,N-1} \sim -\frac{\eta_j}{2\pi } \, , \qquad
 \kappa_{j, N-1}\sum \limits_{j=1}^{2N+1}\lambda_j + 2 \kappa_{j,N-2} \sim
\frac{\eta_j^3}{\pi }\, .
\ee 

In view of \eqref{Bij}, \eqref{ca}   the balance of terms in  the dispersion relations \eqref{kdv_nonlin_disp1}  necessitates the following scaling for the component of the wavenumber and frequency $N$-vectors (cf.~\eqref{kom_scale}):
 \be \label{komN}
 k_j\sim \frac{\varkappa(\eta_j)}{ N}, \quad  \omega_j \sim \frac{\nu(\eta_j)} {N},  \quad \ \  N \gg 1,
 \ee
 where $\varkappa(\eta) \ge 0$ and $\nu(\eta)$ are smooth functions interpolating $\varkappa_j$, $\nu_j$ on $[0,1]$.

\medskip
Substituting \eqref{Bij}, \eqref{ca}, \eqref{komN},  in \eqref{kdv_nonlin_disp1} we arrive at 
two algebraic systems for $\varkappa(\eta_k)$, $\nu(\eta_k)$,  $k=1, \dots , N$:
\be \label{sum1}
\sum \limits^{N}_{j=1}\frac{1}{N}\ln \left| \frac{\eta_k+\eta_j}{\eta_k-\eta_j}\right| \varkappa(\eta_j)
+\tau(\eta_k)\varkappa(\eta_k)= 2 \pi  \eta_k \, ,
\ee
\be \label{sum2}
\sum \limits^{N}_{j=1}\frac{1}{N}\ln \left| \frac{\eta_k+\eta_j}{\eta_k-\eta_j}\right| \nu(\eta_j)
+\tau(\eta_k)\nu(\eta_k)= 8 \pi \eta_k^3\, .
\ee 
 Passing to the continuum limit as $N \to \infty$ we obtain the nonlinear dispersion relations for soliton gas:
 \be \label{inta}
\int\limits_0^1 \ln \left|\frac{\eta + \mu}
{\eta-\mu}\right| f(\mu)\rmd\mu  + \sigma(\eta) f(\eta)
=  \eta\, ,
\ee
\be \label{intat}
\int\limits_0^1 \ln \left|\frac{\eta + \mu}
{\eta-\mu}\right|v(\mu) \rmd \mu  + \sigma(\eta)v(\mu)
=4\eta^3 \, ,
\ee
where
 \be
 f(\eta)= \frac{1}{2\pi}\varphi(\eta) \varkappa(\eta); \quad v(\eta)= \frac{1}{2 \pi}\varphi(\eta) \nu(\eta), \quad \sigma(\eta)=\frac{\tau(\eta)}{\varphi(\eta)} >0. \ee
For a given function $\sigma(\eta)$, which encodes  the Lax spectrum in the thermodynamic limit, the integral equations \eqref{inta}, \eqref{intat}  specify two functions $f(\eta)$ and $v(\eta)$, which we identify below as the DOS and the spectral flux density of the soliton gas respectively.

 Consider a partial sum $K_M=\frac{1}{2 \pi}\sum \limits_{j=1}^M k_j $ over the spectral lattice $\{\eta_1, \eta_2, \dots, \eta_N \}$, where  $1\leq M \leq N$. Given that $2 \pi k_j$ is the spatial density of waves (which we treat as quasiparticles) associated with the $j$-th spectral  band,  the quantity $K_M$ has the meaning of the integrated density of states \cite{lifshits_introduction_1988}, \cite{pastur_spectra_1992}. Invoking the scaling  \eqref{komN}  and passing to the continuum limit $N \to \infty$, $\eta_M \to \eta$, we obtain 
\begin{equation}\label{integrated_dos}
K_M= \frac{1}{2 \pi}\sum _{j=1}^{M} \frac{\varkappa(\eta_j)}{N} \to \frac{1}{2 \pi}\int \limits_{0}^{\eta} \varkappa(\mu) \varphi(\mu) \rmd \mu \equiv \mathcal{K}(\eta).
\end{equation}
Then the DOS is given by
\be
\mathcal{K}'(\eta) = \frac{1}{2 \pi} \varkappa(\eta) \varphi(\eta) =f(\eta).
\ee
 Similarly,  for the temporal counterpart of the integrated DOS---the spectral flux---we have  
\be
\Omega_M = \frac{1}{2 \pi}\sum \limits_{j=1}^M \omega_j = \frac{1}{2 \pi}\sum _{j=1}^{M} \frac{\nu(\eta_j)}{N} \to \frac{1}{2 \pi}\int \limits_{0}^{\eta} \nu(\mu) \varphi(\mu) \rmd \mu \equiv \mathcal{V}(\eta),\ee
so that the spectral flux density in a soliton gas is given by 
\be \label{veta}
 \mathcal{V}'(\eta) = \frac{1}{2 \pi} \nu(\eta) \varphi(\eta) =v(\eta).
\ee

\subsubsection{Equation of state and spectral kinetic equation}
\label{sec:kin_eq_KdV}

Eliminating $\sigma$ from the nonlinear dispersion relations \eqref{inta}, \eqref{intat}  we obtain for $s(\eta)=v(\eta)/f(\eta)$
\begin{equation}\label{eq_state_kdv1} 
s(\eta)=4\eta^2+\frac{1}{\eta}\int \limits_{0}^1 \ln
\left|\frac{\eta + \mu}{\eta-\mu}\right|f(\mu)[s(\eta)-s(\mu)]d\mu\,   ,
\end{equation}
which is exactly the equation of state \eqref{eq_state_kdv} obtained in  Section~\ref{sec:kin_phenomen} 
under  the collision rate assumption \eqref{collision_rate}. Hence this assumption is now justified.

As suggested by the phenomenological derivation in Section~\ref{sec:kin_phenomen}  the function $s(\eta)$ in \eqref{eq_state_kdv1} has the meaning of the effective soliton  velocity (i.e. the transport velocity) in a soliton gas.  We will now show how this interpretation is justified within the thermodynamic limit framework. To this end we consider non-equilibrium soliton gas with $f(\eta) \equiv f (\eta, x, t)$, $s(\eta) \equiv s(\eta, x, t)$ and  derive the evolution equation for the DOS.  
We go back to the original, discrete wavenumber and frequency components
$k_j({\bs \lambda})$, $\omega_j ({\bs \lambda})$ of the finite-gap potential, defined in terms of the fixed branch points ${\bs \lambda}$ of the Riemann surface $\mathcal{R}$ of \eqref{RS}. Let us now consider a slowly modulated finite-gap potential with  ${\bs \lambda }={\bs \lambda} (x,t)$.
The modulation system  describing the evolution of  $2N+1$ parameters $\lambda_1 (x,t), \lambda_2(x,t), \dots \lambda_{2N+1}(x,t)$ has been derived in  \cite{flaschka_multiphase_1980}. This  system admits  an infinite number of hyperbolic conservation laws, that include a  finite subset of $N$ wave conservation laws \eqref{nlin_kinem}, which can be manipulated into the equivalent system
\be
\partial_t K_M + \partial_x \Omega_M =0, \quad M =1, \dots, N \label{wc111},
\ee 
where  $K_M=\frac{1}{2 \pi}\sum \limits_{j=1}^M k_j $ and $\Omega_M = \frac{1}{2 \pi}\sum \limits_{j=1}^M \omega_j $ as defined in the previous section.  Applying  the thermodynamic limit \eqref{integrated_dos} -- \eqref{veta} to \eqref{wc111} yields the transport equation for DOS
\be \label{kineq_kdv}
f_t + (fs)_x=0, 
\ee
where $s=v/f$ is identified as the transport velocity of soliton gas, as expected. Thus we have derived the kinetic equation for the KdV soliton gas  as the thermodynamic limit of the multiphase Whitham modulation system.

Integrating equation \eqref{kineq_kdv} over the spectral interval $[0,1]$ we obtain the  conservation equation
\be
\beta_t + \zeta_x=0
\ee
for the total integrated density of solitons in the gas  $\beta(x,t)=\int_0^1f(\eta, x, t) \rmd \eta$.  The function $\zeta(x,t)= \int_0^1v(\eta, x, t)\rmd \eta$ has the meaning of the soliton gas frequency.
Generally, multiplying \eqref{kineq_kdv} by an arbitrary nonsingular function $p(\eta)$ and integrating 
over spectrum we obtain the conservation law for $M(x,t)=\int_0^1 p(\eta) f(\eta,x,t) \rmd\eta$. In particular, choosing 
$p(\eta)=C_n \eta^{2n-1}$, $n \in \mathbb{N}$ with $C_n=\frac{2^{2n}}{2n-1}$ we obtain the series of averaged  conservation laws for the KdV equation -- the Whitham equations for soliton gas \eqref{stoch_Whitham}, where the ensemble averages of the polynomial conserved densities (the  Kruskal integrals, see Section~\ref{sec:ensemble_aver}) are expressed in terms of the DOS as \cite{el_thermodynamic_2003} \cite{el_critical_2016}
\be
\langle P_n[u ] \rangle = \frac{2^{2n}}{2n-1}\int \limits_0^1 \eta^{2n-1} f(\eta) \rmd \eta.
\ee
In particular,  for $n=1,2$ the Kruskal integrals coincide with the respective statistical moments  of $u$---the observables of the KdV soliton gas field---and have the form
\begin{equation}\label{mean}
\begin{split}
\langle u \rangle(x,t)=  4 \int \limits_{0}^{1}\eta f(\eta, x,t)\rmd\eta \,, \quad
\langle u^{2} \rangle(x,t)
=\dfrac{16}{3}\int \limits_{0}^{1 }\eta ^{3}f(\eta, x, t)\rmd\eta  \, ,
\end{split}
\end{equation}
in full agreement with the result \eqref{moments_KdV} obtained by the heuristic `windowing' procedure in Section \ref{sec:ensemble_aver}.

Concluding this section we note that the presented construction of soliton gas it was assumed that soliton propagate on a fixed (zero) background, which was achieved by fixing the endpoint $\lambda_{2N+1}=0$ of the spectrum \eqref{eq:27}.  A generalisation to a slowly varying background is possible following the modulation construction of solitonic dispersive hydrodynamics in \cite{maiden_solitonic_2018}, \cite{sande_dynamic_2021}.
Such a generalisation could provide interesting insights into new soliton gas phenomena.

\subsubsection{Poisson distribution for position phases}
\label{sec:poisson}

Having defined the thermodynamic limit for the spectrum of finite-gap potentials $F_N({\bs \theta})$,
${\bs \theta} \in \mathbb{T}^N$, we now need to determine what happens  with the  phases $\theta_j=k_jx - \omega_jt+\theta_j^{(0)}$, $j=1, \dots, N$ in this limit. 

As discussed in Section~\ref{sec:therm_lim_kdv}  we adopt the Random Phase Approximation  in which the initial phases $\theta_j^{(0)}$
are assumed to be $N$ independent random values uniformly distributed on $[0, 2\pi)$. 
Under the random phase approximation the finite-gap potential $u(x,t)$ transforms into an ergodic random process \cite{pastur_spectra_1992}, both as function of $x$ and $t$. We now fix $t=t^*$  and represent the cyclic phases 
\ $\theta_j(\mod 2\pi)$ in the form   $\theta_j=k_j(x - x_j)$, where  the position phases $x_j= x_j^0 + c_j t^*$ with   $c_j = \omega_j/k_j$ being the phase velocities.  The initial position phases $x_j^0=-\theta_j^{(0)}/k_j  $ are independent random values, each distributed uniformly on the respective period interval $[0,  2\pi/k_j)$.  The deterministic shifts 
$c_j t^*$ can be absorbed into the random initial positions so  that we shall  replace $ x^{(0)}_j \to x_j$.

The thermodynamic spectral scaling \eqref{exp_scale} implies that in the  limit $N \to \infty$ all  wavenumbers  vanish,  $k_j \to 0$, i.e. all spatial  periods of the finite-gap solution become infinite.   We now demonstrate  that, upon  the thermodynamic limit  \eqref{therm_k} the uniform distribution of $\boldsymbol {\theta}^{(0)}$ on $\mathbb{T}^N$ transforms into the Poisson distribution for the position phases $x_j \in \mathbb{R}$.

Let $\Delta$ be an arbitrary finite interval on the line. We assume that for sufficiently small $k_j$'s we have $\Delta \subset [0, 2\pi/k_j)$ $\forall j = 1, 2, \dots, N$. We introduce a random value $\xi_j = \chi_{\Delta}(x_j)$, where 
\begin{equation}\label{shock_heaviside}
\xi_j=\left\{\begin{array}{ll}
1 \quad & \text{if} \quad  x_j \in \Delta \\[6pt]
0 & \text{if} \quad x_j \notin \Delta,
\end{array}
\right.\end{equation}
with the probabilities $p(\xi_j=1)= (k_j/2\pi) \Delta$, $p(\xi_j=0)=1- (k_j/2\pi) \Delta$.
The probability-generating function $\Phi_j(z)$ for $\xi_j$ is
\be
\Phi_j(z) = (1-p_j) + z p_j.
\ee
Now let $\xi^{(N)} \equiv \sum \limits_{j=1}^N \xi_j$, which is the number of points $x_j$ that have fallen into the interval  $\Delta$. Since $\xi_j $ are independent random variables the probability-generating function for $\xi^{(N)}$ is
\be
\Phi^{(N)}(z) = \prod \limits_{j=1}^N \Phi_j(z)= \prod \limits_{j=1}^N (1 + (z-1)p_j) = 
\prod \limits_{j=1}^N \left(  1+ \frac{(z-1)k_j |\Delta|}{2 \pi}\right).
\ee 
 For $N \gg 1$ the thermodynamic scaling implies $k_j \sim N^{-1} \ll 1$ and we have 
 \be
 \ln \Phi^{(N)} = (z-1) \frac{|\Delta|}{2 \pi} \sum \limits_{j=1}^N k_j + \mathcal{O} (N^{-1}).
 \ee
 Taking the thermodynamic limit 
 \be
  \mathfrak{T}\text{-}\lim \Phi^{(N)}(z) = \exp [(z-1) \beta |\Delta|] = e^{-\beta |\Delta|} e^{\beta |\Delta| z} = \sum \limits_{n=0}^{\infty} z^n 
 \left( e^{-\beta |\Delta|} \frac{(\beta |\Delta|)^n}{n !} \right), 
 \ee
 where 
 $$
\beta =  \mathfrak{T}\text{-}\lim \left\{ \frac{1}{2\pi}\sum\limits_{j=1}^N k_j \right\} = \int \limits_0^1 f(\eta) \rmd \eta. 
$$

Hence the probability of having $n$ points $x_i$ in the interval $\Delta$ is given by the Poisson distribution
\be \label{poisson}
P_\Delta(n, \beta) = e^{-\beta |\Delta|} \frac{(\beta |\Delta|)^n}{n !}
\ee
with parameter $\beta$, the total integrated DOS in the soliton gas. The position phases $x_j$ coincide with the soliton centres in the  stochastic soliton lattice model \eqref{rar_u} of a rarefied gas. The random process \eqref{rar_u} can then be associated with a compound Poisson process \cite{feller_introduction_1968}, a stochastic process with jumps. The jumps, associated with solitons,  are distributed on the line randomly according to a Poisson distribution with probability \eqref{poisson} and the size of the jumps (the soliton amplitudes $a_i=2\eta_i^2$) is also random, where the spectral parameter $\eta$  has  the probability distribution $\varphi(\eta)= \beta^{-1} f(\eta)$.

\medskip

\subsection{Focusing NLS equation: soliton and breather gases}
\label{sec:FNLS}
The spectral theory of soliton gas for the KdV equation was generalised in \cite{el_spectral_2020} to the case of the focusing NLS equation \eqref{eq:fNLS}. As was mentioned in Section~\ref{sec:kin_nls} the essential difference between the KdV and focusing NLS equations at the level of the IST is that the Lax (Zakharov-Shabat) operator \eqref{ZS} for  the focusing NLS has  a complex-valued  spectrum satisfying the Schwartz symmetry  property. This results in the availability of the qualitatively new, compared to KdV, families of localised wave solutions and, consequently, to new types of soliton gases. Along with the individual fundamental solitons propagating on zero background  the focusing NLS equation supports a family of  the so-called bound state $N$-soliton solutions, the  complexes of strongly interacting, co-propagating (i.e. having the same, possibly zero, velocity) solitons. Another important family of localised solitonic structures supported by the focusing NLS equation are  breathers, that can be viewed as solitons on finite background. All these localised  solutions can serve as quasiparticles in the respective  soliton or breather gases, which we consider in this section.

\subsubsection{Solitons, breathers and finite-gap spectral solutions}
\label{sec:FNLS_sol}

{\it Solitons and breathers}

\begin{figure}[h]
\centering
\includegraphics[width= .4 \linewidth]{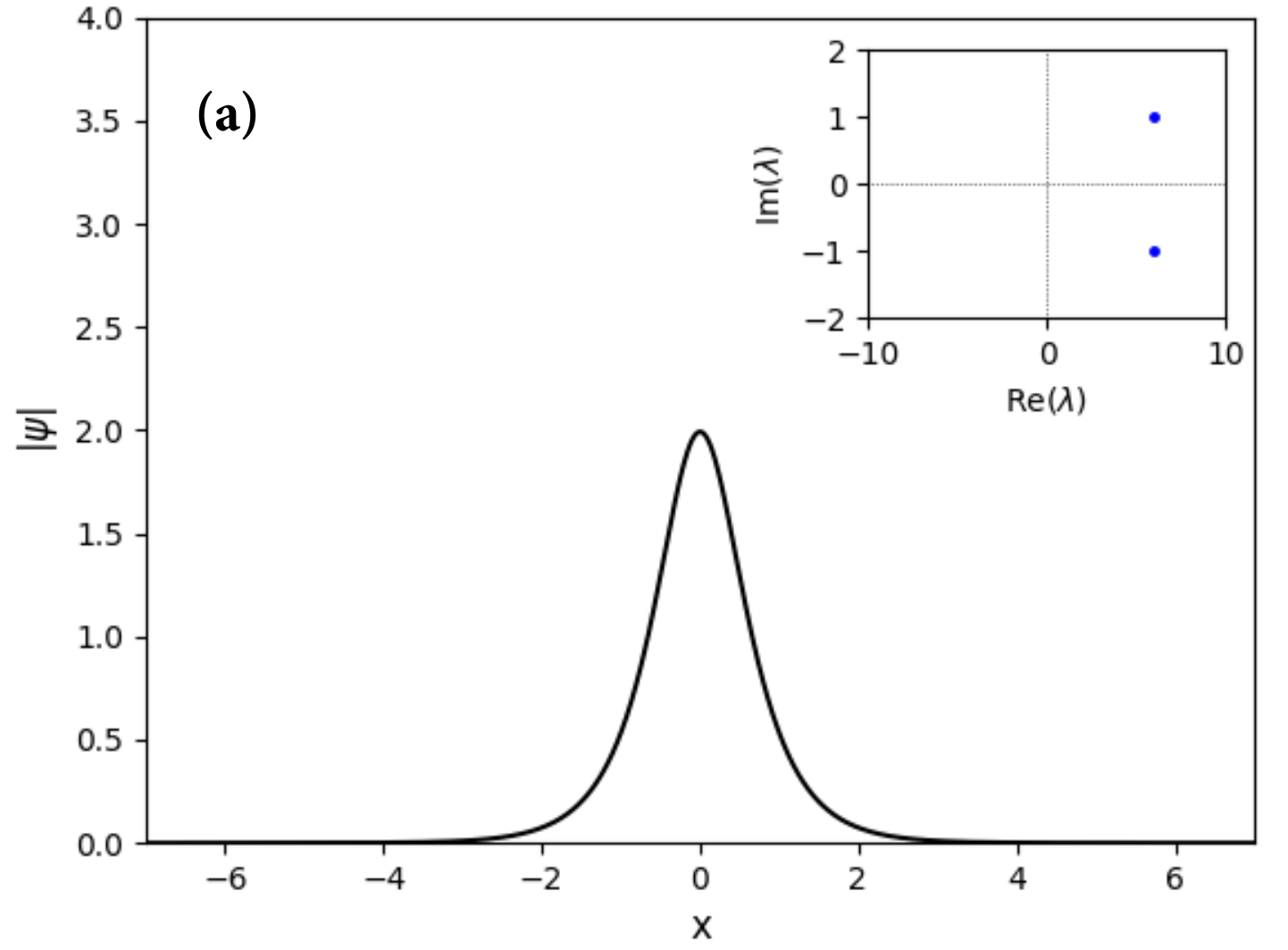} \quad \includegraphics[width= .4 \linewidth]{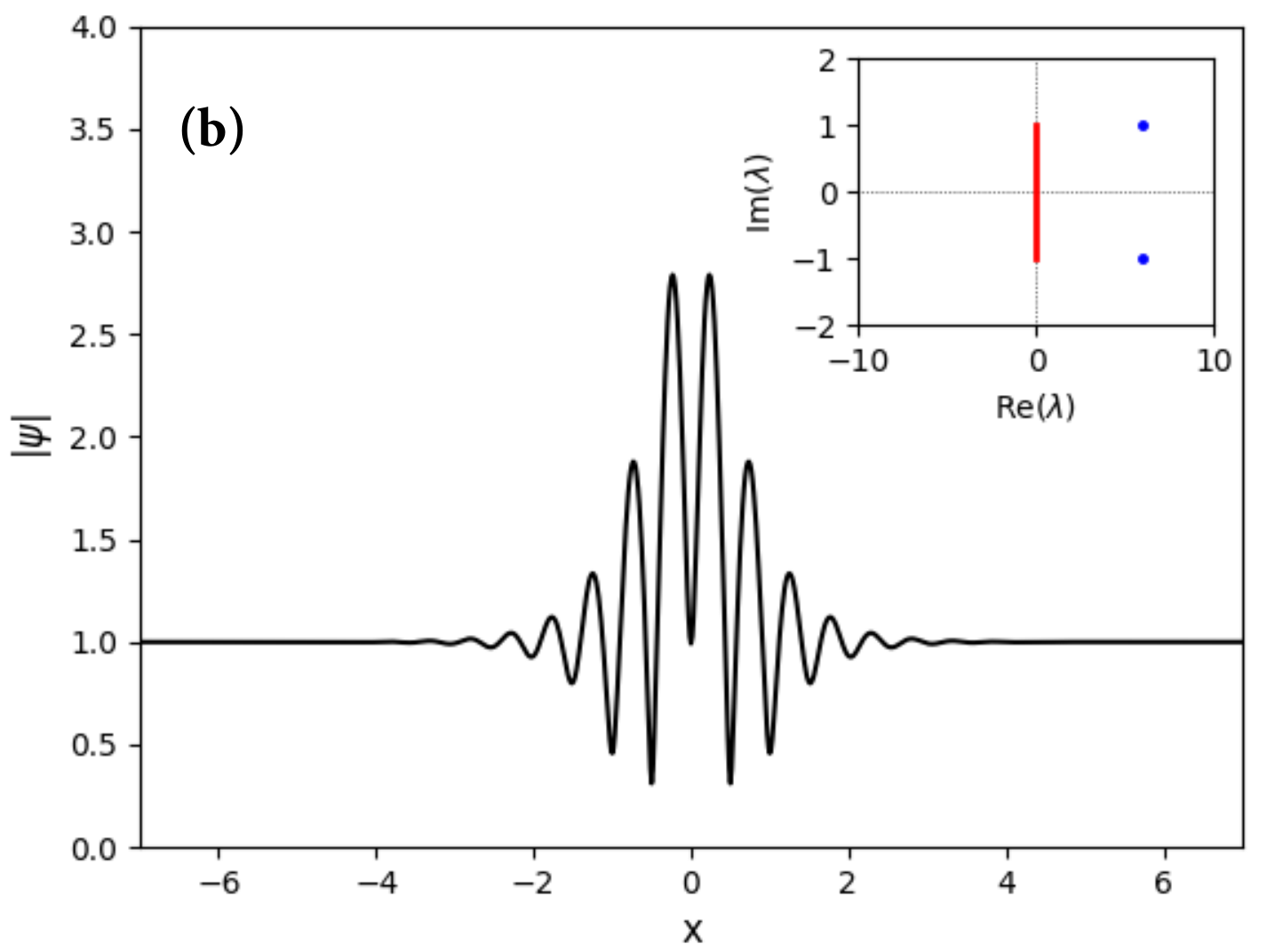}
 \caption{Fundamental soliton (a) and Tajiri-Watanabe (TW) breather (b) solutions of the focusing NLS equation. Shown is $|\psi(x,0)|$ and the  Lax spectrum (IST spectral portrait) (inset).}
 \label{fig:sol_TW}
\end{figure}

\medskip
The fundamental  soliton solution of the focusing NLS equation is given by the formula \eqref{fnls_soliton} (see Fig.~\ref{fig:sol_TW}(a) for the profile of $|\psi(x)|$ in such soliton). The inherent feature of the fundamental solution is that it can only exist on the zero background, which  makes it very different from KdV solitons and solitons in the  defocusing NLS/resonant NLS dispersive hydrodynamics. The IST spectral portrait of the fundamental soliton consists of  two complex conjugate points $\lambda \in \{ \lambda_1,\bar{\lambda}_1 \}$ (see the inset in Fig.~\ref{fig:sol_TW} (a)). 
Apart from  the fundamental solitons \eqref{fnls_soliton} the focusing NLS equation supports a more general family of localised  solutions, called breathers,  that can be viewed as solitons on finite background.  The Lax spectrum of a  breather consists of a vertical band $\gamma_0 = [-iq, iq]$, $q>0$ complemented  by two complex conjugate solitonic discrete points $ \lambda_1,\bar{\lambda}_1$. In physical space such a breather, sometimes called the Tajiri-Watanabe (TW) breather,  represents a localised nonlinear wavepacket propagating on a uniform nonzero background described at $x \to \pm \infty$ by the plane wave solution $\psi =q e^{i 2q^2 t}$ (see Fig.~\ref{fig:sol_TW} (b) for the typical behaviour of $|\psi|$ in the TW breather along with the spectral portrait in the inset). The analytical expression for the TW breather solution is available elsewhere (see e.g. \cite{tajiri_breather_1998}, \cite{slunyaev_nonlinear_2002}, \cite{gelash_formation_2018}), here we only present its group (envelope) and phase (carrier wave) velocities: 
\begin{equation}\label{TW_speed}
c_{g}=-2\frac{\text{Im} [\lambda_1 R_0(\lambda_1)]}{\text{Im} [R_0(\lambda_1)]} ,  \qquad c_p = - \frac{2\text{Re}[\lambda_1 R_0(\lambda_1)]}{\text{Re} [R_0(\lambda_1)]},  
\end{equation} 
where  $\lambda_1$ is the IST solitonic spectral point characterising the TW breather and $R_0(\lambda)=\sqrt{\lambda^2 + q^2}$.
The transition from the TW breather solution to the fundamental soliton \eqref{fnls_soliton} is achieved by vanishing the background,  $q \to 0$.
\begin{figure}[h]
\centering
\includegraphics[width= .8 \linewidth]{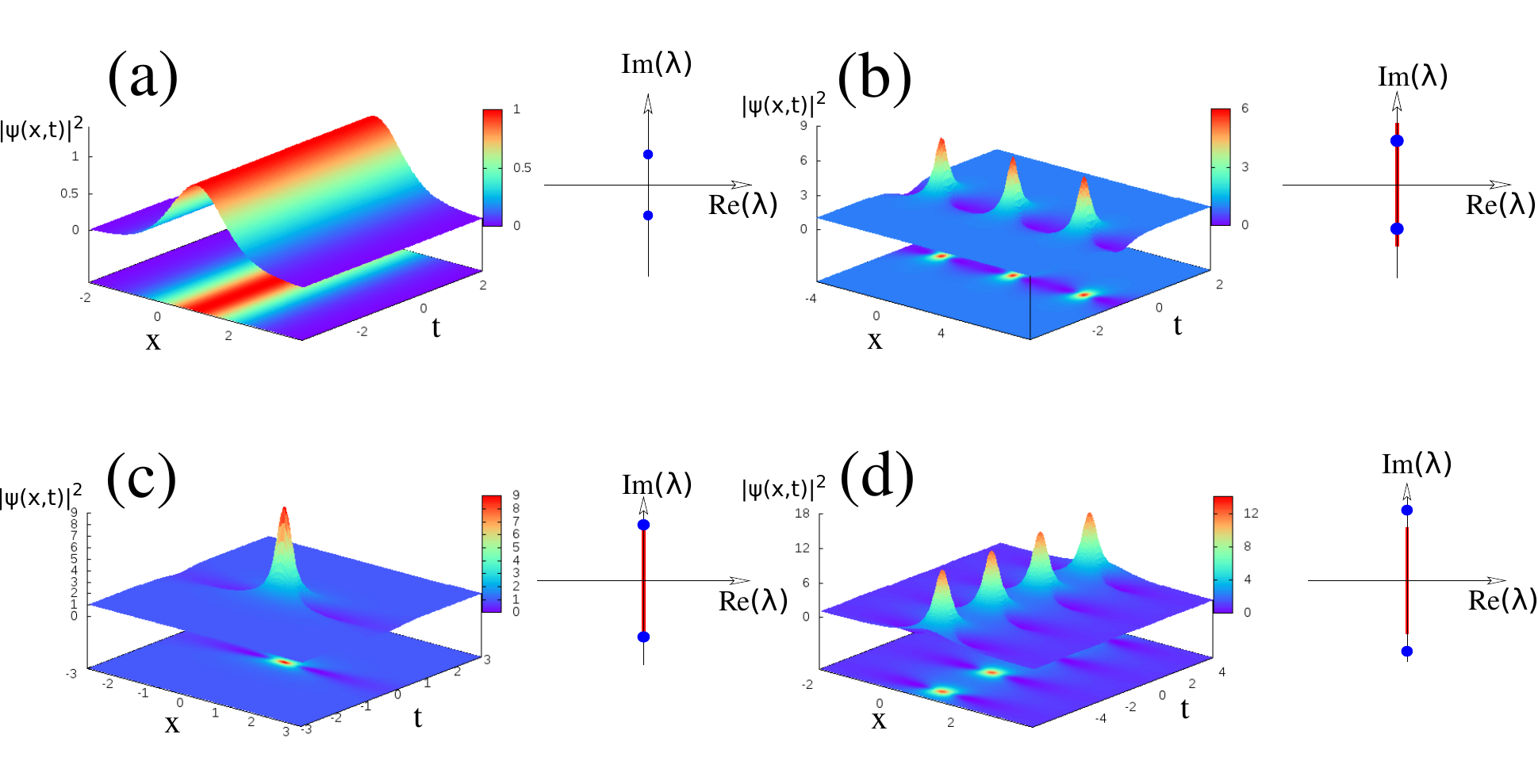} 
 \caption{Adapted from \cite{randoux_inverse_2016}. Solitonic solutions of the FNLS equation along with their IST spectral portraits: (a) fundamental soliton; (b) Akhmediev breather (AB); (c) Peregrine soliton (PS); (d) Kuznetsov-Ma (KM) soliton. }
 \label{fig:breathers}
\end{figure} 

The TW breather has three special reductions, spectrally realised by placing the soliton discrete eigenvalues $ \lambda_1,\bar{\lambda}_1$ on the imaginary axis.   These reductions are
 the Akhmediev breather (AB), the Kuznetsov-Ma (KM) breather and the Peregrine soliton (PS) and they, particularly the PS,  are often considered as `analytical prototypes' for rogue, or anomalous waves (see \eg \cite{akhmediev_rogue_2009}, \cite{kharif_rogue_2009}, \cite{onorato_rogue_2013}, \cite{onorato_rogue_2016} and references therein). All these three special families of breathers possess specific localisation properties: the AB is localised in time and is periodic in space, the KM breather is localised in space and is periodic in time, and finally, the PS exhibits both temporal and spatial localisation. Their IST spectral characterisation  is as follows (see \eg \cite{randoux_inverse_2016}). 
 
 Let the spectral band corresponding to the plane wave background of the breather be  $\g_0 = [-i q; i q]$ for some $q>0$, and  the solitonic discrete spectrum points $\lambda_1= ip$, $\bar \lambda_1= -ip$, $p>0$. Then  $p<q$ corresponds to AB, $p>q$ to KM breather and $p=q$ to PS.  The behaviours of $|\psi(x,t)|$ in the AB, KM and PS breathers, along with their spectral portraits, are displayed in Fig.~\ref{fig:breathers}.

\medskip
{\it Finite-gap solutions}

\begin{figure}[h]
\centering
\includegraphics[width= .4 \linewidth]{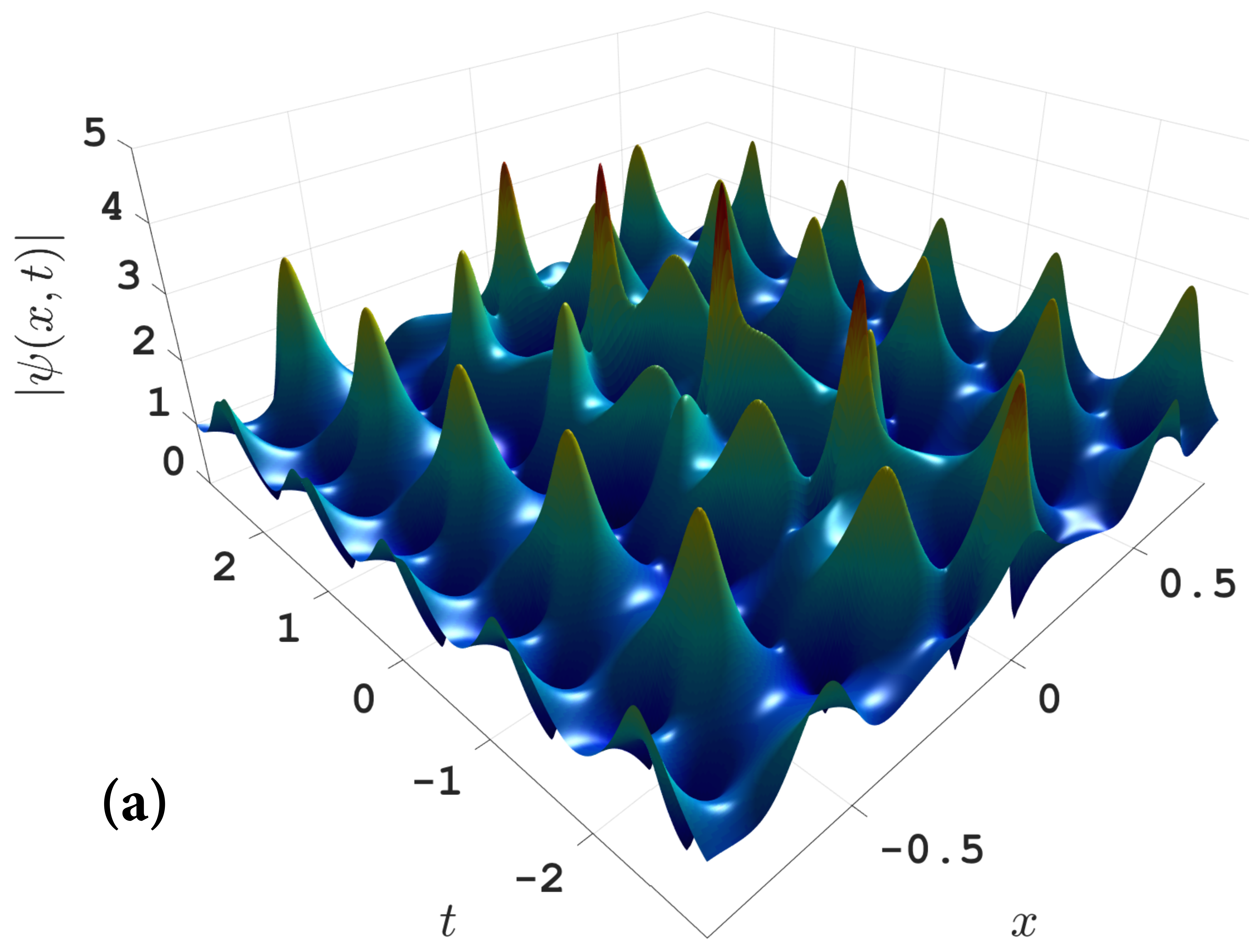} \qquad  \includegraphics[width= .4 \linewidth]{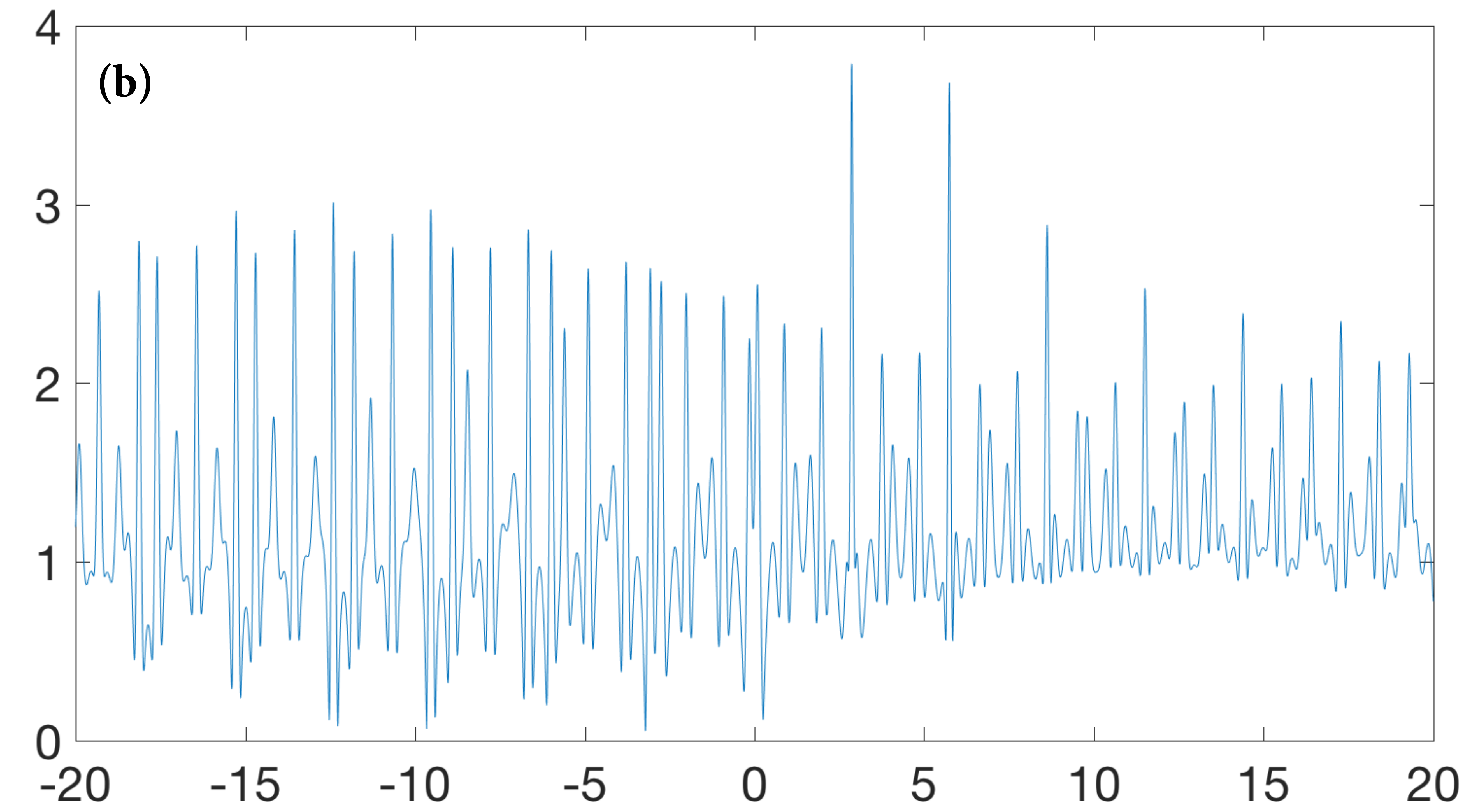} 
 \caption{ Adapted from \cite{bertola_rogue_2016}. Genus $4$ multiphase focusing NLS solution with the endpoints of the spectral bands 
 $\bs\lambda=(-0.39271 +i,   -0.21336 +i,   0.010556+i,     0.20525+i  ,   0.39027+i)$;
 a) Plot of $|\psi_4 (x,t)|$; b) Plot $|\psi_4(x)|$ for a fixed $t$.}
 \label{fig:FG_nls}
\end{figure}

\medskip
Similar to the KdV equation, the focusing NLS equation \eqref{eq:fNLS} supports  finite-gap solutions which transform into the solitonic solutions when the spectral bands collapse. 
An   $n$-gap solution $\psi=\psi_n(x,t)$ of the focusing NLS equation \eqref{eq:fNLS} is defined by a fixed set of  $2n+2$ endpoints $\{\lambda_{ j} , {\bar \lambda}_{ j},  j= 0, 1, 2, \dots, n\}$  of spectral bands  $\g_j$, $j=0,1, \dots, n+1$,  and depends on $n$ real phases 
$\bs{\theta}(x,t)=\bs{k} x-\bs{\o} t +\bs{\theta}^0$ with the initial phase vector $\bs{\theta}^0  \in \T^n$, so that   $|\psi_n(x,t)| = F_n(\bs{\theta}(x,t))$, where $F_n$ is a multi-phase (quasiperiodic) function in both $x$ and $t$, that can be expressed in terms  of the Riemann theta-functions  \cite{its_explicit_1976}. The $n$-component wavenumber $\bs k $  and the frequency $\bs \o $ vectors depend on the endpoints $\{\lambda_{ j} ,  j= 0, 1, 2, \dots, n\} $ of the spectral bands, which define 
a hyperelliptic Riemann surface $\Rscr$ of genus $n$  given by (cf. equation \eqref{RS} for the KdV equation): 
\begin{equation}
\label{rsurf}
R(z)=\prod_{j=0}^{n}(z-\lambda_j)^\hf(z-\bar\lambda_j)^\hf,\quad
\lambda_j = a_j + i b_j, \  \ b_j>0,  \end{equation}
$z \in \C$ being a complex spectral parameter in the Zakharov Shabat scattering problem; ${R_n(z)} \sim z^{n+1}$ as $z \to \infty$. 
The branch cuts of $R(z)$ are made along spectral bands which will be specified below.   An example of the behaviour of $|\psi(x,t)|$ in a genus $4$ solution is shown in Fig.~\ref{fig:FG_nls}. 

There are two qualitatively different types of spectral bands characterising finite-gap NLS solutions: (i) the bands that cross the real axis in the complex spectral plane $z \in \mathbb{C}$ and connect the complex conjugate spectral points $\lambda_j$ and $\bar \lambda_j$ (these are often called the Stokes bands, see \eg \cite{osborne_nonlinear_2010}); (ii) the Schwartz symmetric bands that do not cross the real axis (we shall call them the solitonic bands). By manipulating the spectral bands  one can achieve various wave configurations for the field $\psi_n(x,t)$. E.g. collapsing a pair of solitonic bands into two complex conjugate double points of the spectrum gives rise to a localised mode in the solution, a soliton on the finite-gap background, a generalised breather (see e.g. \cite{chen_rogue_2019}). Collapsing a Stokes band into a double point on the real axis implies vanishing of one of the background plane wave modes. By collapsing all Stokes and solitonic bands into double points of the spectrum  a finite-gap solution is transformed into a multi-soliton solution. Keeping one of the Stokes bands finite, but collapsing all other bands implies a multi-breather solution. 

The finite-gap analogues of the TW breather solutions and their reductions (AB, KM and PS) have genus $n=2$ and contain one Stokes band in their Lax spectrum.  We shall consider a generalisation of these solutions for an arbitrary genus and, having in mind the construction of soliton and breather gas,  make two simplifying assumptions: 

\smallskip
{\it (i) assume an even genus $n=2N$ of the spectral Riemann surface; }

\begin{figure}[h]
\centering
\includegraphics[width= .5 \linewidth]{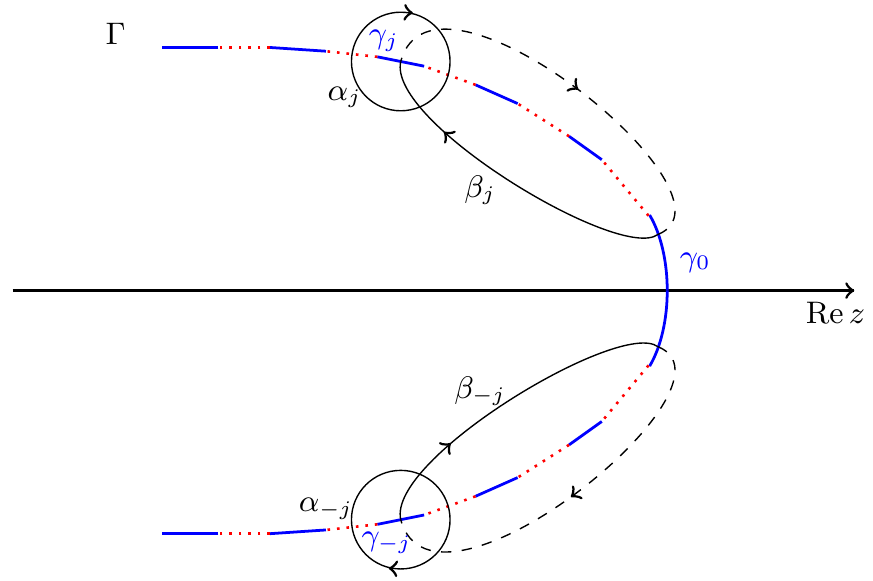}
 \caption{The spectral geometry of the finite-gap solution of the focusing NLS equation and the canonical homology basis used in the construction of breather and soliton gases. The branch cuts (bands) $\gamma_{\pm j}$  of the hyperelliptic Riemann surface $\Rscr$  and the gaps $c_{\pm j}$ (shown by red dotted lines) are located on the 1D Schwarz symmetric curve $\Gamma$.}
 \label{Fig:Cont}
\end{figure}

{\it (ii) assume that the Lax spectrum of the finite-gap potential is located on a 
Schwarz symmetric, simply connected 1D curve $\G\subset \C$, see Fig.~\ref{Fig:Cont}.} 

\smallskip
Thus we have one  Stokes band $\gamma_0$ between $\lambda_0$ and $\overline \lambda_0$ and $2N$ solitonic bands: $\gamma_j$, $j=1, \dots, N$  between $\lambda_{2j-1}$ and $\lambda_{2j}$, and their Schwarz symmetric counterparts $\gamma_{-j}$,  $j=1, \dots, N$ between  $\overline \lambda_{2j-1}$ and $\overline \lambda_{2j}$ in the lower half-plane. We note that the described spectral geometry also covers the case of an odd genus (achieved by collapsing the Stokes band)  and a `bound state' configuration, when $\G$ lies on a  vertical line so that all the solitons or breathers corresponding to the collapsed bands  have the same velocity.
Due to the symmetry of the curve $\G$   it is sufficient to consider only the upper complex half-plane {($\C^+$)} part  of it, which we denote $\G^+$ {(so that $\G^+=\G\cap\C^+$)}

While appearing quite restrictive, the described 1D spectral geometry provides a major insight into the  properties of breather and soliton gases and admits a straightforward generalisation in the more  physically realistic  case, where  the  spectral bands $\g_j$  are located in some (Schwarz symmetric) 2D region $\L \subset \C$.  It also admits  generalisation to the case of more than one Stokes bands.

\medskip

{\it Wavenumbers and Frequencies}

\medskip

The wavenumber and frequency vectors, $\bs k$ and $\bs \o$ respectively,  associated with a given finite-gap  solution $\psi_{2N}(x,t)$ of the focusing NLS equation, are not uniquely defined since any  linear combination  of the wavenumber (frequency) vector components with integer coefficients is also a wavenumber (frequency).  
The construction of the thermodynamic limit outlined in Section \ref{sec:therm_lim_kdv} and realised in Section \ref{sec:spectral_theory_kdv} for the KdV equation relies on the availability of  the wavenumbers and frequencies that vanish in the limit when the relevant bands collapse into double points. Such wavenumbers and frequencies automatically appear in the finite-gap KdV theory \cite{flaschka_multiphase_1980}.  The situation with the focusing NLS equation is different since the sets of the wavenumbers  and frequencies that arise `naturally' in the focusing NLS finite-gap theory (see \eg \cite{dafermos_geometry_1986}, \cite{tracy_nonlinear_1988}, \cite{tovbis_semiclassical_2016}) do not necessarily possess the requisite property.

The special, {\it fundamental}  wavenumber and frequency vectors   
\be
{\bs k} = (k_1, \dots, k_N, \tilde k_1, \dots, \tilde k_N) \ \ \hbox{and}    \quad {\bs \o} = (\o_1, \dots, \o_N, \tilde \o_1, \dots, \tilde \o_N),
\ee
that possess the  properties necessary for the application of the thermodynamic limit have been identified  in \cite{el_spectral_2020}. The definitions of the fundamental vectors ${\bs k}$ and $\bs \o$ can be found in the Appendix (see \eqref{kdp}, \eqref{omdq}). Here we only highlight their main properties and the principal differences between the components $\{k_1, \dots, k_N\}, \{ \o_1, \dots, \o_N\}$ and $\{\tilde k_1, \dots, \tilde k_N\}, \{\tilde \o_1, \dots, \tilde \o_N\}$.

Following the KdV construction in Section \ref{sec:spectral_theory_kdv} we introduce two new quantities characterising the finite-gap potentials instead of the endpoints of spectral bands (we consider only the upper half-plane with posistive indices of spectral bands $\gamma_j$ and gaps $c_j$  for convenience):
  \be\label{eta_delta}
\eta_j=\frac12(\lambda_{2j-1}+\lambda_{2j}),~~~~\delta_j=\frac{1}{2}(\lambda_{2j}-\lambda_{2j-1}),
\ee
where $j=1,\dots, N$.  We shall call  the point $\eta_j$  the centre of the $j$-th band $\g_j$ and $2 |\delta_j|$ the $j$-th band width.   Also for the Stokes band we have $\delta_0 =  i \text{Im} \lambda_0$.
 Note that the notations and the band numeration we are using here are slightly different from those used in \cite{el_spectral_2020}. 

 It then follows that the fundamental wavenumbers and frequencies defined by  \eqref{kdp} and \eqref{omdq} have drastically different asymptotic properties in the soliton/breather limit, when one of the solitonic spectral bands collapses into a double point, $\lambda_{2i-1}, \lambda_{2j} \to \eta_j $. Namely,
\bea
&& \delta_j \to 0
  \ \  \implies 
    k_j, \o_j \to 0,  \quad \tilde k_j, \tilde \o_j = \mathcal{O}(1),  \label{breather_lim} \\ 
&& j=1, \dots, N. \nonumber
\eea 
In particular, for  $N=1$ (genus 2), the limit  \eqref{breather_lim} ($k_1 \to 0$, $\omega_1 \to 0$) with   non-zero Stokes band $\g_0$  (i.e.  $\delta_0 \ne 0$) corresponds to the TW breather limit 
of the  two-phase nonlinear wave solution. The remaining wavenumber and frequency
$\tilde k_1= \mathcal{O}(1)$,  $\tilde \o_1 = \mathcal{O}(1)$ correspond to the ``carrier'' wave of the TW breather (see Fig.~\ref{fig:sol_TW}).  If we further tend $\delta_0 \to 0$, then $\tilde k_1 \to 0$, $\tilde \o \to 0$ and the TW breather transforms into a fundamental soliton.
Motivated by these properties for $N=1$ we shall  call the components $k_j$,  $\o_j$  of the wavenumber and the frequency vectors   $\bs k$ and $\bs \o$  the {\it solitonic components} and the components $\tilde k_j$, $\tilde \o_j$---the {\it carrier components}. The solitonic components $k_j,\ \o_j$ are the focusing NLS counterparts of the conventional KdV wavenumbers and frequencies specified by \eqref{kdv_nonlin_disp} while the carrier components  $\tilde k_j$, $\tilde \o_j$ do not have analogues in the KdV theory. 

Generally the limit \eqref{breather_lim} for finite $N$ corresponds to $N$-breather solution if $\delta_0 \ne 0$ and to $N$-soliton solution if $\delta_0=0$.

\bigskip
{\it Nonlinear dispersion relations}

\medskip

It was shown in \cite{el_spectral_2020} that the solitonic components $k_m$, $\o_m$  satisfy the following nonlinear dispersion relations (again, considering only the upper half plane):
\bea 
&&\sum_{m=1}^{N} k_m \ \text{Im}\oint_{\beta_m}{{P_j(\zeta)d\zeta}\over{R(\zeta)}}=\pi \text{Re}\, \kappa_{j,1},\nonumber \\
&&\sum_{m=1}^{N} \o_m \ \text{Im}\oint_{\beta_m}{{P_j(\zeta)d\zeta} 
\over{R(\zeta)}}=2\pi \text{Re} (\kappa_{j,1}\sum_{k=0}^{2N}\text{Re} \, \lambda_k +\kappa_{j,2}), \nonumber \\
&& j= 1,\dots, N, \label{WFR}
 \eea
 where 
\be\label{Pj0}
P_j(z)=\kappa_{j,1}z^{2N-1}+\kappa_{j,2}z^{2N-2}+ \dots +\kappa_{j,2N},
\ee
and $\kappa_{i,j}$ are the coefficients of the normalized holomorphic differentials $w_j$  defined by: 
\be\label{D-P0}
w_j=[P_j(z)/R(z)] dz, \quad \oint_{\alpha_i}w_j = \delta_{ij}, \quad i, j= 1, \dots,  N.
\ee 
The contours $\alpha_j$, $\beta_j$   on the Riemann surface $\Rscr$ are defined as follows (see Fig.~\ref{Fig:Cont}): the  $\alpha_j$-contours  surround the bands $\g_j$  clockwise, and  the  $\beta_j$-contours  start at the Stokes  band $\g_0$,  go to the  band $\g_j$  on the upper sheet clockwise and return to the  band  $\g_0$ on the lower sheet of the Riemann surface.  The derivation of \eqref{WFR} is outlined in  Appendix.

 Relations \eqref{WFR}  are the analogs  of the nonlinear dispersion relations \eqref{kdv_nonlin_disp} for the finite-gap potentials of the KdV equation. 
 Similar relations are  available for the carrier components   \cite{el_spectral_2020} 
 but we do not present them here as they play a secondary role in the thermodynamic limit  construction.

\subsubsection{Thermodynamic spectral scalings}
\label{sec:therm_fnls}

We now fix the endpoints of the band spectrum, $\lambda_1=a$, $\lambda_{2N}=b$, and assume that for $N \gg 1$ the centres  
$\eta_j$ of the bands $\g_j$, $j=1,\dots,N,$ are distributed  along  $\text{arc}(a,b) \subset \Gamma^+$
with some limiting density $\varphi(\la)>0$,  $\la \in\G^+$,
that is smooth on  $\G^+$, so that $\int_a^b \varphi(\mu) |\rmd \mu| =1$. It then follows that $|\eta_j - \eta_{j+1}|  \sim 1/N$.

 As for the scaling of the band widths, we consider the following options (cf. eq. \eqref{exp_scale} for KdV): 
 \begin{itemize}
  \item 
 [(i)] {\it exponential spectral scaling}: the band widths $|\delta_j|$   are exponentially small in $N$:
 \be\label{exp_scal}
 |\delta_j|\sim e^{-N \tau{(\eta_j)}}, \quad j= 1, \dots, N,
 \ee
 where $\tau(\la)$ is a smooth {positive}
 function on $\G^+$ having the meaning of the normalised logarithmic  band width ($\tau(\eta_j) \sim -\ln |\delta_j|/N)$. 
  \item 
 [(ii)] 
 {\it sub-exponential spectral scaling}: for any $a>0$
\be\label{sub-exp_scal}
e^{- a N } \ll  |\delta_j|  \ll \frac{1}{N}, \quad j= 1, \dots, N,
 \ee
  It is clear that in this limit $\tau(\lambda) \to 0$.
\item 
 [(iii)] 
 {\it super-exponential spectral scaling}: for any $a>0$
\be \label{sup-exp_scal}
e^{-a N } \gg  |\delta_j|,  \quad j= 1, \dots, N. 
\ee
In this limit $\tau(\la) \to \infty$. 
\end{itemize}

 Note that in all three scalings $|\delta_j| \ll |\eta_j-\eta_{j+1}|$, hence the gap width  $|c_j| \sim N^{-1}$, $j=1, 2, \dots, N$ and  so $|\delta_j|/|c_j| \to 0$ as $N \to \infty$. We then say that in the limit each  collapsed band $\g_j \to \eta_j$ corresponds to a soliton (breather) state within a soliton (breather) gas. 
 We remind that for breather gas $\delta_0 \ne 0$ and for soliton gas $\delta_0=0$.
 
 We also note that the exponential and sub-exponential spectral scalings have the `thermodynamic' property in the sense that they preserve finiteness of the total density  of waves  $K_N = \sum_{j=1}^N k_{j}$  in the limit $N\to \infty$  so that $\lim_{N \to \infty} K_N= \beta $,  where $0 < \beta <\infty$.  Note that for the super-exponential scaling $\beta \to 0$.

\subsubsection{Nonlinear dispersion relations and kinetic equation }
We first present the results for the general case of breather gas by evaluating the thermodynamic limit of the nonlinear dispersion relations \eqref{WFR}  for  the exponential spectral scaling \eqref{exp_scal}. Without much loss of generality we assume that the Stokes band lies on the imaginary axis, $\gamma_0 = [-iq, iq]$, $q > 0$ (so that $\delta_0=iq$) and that $\eta_i \in \text{arc}(a,b) \subset \Gamma^+$, and introduce the simultaneous  scaling
for the solitonic wavenumbers and frequencies (cf. \eqref{komN})
\begin{equation}\label{scal_k_nu}
k_j \sim \frac{\varkappa_j}{N}, \qquad \o_j \sim \frac{\nu_j}{N}, \qquad N \gg 1,
\end{equation}
so that  $\varkappa_j=\varkappa(\eta_j)$ and $\nu_j=\nu(\eta_j)$, where  the functions $\varkappa(\la)\geq 0$ {and  $\nu(\la)$} are smooth interpolations of $\varkappa_j,\nu_j$. Similar to the KdV case, the scaling \eqref{scal_k_nu}
provides a balance of terms in  relations \eqref{WFR} for $N \gg 1$. 

 The resulting nonlinear dispersion relations for breather gas have the form (we refer the reader to \cite{el_spectral_2020} for  details of the derivation): 
\bea\label{dr_breather_gas1}
 \int_{\G^+}
D(\lambda, \mu)   f(\m) |\rmd\m|
+\sigma(\lambda)f(\la) 
= \text{Im} [R_0(\la)], \\
 \int_{\G^+}
D(\la, \mu) v(\mu) |\rmd\m|
+\sigma(\la)v(\la)
= - 2\text{Im} [\la R_0(\la)] \label{dr_breather_gas2},
\eea
where $R_0(z)=\sqrt{z^2 + q^2}$ (with the branch cut $[ -iq, iq]$,  and the branch of the radical  defined by $R_0(z) \to z$ as $z \to \infty$), and with a slight abuse of notation, we denoted $ \int_{a}^{b} \dots |d\mu| \equiv \int_{\G^+} \dots |d \mu|$. Further, 
\be
D(\la, \mu)=\left[\ln\left| \frac{\mu-\bar\la}{\mu-\la}\right|+ \ln\left|\frac{R_0(\la)R_0(\mu)+\la \mu + q^2}
{R_0(\bar\la)R_0(\m)+\bar\la \m + q^2}\right|\right],
\ee
\begin{equation}\label{dens_states}
f(\la)= \frac{1}{2\pi} \varkappa(\la) \varphi(\la),\ \  v(\la)= \frac{1}{2\pi}\nu(\la) \varphi (\la), \ \ \sigma(\la)=\frac{2\tau(\la)}{\varphi(\la)}.
\end{equation}
Equations \eqref{dr_breather_gas1}, \eqref{dr_breather_gas2} are the focusing NLS counterparts of the nonlinear dispersion relations \eqref{inta}, \eqref{intat} for the KdV soliton gas with $f(\la)$ and  $v(\la)$ having the meanings of the DOS  and the spectral flux density respectively, and the function $\sigma(\la)$ encoding the  Zakharov-Shabat spectrum of the finite gap potentials in the thermodynamic limit. 

The soliton gas limit in \eqref{dr_breather_gas1}, \eqref{dr_breather_gas2}  is achieved by vanishing the Stokes band, $q\to 0$, resulting in
\bea
 \int _{\G^+}\ln \left|\frac{\m-\bar\la}{\m-\la}\right|
f(\m)|\rmd\m|+\sigma(\la)f(\la)&= &\text{Im} \la, \label{dr_soliton_gas1} \\
 \int _{\G^+}\ln \left|\frac{\m-\bar\la}{\m-\la}\right| v(\m)|\rmd\m|+  \sigma(\la) v(\la) &=& -4 \text{Im} \la \, \text{Re}\la. \label{dr_soliton_gas2} 
\eea

Eliminating the function $\sigma(\lambda)$ from the nonlinear dispersion relations \eqref{dr_breather_gas1}, \eqref{dr_breather_gas2} and \eqref{dr_soliton_gas1}, \eqref{dr_soliton_gas2} we obtain the equation of state
\begin{equation}\label{eq_state_fnls1}
s(\la) = s_0(\la) + \int_{\G^+}G(\la, \mu)[s(\la) - s(\mu)] f(\mu) |\rmd \mu|,
\end{equation}
where  $s(\la)=v(\la)/f(\la)$ is the effective velocity of the tracer soliton (breather) in the gas, and  $s_0(\la)$, $G(\la, \mu)$ are defined as follows. 

For breather gas: 
\begin{equation}\label{eq_state_breather}
s_0(\la) =  -2  \frac{\text{Im}[\la R_0(\la)]}{\text{Im} [R_0(\la)]}, \qquad G(\la, \mu) =\frac{1}{\text{Im}[R_0(\la)]} D(\la, \mu),
\end{equation}
and for soliton gas:
\begin{equation}\label{eq_state_sol}
s_0(\la) = -4 \text{Re} \la, \quad G(\la, \mu) = \frac{1}{ \text{Im} \la}\ln \left|\frac{\m-\bar\la}{\m-\la}\right|.
\end{equation}

\medskip
One can see that $s_0(\la)$ in \eqref{eq_state_breather} coincides with the group velocity $c_g(\la)$ of an isolated TW breather  \eqref{TW_speed} and $s_0(\la)$ in \eqref{eq_state_sol}  coincides with the group velocity of the fundamental  soliton of the focusing NLS equation. Furthermore, the integral kernel $G(\lambda, \mu)$ in \eqref{eq_state_sol} coincides with the  absolute value of the position shift \eqref{fnls_phase_shift} in the  two-soliton collision for the focusing NLS equation. As one may expect,   the expression for $G(\la, \mu)$ in  eq. \eqref{eq_state_breather} is identified with  the position shift in two-breather collisions \cite{li_soliton_2018}, \cite{gelash_formation_2018} (see the proof in \cite{roberti_numerical_2021}).
Thus the outlined derivation provides the justification of the collision rate assumption \eqref{collision_rate} for dense soliton and breather gases of the focusing NLS equation.  

The integral equations \eqref{dr_soliton_gas1}, \eqref{dr_soliton_gas2} have been recently studied  in a rather general spectral geometry \cite{kuijlaars_minimal_2021}. The existence and uniqueness  of their solutions  was investigated and the non-negativity of the solutions for the DOS $f(\la)$ was proved.

\medskip
{\it Spectral kinetic equation for nonequilibrium  gas}

\medskip
The derivation of the transport equation for non-equilibrium soliton and breather gases for the focusing NLS equation follows the general modulation construction outlined in Section \ref{sec:big_pic} and realised for the KdV equation in Section~\ref{sec:kin_eq_KdV}, albeit with some important differences. 

The multiphase Whitham modulation theory for $n$-gap solutions of the focusing NLS equation \cite{dafermos_geometry_1986}, \cite{tovbis_semiclassical_2016} includes the system of $n=2N$ conservation laws ${\bs k}_t + {\bs \o}_x=0$, which is split into two distinct subsystems
\begin{eqnarray}
&\partial_t k_j({\bs \la})+ \partial_x \omega_j ({\bs \la})=0, \quad j=1, \dots, N \label{wc11}, \\
&\partial_t \tilde k_j ({\bs \la})+ \partial_x \tilde \omega_j({\bs \la})=0, \quad j=1, \dots, N  \label{wc_tilde11}
\end{eqnarray}  
 for the solitonic and carrier components  of the fundamental wavenumber and frequency vectors (see Section~\ref{sec:FNLS_sol}). These subsystems are complemented by the respective nonlinear dispersion relations
 \bea
 & { k}_j = K_j({\bs \la}), \quad \o_j = \Omega_j({\bs \la}), \quad j=1, \dots, N, \label{nlindr1}  \\
& \tilde { k}_j = \tilde   K_j({\bs \la}), \quad \tilde  {\o}_j = \tilde  \Omega_j({\bs \la}), \quad j=1, \dots, N, \label{nlindr2}
 \eea
obtained by taking the imaginary and the real part of the basic system \eqref{WUPjM} (see Appendix).  In particular, the dispersion relations for the solitonic components are explicitly given by the system \eqref{WFR}.  
 
 Application of the thermodynamic limit to the  modulation system \eqref{wc11}, \eqref{nlindr1}  for the solitonic wavenumbers (see  Section~\ref{sec:kin_eq_KdV} for the similar derivation in the KdV context) yields the spectral transport equation 
\be \label{kineqnls}
f_t +(fs)_x=0
\ee
for the DOS $f(\la, x, t)$. The transport equation \eqref{kineqnls} is complemented by the equation of state \eqref{eq_state_fnls1},  \eqref{eq_state_breather}  for breather gas (or \eqref{eq_state_fnls1}, \eqref{eq_state_sol} for soliton gas). 
One can see that  the kinetic equation for soliton gas agrees with equation \eqref{FNLS_kin} derived via the phenomenological approach based on the collision rate assumption \eqref{collision_rate}.

The carrier wave modulation system \eqref{wc_tilde11}, \eqref{nlindr2} in the thermodynamic limit becomes (see \cite{el_spectral_2020} for details)
 \be
  \tilde f_t + (\tilde f \tilde s)_x=0, \quad \tilde s(\la) = \tilde{ \mathcal{S}}[f(\la)], 
\ee
where $\tilde f(\lambda, x, t)$ and $\tilde f \tilde s=\tilde v(\lambda, x,t)$ are some smooth functions on $\G^+$ interpolating $\tilde k_j, \tilde \o_j$, that is,
$\tilde f(\eta_j)=\tilde k_j$, $\tilde v(\eta_j)=\tilde \o_j$, $j=1,\dots,N$, where $\eta_j$ are the centres of the bands \eqref{eta_delta}. The dependencies of $\tilde s(\la)$ on the DOS $f(\la)$ for soliton and breather gases are given by somewhat lengthy expressions which we do not present  here. We only mention that in the limit of zero density, $f(\la)\to 0$, the  quantity $\tilde s$ transforms into the phase velocity $s_p(\la)= -{2 \text{Re} [\la^2]}/{\text{Re} \la}$ of the fundamental soliton \eqref{fnls_soliton} if $q=0$ or the phase velocity $c_p (\la)= - {2\text{Re}[\lambda R_0(\lambda)]}/{\text{Re} [R_0(\lambda)]}$ \eqref{TW_speed} of the TW breather  ($q \ne 0$).

\bigskip
{\it Generalisation to {2D} Case}

\medskip
The above derivation  of the spectral kinetic equation for the focusing NLS soliton and breather gases involves the basic assumption that the DOS $f(\lambda)$ is supported on a 1D symmetric curve $\Gamma$ in the complex plane. This restriction can be readily removed as we show below.

 In the case  when the shrinking  bands $\g_j$, $j>0$ fill a compact 2D region $\L^+$ of the upper complex half-plane,
 the  counterpart of the exponential scaling \eqref{exp_scal} is
 \begin{equation}\label{exp_2D_scal}
 |\delta_j|\sim e^{-N^2 \tau{(\eta_j)}},
 \end{equation}
where $\tau(\la)$ is a positive smooth function
 on $\L^+$. The scaling of the gaps remains 
 $\mathcal{O}(1/N)$, where by the gap width we understand the closest distance  between the bands. 
 In this case  $\varphi(\la)>0$ is the 2D density of bands
(and we also distinguish the cases of exponential,
 sub-exponential and super-exponential scalings of bands, similarly to the 1D case). 

We now assume one of the 2D spectral thermodynamic spectral scalings (exponential, sub-exponential, and super-exponential) when the shrinking  bands $\g_j$ fill a 2D region $\L$ of the complex plane, see  Section III B. 
For the wave numbers and frequencies instead of \eqref{scal_k_nu} we introduce
\begin{equation}\label{scal_k_nu2}
k_j=\frac{\varkappa_j}{N^2}, \qquad \o_j=\frac{\nu_j}{N^2}, \qquad N \gg 1,
\end{equation}
where $\varkappa_j=\varkappa(\eta_j)$ and $\nu_j=\nu(\eta_j)$, and the interpolating  functions $\varkappa(\lambda) \geq 0$, $\nu(\lambda) $ are assumed to be smooth  on $\L^+$.

Then the  2D thermodynamic limit of the nonlinear dispersion relations \eqref{WFR} leads to the same
integral equations \eqref{dr_breather_gas1}-\eqref{eq_state_fnls1} but with the line integration along $\G^+$ replaced by the integration over a 2D compact domain $\L^+$:
\begin{equation}\label{2D_int}
\int \limits_{\G^+} \dots |\rmd \mu| \to \iint \limits_{\Lambda^+}\dots \rmd \xi \rmd\zeta\, ,
\end{equation}
where $\mu = \xi +i \zeta$.

For convenience  of the exposition we shall be using the  notation $\int _{\G^+} \dots |\rmd \mu|$ in both 1D and 2D cases
keeping in mind that in {the 2D} case the meaning of the integral is given by Eq. \eqref{2D_int}.

\subsubsection{Rarefied soliton gas and soliton condensate}
\label{sec:sol_cond}

The dispersion relations and the equations of state for soliton and breather gases were derived under the assumption of the exponential spectral scaling \eqref{exp_scal} (or \eqref{exp_2D_scal} in 2D case). 
We now consider the two other scalings of interest: the superexponential scaling \eqref{sup-exp_scal} and subexponential scaling \eqref{sub-exp_scal}. We will restrict our exposition  to soliton gas but will indicate when the results can be  extended to breather gas.

 It is convenient to characterise the spectral scalings and the corresponding soliton gases in terms of the function $\sigma(\lambda)$ parametrising the nonlinear dispersion relations \eqref{dr_breather_gas1} and \eqref{dr_soliton_gas1}. 
 From Eq.~\eqref{dr_breather_gas1} we have:
\begin{equation}\label{sigma_bg}
 \sigma (\la)=\frac{\text{Im} \la -\int_{\G^+} \ln \left|\frac{\m-\bar\la}{\m-\la}\right|
 f(\mu) |\rmd\mu|}{f(\la)} \ge 0,
\end{equation}
(there is a similar expression for breather gas which we do not present here).
 For the exponential scaling $\sigma(\lambda)=\mathcal{O}(1)$, while the limiting cases $\sigma \to \infty$ and $\sigma \to 0$ correspond to the super- and sub-exponential spectral scalings respectively. 
 
 \newpage
 {\it Rarefied soliton gas}
 
 \medskip
 Rarefied soliton gas represents an infinite random ensemble of  weakly interacting solitons characterized by a small DOS, $f \ll 1$, and therefore, $\sigma \gg 1$ by \eqref{sigma_bg}.  We shall refer to the limit $f \to 0$,  $\sigma \to \infty$,  $f \sigma = \mathcal{O}(1)$ as the {\it ideal gas limit} as it corresponds to the gas of non-interacting breathers (solitons). Spectrally this limit corresponds to the super-exponential spectral scaling \eqref{sup-exp_scal}.  
 
For a rarefied gas the interaction (integral) term in the equation of state \eqref{eq_state_fnls1} is sub-dominant so the leading order term $s(\lambda) = s_0(\lambda)$ describes the group velocity distribution in an ideal breather (soliton) gas. Then the  first correction to the ideal gas velocity $s_0(\lambda)$ is readily computed to give:
\begin{equation}\label{rarefied_group}
s(\lambda) \approx  s_0(\lambda) + \int \limits_{\G^+} G (\lambda, \mu)[s_0(\lambda) - s_0(\mu)] f(\mu) |\rmd \mu|.
\end{equation}
Eq.~\eqref{rarefied_group} represents the  focusing NLS counterpart of the equation \eqref{s1} for the effective soliton velocity in a rarefied KdV soliton gas introduced by Zakharov \cite{zakharov_kinetic_1971}. 

All the above results apply to breather gas as well.
 
\bigskip
 {\it  Soliton condensate}

\medskip
The  inequality $\sigma(\la)  \ge 0$  in \eqref{sigma_bg} imposes a fundamental constraint 
on the DOS  $f(\la)$. As follows from the discussion in Sec.~\ref{sec:therm_fnls}, the critical value $\sigma(\eta)=0$  corresponds to the sub-exponential spectral scaling \eqref{sub-exp_scal}.  One can see from the nonlinear dispersion relations \eqref{dr_soliton_gas1}, \eqref{dr_soliton_gas2}  that in this case the gas properties are fully determined by the interaction (integral) terms, while the information about the individual quasi-particles  (described by the non-integral, secular, terms) is completely lost.  By analogy with Bose-Einstein condensation we shall call the soliton gas at $\sigma=0$ the  {\it soliton condensate}. From   \eqref{dr_soliton_gas1}, \eqref{dr_soliton_gas2} we obtain  the dispersion relations for the focusing NLS soliton condensate
 \begin{equation}\label{crit_u}
 \int_{\G^+} \ln \left|\frac{\m-\bar\la}{\m-\la}\right| 
f(\mu) |\rmd \mu| = \text{Im} \la, \quad  \int _{\G^+}\ln \left|\frac{\m-\bar\la}{\m-\la}\right| v(\m)|\rmd\m| = -4 \text{Im} \la \, \text{Re}\la.\end{equation}
  Equations~\eqref{crit_u} represent integral equations (the Fredholm equations of the first kind) for the critical density of states $f=f_{\rm c}(\la)$ and the corresponding  spectral flux density $v=v{\rm_c}(\la)$. 
 The existence and uniqueness of their solutions depend on the geometry of the spectral locus domain $\G^+$ (generally 2D).  Below we present a particular case where an explicit solution is available.

\begin{figure}[h]
\centering
\includegraphics[width= .8 \linewidth]{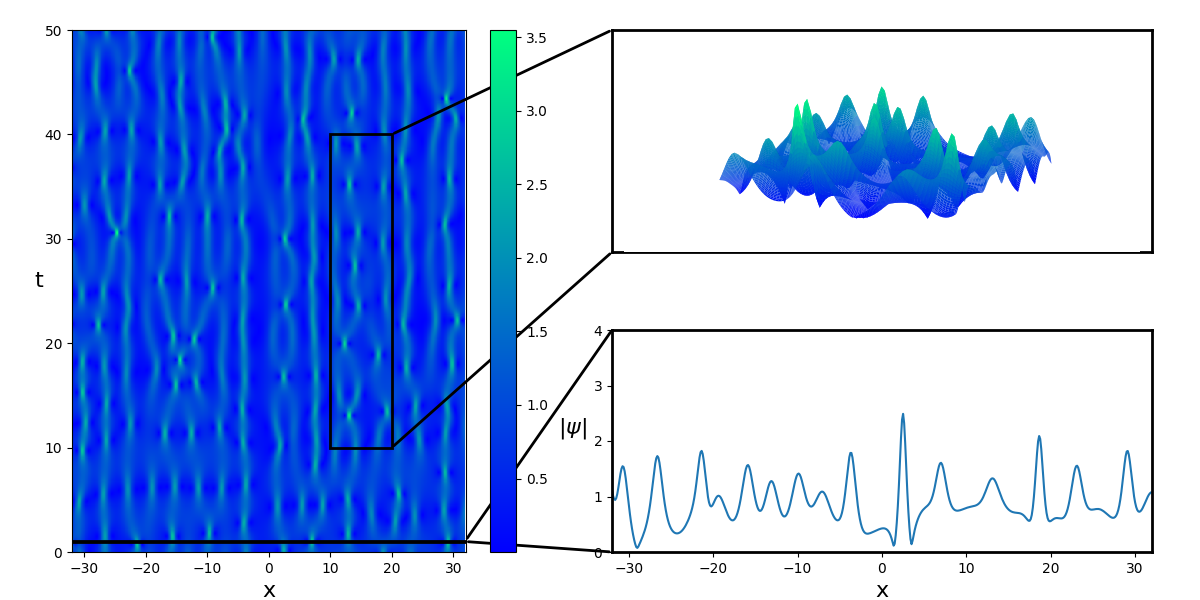}
 \caption{Numerical realisation of the bound state soliton condensate for the focusing NLS equation. Shown is the central portion of a random $100$-soliton ensemble.}
 \label{fig:nks_condensate}
\end{figure}

If the spectral support $\G$ of the DOS belongs to a vertical line,  $\G^+ \subset i \mathbb{R}^+$, the second equation   \eqref{crit_u} implies  $v(\lambda)=0$ and, hence, $s(\lambda)=0$. We shall generally call such a non-propagating  gas a {\it bound state soliton gas}, and it will be the bound state soliton condensate in the present context.  Let $\G=[-iq,iq]$ for some $q>0$ and assume  that $f(\bar\m)=-f(\m)$ for all $\m\in[0,iq]$ (an odd extension of $f(\m)$ onto $[-iq, 0]$). Then, as was shown in \cite{el_spectral_2020}  the first equation \eqref{crit_u} for the DOS reduces to  
\be
\pi H[\hat f](\xi):=\int_{-q}^{q}\frac{\hat f(y)\rmd y}{y-\xi} = 1,
 \ee
 where $\hat f(y) = f(iy)$, $\xi= \text{Im}\la$, and $ H[\hat f]$ denotes the finite Hilbert transform (FHT) of $\hat f$ over $[-q,q]$  \cite{tricomi_finite_1951}, \cite{okada_finite_1991}.    Inverting the FHT $H$ subject to the additional constraint
 $H[\hat f](0)=0$, we obtain  the DOS $f=f_{\rm c} (\la)$ for the bound state soliton condensate
 \be\label{Weyl}
 f_{\rm c}(\la)=\frac {-i \la }{ \pi\sqrt {\la^2+q^2}} , ~~~\la \in (-iq,iq).
  \ee
 One can observe that the DOS \eqref{Weyl}   coincides with the appropriately normalised semi-classical distribution of discrete Zakharov-Shabat spectrum for the potential $\psi \in \mathbb{R}$ in the form of a broad rectangular barrier. Eq.~\eqref{Weyl} (up to a norming factor) is  obtained as a derivative of the corresponding Weyl's law following from the Bohr-Sommerfeld quantisation rule  for the Zakharov-Shabat operator \cite{Zakharov:72}. A numerical realisation of the bound state soliton condensate is shown in Fig.~\ref{fig:nks_condensate}. It is achieved by building a dense (as dense as possible) ensemble of $100$  strongly interacting  solitons  with the discrete spectrum eigenvalues distributed according to the Bohr-Sommerfeld rule and random phases of the norming constants (see \cite{gelash_strongly_2018} for the details of the effective numerical implementation of dense $N$-soliton ensembles with $N$ large). 
 
It was shown in \cite{gelash_bound_2019} that the bound state soliton condensate dynamics reproduces with remarkable accuracy the statistical characteristics of stationary integrable turbulence generated at the nonlinear stage of the development of the noise-induced modulational instability \cite{Agafontsev:15}, \cite{kraych_statistical_2019}, a fundamental and ubiquitous physical phenomenon in focusing nonlinear media \cite{BF:67a},  \cite{Zakharov:09a}, \cite{osborne_nonlinear_2010}.    Furthermore, it was  demonstrated in \cite{agafontsev_rogue_2021} that the nonlinear wave field in the bound state soliton condensate exhibits a spontaneous emergence of rogue waves, offering a new perspective on the dynamical and statistical mechanisms underlying this  phenomenon observed in water waves, nonlinear optics and matter waves (see \cite{onorato_rogue_2013} and references therein).

The bound state soliton condensate  DOS \eqref{Weyl} is a particular solution of the dispersion relations \eqref{crit_u}. Another explicit solution of the soliton condensate dispersion relations \eqref{crit_u} is available  in the special case when the spectral  support  $\G$ of the DOS represents a circle or a circular arc in the complex plane. This  solution with nonzero spectral flux density $v(\la)$ obtained in \cite{el_spectral_2020} corresponds to a dynamic (non-bound state) soliton condensate. 
 
In conclusion we note that the soliton condensate regime is also available for the KdV soliton gas. Indeed, substitution of  $\sigma=0$  in the KdV soliton gas dispersion relations \eqref{inta}, \eqref{intat} yields
 \be \label{inta1}
\int\limits_0^1 \ln \left|\frac{\eta + \mu}
{\eta-\mu}\right| f(\mu)\rmd\mu =  \eta\, , \quad  \int\limits_0^1 \ln \left|\frac{\eta + \mu}
{\eta-\mu}\right|v(\mu) \rmd \mu  =4\eta^3 \, .
\ee
One can see that the soliton condensate dispersion relations \eqref{crit_u} and \eqref{inta1} represent particular cases  of the general  conditions \eqref{cond_kdv} for singular solutions of the soliton gas' equation of state \eqref{eq:state}.
 
 \subsubsection{Special breather gases}
 
 In Section \ref{sec:FNLS_sol} we presented some properties of breather solutions to the focusing NLS equation and indicated three special cases: the Akhmediev breather (AB), the Kuznetsov-Ma (KM) breather and the Peregrine soliton (PS), which are often considered as mathematical models for rogue waves in fluids and nonlinear optical media.  These breathers can be viewed as quasiparticles of the special, `rogue wave breather gases'.  
 
 \begin{figure*}[h]
  \center
  \includegraphics[width=1.\textwidth]{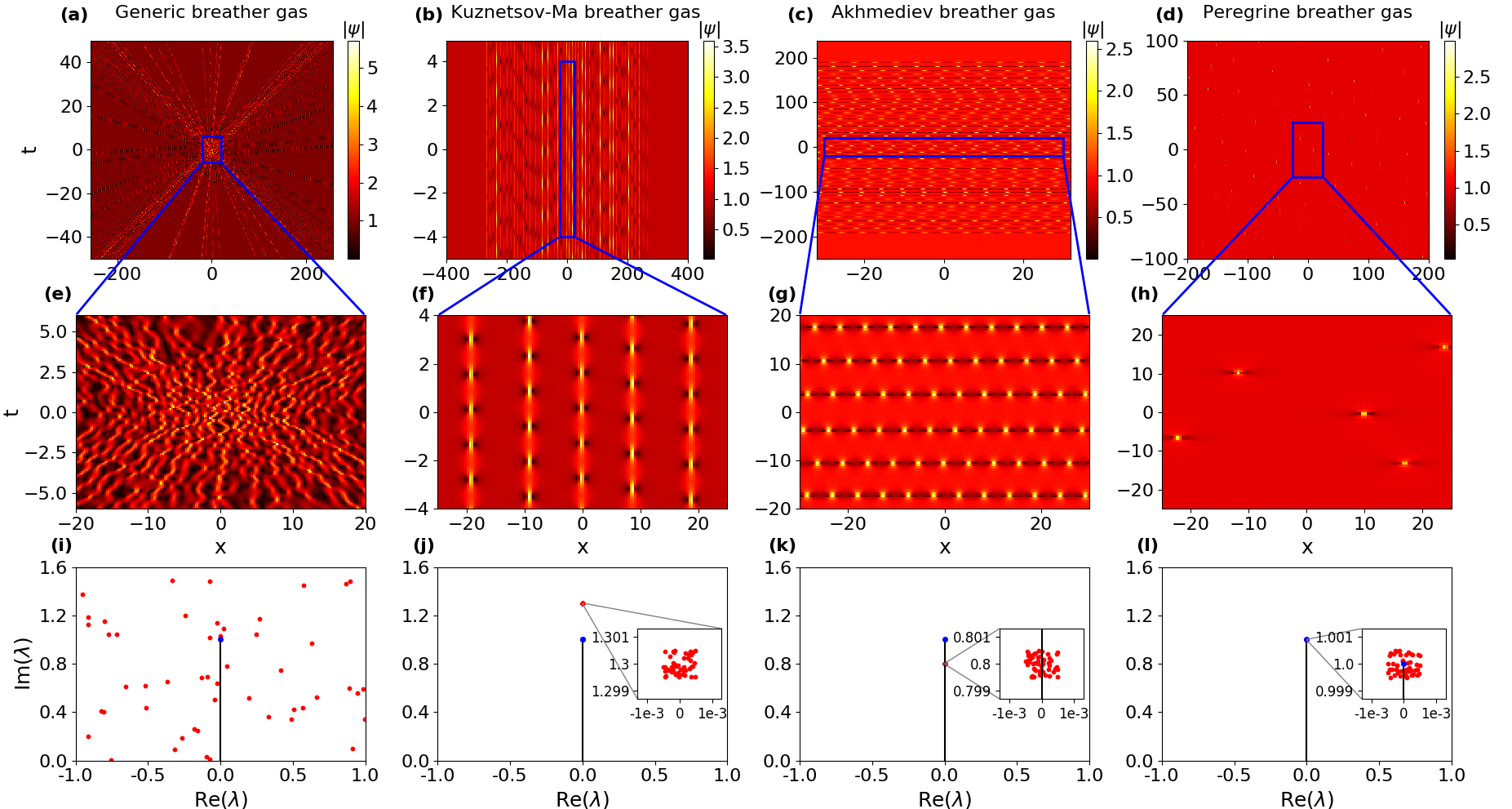}
  \caption{Adapted from \cite{roberti_numerical_2021}. Numerical synthesis of a generic breather gas   (left column (a), (e), (i)) and
    of three single-component breather gases:  KM  gas (second column (b), (f), (j)),
    AB  gas ( third column (c), (g), (k)) and PS gas (fourth column
    (d), (h), (i)). The four breather gases are parametrised by $N=50$ complex eigenvalues $\lambda_n$,
    see bottom row.
    The first row (a)--(d) represents the space-time
    evolution of the breather gases, with the second row (e)--(h) being an enlarged view of some
    restricted region of the $(x-t)$ plane. The third row (i)--(l) represents the
    spectral portraits of each breather gas with the vertical line between $0$ and $+i$ being
    the branch cut (Stokes band) associated with the plane wave background. Each point in the upper
    complex plane in (i), (j), (k), (l) represents a discrete eigenvalue in the
    IST problem with non-zero boundary conditions. The eigenvalues parametrising the single-component breather gases
    are densely placed in a small square region which is centred around a point
    $\lambda_0$ of the imaginary vertical axis and which is strongly enlarged in the insets
    shown in (j), (k), (l). The individual breather `positions'  $x_j$ are uniformly distributed in the range
    $[-1,1]$ for the generic breather gas (a) and for the PS gas (d) while they are uniformly distributed
    in the range $[-32,32]$ for the KM gas (b) and the AB gas (c), see \cite{roberti_numerical_2021} for details.
  } \label{fig:BG}
\end{figure*}

Assuming the Stokes band $\g_0=[-iq, iq]$, $q>0$  the AB and KM  breather gases are  characterised by the DOS $f(\lambda)$ that is supported on the appropriate intervals of the imaginary axis,  $\lambda \in \Gamma^+$: $\G^+ = [0, ip]$, $p<q$ for AB gas and $\G^+ =[ip, ib]$, $b>p>q$ for KM gas. For PS gas the DOS is given by $f(\la)=w\delta(\lambda - iq)$, where $w>0$ is the PS gas density. The three types of the rogue wave breather gases were realised numerically in 
\cite{roberti_numerical_2021} by building a random ensemble of $N \sim 50$ breathers via the  Darboux transform recursive scheme in high precision arithmetic, see Fig.~\ref{fig:BG}. The interaction properties of the constructed breather gases were investigated by propagating through them a trial generic breather (TW) and comparing the effective (mean) propagation velocity  with the  predictions  following from the equation of state \eqref{eq_state_fnls1}. One of these predictions is that the propagation through the PS breather gas is ballistic, i.e. the PS breather gas does not alter the  velocity $s_0(\lambda)$ \eqref{eq_state_breather} of the trial breather due to vanishing of the logarithmic interaction kernel $G(\lambda, \mu)$ \eqref{eq_state_breather}  evaluated at $\mu=iq$ \cite{el_spectral_2020}.  Other predictions, confirmed by detailed quantitative comparisons with numerical simulations in \cite{roberti_numerical_2021},  are that the effective velocity of propagation $s(\lambda)$ of a trial TW breather through a rogue wave gas  is greater  than  $s_0$ for the KM gas and less than $s_0$ for the AB gas.

\section{Hydrodynamic reductions and integrability}
\label{sec:hyd_red}

In this section we review some general mathematical properties of the kinetic equation for soliton gas studied in \cite{el_kinetic_2011}, \cite{pavlov_generalized_2012}.
For convenience we reproduce here the generalised kinetic equation \eqref{kin_eq0}, \eqref{eq:state} for  unidirectional/bidirectional isotropic soliton gas:
\begin{equation}\label{gen_kin_eq1}
f_t + (fs)_x=0, \quad s(\la, x,t)= s_0(\la) + \int_{\Gamma}
G(\la, \mu) f(\mu, x,t)[s(\la,x,t) - s(\mu, x,t)] \rmd \mu ,
\end{equation}
where $f(\la, x, t), s(\la, x,t), s_0(\la)$ and $G(\la, \mu)$ are real functions and the  parameters $\lambda, \mu$ can be real or complex.  For unidirectional gas the spectral support $\G$ of the DOS $f(\la, x, t)$  is an interval  in $\R$. For bidirectional isotropic  gas $\G$ is either a symmetric interval $[-a, a] \subset \R$ (cf. \eqref{eq:s_DNLS}) or a Schwarz symmetric compact domain  in  $\C$ (1D or  2D, see \eqref{2D_int} for the formal correspondence between the two cases). 

Consider a soliton gas
 composed of a finite ($M \in \N$) number of distinct components,
termed monochromatic, or cold, components modelled by the DOS in the form of a linear combination of the Dirac delta-functions centred at distinct spectral points $\zeta_j \in \Gamma$, 
 \be\label{u_delta1}
 f(\la, x, t) = \sum \limits_{j=1}^{M} w_j (x,t)\delta(\la - \zeta_j),
 \ee
 where $w_j(x,t)>0$ are the components' weights, and  
 $\{\zeta_{j}\}_{j=1}^M \subset \Gamma$, $(\zeta_j \ne \zeta_k \iff  j \ne k)$.  Substitution of  \eqref{u_delta1} into the kinetic equation  \eqref{gen_kin_eq1}  reduces it to a system of hydrodynamic conservation laws
 \begin{equation}\label{cont}
(w_i)_{t}+(w_{i}s_{i})_{x} =0,\qquad i=1,\dots ,M\,,  
\end{equation}%
 where the component densities $w_i(x,t)$ and the transport velocities $s_{j}(x,t)\equiv s(\zeta_j, x, t)$ are related 
algebraically:
 \begin{equation}\label{s_alg}
 s_{j}= s_{0j} + \sum \limits_{m=1, m \ne j}^M G_{jm} w_m(s_{j}- s_{m}), \quad j=1, 2, \dots M.
 \end{equation}
Here we used the notation
 \be
  s_{0j} \equiv s_0(\zeta_{j}) , \quad G_{jm} \equiv G(\zeta_{j}, \zeta_{m}), \quad j \ne m.  
 \ee

We note that, although we derived the hydrodynamic reduction \eqref{cont}, \eqref{s_alg} in the context of unidirectional/isotropic soliton gas, it can be readily generalised to the bidirectional anisotropic case, see \cite{congy_soliton_2021}. In the latter case the ansatz \eqref{u_delta1} is introduced separately for the slow and fast soliton branches so that  the pair of distributions
 \begin{equation}
\label{eq:F0}
f_-(\lambda,x,t) = \sum_{i=1}^{M_-} w_i(x,t)
\delta(\lambda-\zeta_i),\quad
f_+(\lambda,x,t) = \sum_{i=M_-+1}^{M_-+M_+} w_i(x,t)
\delta(\lambda-\zeta_i),
\end{equation} 
is transformed into a $M$-dimensional vector
${\bs w} = (w_1, \dots, w_M)$, where $M=(M_-+M_+)$. 
  
 \medskip
For $M=2$ system \eqref{s_alg} can be solved  to give explicit expressions for $s_{1,2}(w_1, w_2)$:
  \be \label{s_12}
 s_1= s_{01} + \frac{ G_{12} w_2 (s_{01}-s_{02})}{1-(G_{12} w_2+ G_{21} w_1)},
 \quad s_{2}= s_{02}-\frac{ G_{21} w_1 (s_{01} - s_{02})}{1-(G_{12} w_2+ G_{21} w_1)}. 
\ee

 An important remark is in order on the meaning of the delta-function ansatz \eqref{u_delta1}  for the DOS $f(\lambda)$.  As a matter of fact, the representation \eqref{u_delta1}  is a mathematical idealisation, which has a formal sense in the context of the integral equation of state \eqref{gen_kin_eq1},  but cannot be applied to the original dispersion relations  where it appears  in both the integral and the secular terms (cf. \eqref{inta}, \eqref{intat} for the KdV equation).  In a physically realistic description  the delta-functions in \eqref{u_delta1} should be replaced by some narrow distributions around the spectral points $\zeta_{j}$, {i.e. we first take
the thermodynamic limit $N\to\infty$ and then allow the distributions to become sharply peaked.}  
As a result, the non-condensate condition $\sigma > 0$  in the dispersion relation for the DOS  (equation \eqref{inta} for KdV or equation \eqref{dr_breather_gas1} for focusing NLS) would impose
 a constraint  $\int_{\Gamma } G (\la, \mu) f(\la) \rmd \mu <1$  on $f(\la)$ which, among other things,  implies  that the denominators in \eqref{s_12} must be positive.  It is interesting to note that a variant of hydrodynamic system  \eqref{cont}, \eqref{s_12} has  been recently derived in \cite{medenjak_t_2021} in the context  of $T \bar T$-deformed  conformal field theories out of equilibrium.  
 
 We now  assume that the integral  kernel $G(\lambda, \mu)$ in the equation of state \eqref{gen_kin_eq1} can be represented as
\be \label{sym_G}
G(\la, \mu)= a(\mu) \epsilon(\la, \mu), \ \ \text{where}   \ \ \epsilon(\la, \mu)=  \epsilon(\mu, \la) > 0, \ \ \la \ne \mu
\ee
for  some non-singular real function  $a(\mu) \ne 0$. This structure of the soliton interaction kernel is typical for  integrable dispersive hydrodynamics  and is indeed shared by  the equations of state of the KdV, defocusing and focusing NLS  unidirectional/isotropic soliton gases  (see \eqref{eq_state_kdv1}, \eqref{eq:s_DNLS}, \eqref{FNLS_kin}). E.g. for KdV we have  $a(\mu)=\mu$, $\epsilon(\la, \mu)=\frac{1}{\la \mu}\ln \left|\frac{\la+ \mu}{\la - \mu} \right|$, see \eqref{eq_state_kdv}.
 
 We then have 
\be \label{structure}
G_{jk}=a_k\epsilon_{jk}, \quad  \epsilon_{jk} = \epsilon_{kj}, \ j \ne k,
\ee where 
$a_k=a(\zeta_k)$, $\epsilon_{jk}=\epsilon(\zeta_{j}, \zeta_{k})$. Solving  equations  \eqref{s_12}  for $w_{1,2}$ gives
\begin{equation}  \label{alg2}
a_1w_1=\frac{1}{\epsilon_{12}}\frac{s_{2}-s_{02}}{s_{2}-s_{1}}, \quad a_2w_{2}=\frac{1}{\epsilon_{12}}\frac{s_{1}-s_{01}}{s_{1}-s_{2}}.
\end{equation}
Substituting \eqref{alg2} in \eqref{cont} for $M=2$ we obtain a diagonal system \cite{el_kinetic_2005}:
\begin{equation}
\frac{\partial s_1}{\partial t} + s_{2} \frac{\partial s_{1}}{\partial x}=0, \qquad \frac{\partial s_2}{\partial t} + s_{1} \frac{\partial s_{2}}{\partial x}=0
\label{chap}
\end{equation}
with  $s_{1,2}$ being the Riemann invariants.
One can see that system (\ref{chap}) is \textit{linearly
degenerate} because its characteristic velocities do not depend on the
corresponding Riemann invariants.  System \eqref{chap} represents the diagonal form of the  so-called Chaplygin gas equations,  the system of isentropic gas dynamics with  the equation of state $p= - {A}/{\rho}$, where $p$ is the pressure, $\rho$ is the gas density and $A>0$ is a constant. The Chaplygin gas equations   occur in certain theories of cosmology (see e.g. \cite{bento_generalized_2002}) and is also equivalent to the 1D Born-Infeld equation \cite{born_foundations_1934}, \cite{whitham_linear_1999} arising in nonlinear electromagnetic field theory.

As any two-component quasilinear hyperbolic system, system (\ref{chap}) is integrable
(linearisable) via the classical hodograph transform. However, for any $%
M\geq 3$  integrability of the original system (\ref{cont}), (\ref{s_alg}) is
no longer obvious.  As a matter of fact, a $M$-component hydrodynamic type
system is generally \textit{not integrable} for $M\geq 3$. 
Below we present the theorem, proved in \cite{el_kinetic_2011}, which summarises the mathematical properties  of  the hydrodynamic reductions \eqref{cont}, \eqref{s_alg} subject to the structure \eqref{structure}  of the interaction matrix $G_{jk}$. 
 
\medskip

\textbf{Theorem} \cite{el_kinetic_2011}. \ \textit{  $M$-component  reductions (\ref{cont}), (\ref%
{s_alg}) of the generalized kinetic equation (\ref{gen_kin_eq1}) with the interaction kernel satisfying \eqref{sym_G} are hyperbolic, linearly
degenerate  integrable  hydrodynamic type systems for any $M \in \mathbb{N}$.}

\medskip
Specifically,  it was shown in \cite{el_kinetic_2011}  that the following fundamental properties are satisfied for the system  (\ref{cont}), (\ref{s_alg}):

\medskip
(i) {\it Riemann invariants.}  

\smallskip
Let $u_j=a_j w_{j}$, where the symmetrising coefficients $a_j$ are defined by \eqref{sym_G}, \eqref{structure}.  Then the invertible point transformation ${\bs w} \to {\bs r}$ specified by 
\begin{equation}
r_{k}=-\frac{1}{u_j}\left( 1+ \sum_{m\ne k} u_{m}\epsilon_{mk}\right) , \ k=1,2,...,M, \label{rim1}
\end{equation}
reduces the system (\ref{cont}), (\ref{s_alg}) to the diagonal (Riemann invariant) form 
\be \label{riemann}
\frac{\partial r_i}{\partial t} + V_i({\bs r}) \frac{\partial r_i}{\partial x}=0, \quad i=1, 2, \dots, M,
\ee
where the characteristic velocities $V_i$ are expressed in terms of the Riemann invariants $r_1, r_2, \dots, r_N$ as
\be
V_i({\bs r})=s_i(w_1({\bs r}), w_2({\bs r}), \dots, w_M({\bs r})),  \quad i=1, 2, \dots, M.
\ee
Here $s_i(w_1, \dots, w_M)$ are solutions of the system \eqref{s_alg},

\medskip
(ii) {\it Linear degeneracy.} 

\smallskip
The characteristic velocities $V_j$ of the diagonal system \eqref{riemann} satisfy

\be \label{lin_deg}
{\partial_j V_j} =0,    \quad j=1, 2, \dots, M,  \quad \text{where} \quad \partial _{j}\equiv
\partial /\partial r_{j},
\ee
so that system (\ref{cont}), (\ref{s_alg}) is linearly degenerate in the Lax sense \cite{lax_hyperbolic_1973}.

\medskip
(iii) {\it Semi-Hamiltonian property.}

\smallskip
In addition to \eqref{lin_deg} the characteristic velocities $V_i({\bs r})$ satisfy the overdetermined system
\begin{equation}
\partial _{j}\frac{\partial _{k}V_{i}}{V_{k}-V_{i}}=\partial _{k}\frac{%
\partial _{j}V_{i}}{V_{j}-V_{i}}\text{, \ \ \ }i \neq j \neq k  \label{iden}
\end{equation}%
for each three distinct characteristic velocities ($\partial _{k}\equiv
\partial /\partial r_{k}$). Diagonal hydrodynamic type systems satisfying \eqref{iden} are called semi-Hamiltonian \cite{tsarev_poisson_1985}.

\medskip
 (iv) {\it Generalised Hodograph solution.}

\medskip
A semi-Hamiltonian hydrodynamic type system can be locally integrated via the generalised hodograph transform due to Tsarev \cite{tsarev_poisson_1985}, \cite{tsarev_geometry_1991}.  Its general local solution for $\partial _{x}r_{i}\neq 0$, $%
i=1,\dots , M$ is given by the  formula
\begin{equation}
x-V_{i}({\bs r})t=W_{i}({\bs r})\, , \quad i=1, 2, \dots, M,  \label{hod}
\end{equation}
where functions $W_{i}({\bs r})$ are found from the linear system of
PDEs:
\begin{equation}
\frac{\partial _{i}W_{j}}{W_{i}-W_{j}} = \frac{\partial _{i}V_{j}}{%
V_{i}-V_{j}} \,,\quad i,j=1,\dots ,N,\quad i\neq j.  \label{ts}
\end{equation}
Properties of linearly degenerate semi-Hamiltonian (integrable) hydrodynamic type systems have been studied in \cite{pavlov_hamiltonian_1987}, \cite{ferapontov_integration_1991}. Using the results of \cite{pavlov_hamiltonian_1987}, \cite{ferapontov_integration_1991} it was shown in \cite{el_kinetic_2011} that the general local solution of the hydrodynamic reductions \eqref{cont}, \eqref{s_alg} of the kinetic equation
can be represented in the form
\begin{equation}  \label{fer1}
x- s_{0i}t=\overset{r_{i}}{\int }{\xi \phi _{i}(\xi )} \rmd \xi +%
\underset{m\neq i}{\sum }\epsilon _{im}\overset{r_{m}}{\int }{\phi
_{m}(\xi )} \rmd \xi \, , \quad i=1, 2, \dots, M,
\end{equation}
where $\phi_1(\xi), \dots, \phi_M(\xi) $ are arbitrary functions. Some interesting particular solutions (self-similar, quasiperiodic) following from \eqref{fer1} can be found in \cite{el_kinetic_2011}.
The Hamiltonian structure of the hydrodynamic reductions \eqref{cont}, \eqref{s_alg} was investigated in \cite{bulchandani_classical_2017}. 
The generalised hydrodynamic reductions arising if one allows the discrete spectral points $\zeta_j$ in \eqref{u_delta1} be functions of $(x,t)$ were studied in \cite{pavlov_generalized_2012}.

In conclusion of this section we note that, somewhat counter-intuitevely, the above results on the hyperbolic structure and integrability of hydrodynamic reductions of the kinetic equation for soliton gas apply to the multi-component soliton gas of the focusing NLS equation, whose Whitham modulation system is known to be  elliptic  for a generic set of modulation parameters and for any genus \cite{pavlov_nonlinear_1987}, \cite{dafermos_geometry_1986}. This apparent paradox is resolved by noticing that the  Whitham system  for the focusing NLS equation exhibits real eigenvalues (characteristic speeds) in the soliton limit. E.g. for the genus 1 case, the two pairs of complex conjugate eigenvalues of the modulation matrix collapse into a single, quadruply degenerate real eigenvalue, corresponding to the velocity of the fundamental soliton, see e.g. \cite{el_dam_2016}.

\section{Riemann problem for soliton gas}
\label{sec:riemann}
Having obtained the hydrodynamic description of multi-component soliton gas  it is natural to consider the Riemann problem that plays the fundamental role in classical gas  and fluid dynamics \cite{landau_fluid_1987}. 
The classical Riemann problem consists in finding solution to a system of hyperbolic conservation laws subject to piecewise constant initial conditions exhibiting discontinuity at $x=0$.   The distribution solution of the Riemann problem generally consists of a combination of constant states, simple (rarefaction) waves and strong discontinuities (shocks or contact discontinuities) \cite{lax_hyperbolic_1973}. The discontinuous solutions satisfy the Rankine-Hugoniot jump conditions.
If a hyperbolic system is linearly degenerate it does not support simple waves and classical shocks, and the solution of Riemann problem contains only constant states and contact discontinuities \cite{rozhdestvenskii_systems_1983}. Here, following \cite{carbone_macroscopic_2016}, \cite{congy_soliton_2021}, we present such solutions for different types of soliton gases.

\subsection{General solution}

We consider the hyperbolic  system \eqref{cont}, 
\begin{equation}
\label{eq:F}
(w_i)_t + (s_i(\bs w) w_i)_x = 0, \quad i=1, \dots, n
\end{equation}
subject to the initial data
 corresponding to two $n$-component unidirectional or bidirectional soliton gases prepared in the respective
uniform states $\bs w^{\rm L} \in \mathbb{R}^n$ and
$\bs w^{\rm R}\in \mathbb{R}^n$, that are initially separated:
\begin{equation}
\label{eq:init}
\bs w(x,0) =
\begin{cases}
\bs w^{\rm L}, &\text{if } x<0,\\
\bs w^{\rm R}, &\text{if } x\geq 0.
\end{cases}
\end{equation}
We emphasize here that the  Riemann problem ~\eqref{eq:F}, \eqref{eq:init} is formulated for the
kinetic equation (via the substitution~\eqref{u_delta1}) but not for the original dispersive
hydrodynamic system~\eqref{eq:scalar_dh} or \eqref{eq:disp_Euler}.  For this reason we shall call it 
 the {\it spectral Riemann problem} as it essentially describes the
spatiotemporal evolution of the spectral DOS of the soliton
gas. The spectral Riemann problem  for the KdV
and focusing NLS two-component soliton gases ($n=2$) was investigated in~\cite{el_kinetic_2005}, \cite{carbone_macroscopic_2016}, \cite{el_spectral_2020} and for rather general isotropic  and anisotropic dispersive hydrodynamic soliton gases with an arbitrary number of spectral components in \cite{congy_soliton_2021}. The  Riemann  problem of this type has also been considered  in the context of generalised
hydrodynamics~\cite{castro-alvaredo_emergent_2016}, \cite{bertini_transport_2016}, \cite{doyon_dynamics_2017},
\cite{kuniba_generalized_2020}, \cite{croydon_generalized_2020}, \cite{alba_generalized-hydrodynamic_2021}. 

The  spectral Riemann problem \eqref{eq:F}, \eqref{eq:init}  corresponds to the dispersive hydrodynamic Riemann problem that can be viewed as a soliton
gas shock tube problem, an analog of the  shock tube
problem of classical gas dynamics \cite{landau_fluid_1987}, \cite{whitham_linear_1999}. 
The shock tube problem represents
a good benchmark for the spectral kinetic theory of soliton gas where one can investigate both
overtaking and head-on collisions by choosing the appropriate 
components of solitonic spectra and then comparing the predictions of the  kinetic theory with direct numerical simulations of dispersive hydrodynamics.

Following \cite{carbone_macroscopic_2016}, \cite{congy_soliton_2021} we consider the spectral Riemann 
problem (analytically) and the corresponding soliton gas shock tube problem (numerically) for  $n$-component unidirectional and bidirectional  soliton
gases.
Due to the scale invariance of the problem (the kinetic
equation~\eqref{eq:F} and the initial condition~\eqref{eq:init} are
both invariant with respect to the scaling transformation $x \to Cx$,
$t \to Ct$), the spectral solution is a self-similar distribution ${\bs
w}(x/t)$. Because of the linear degeneracy  \eqref{lin_deg} of the hydrodynamic 
system~\eqref{eq:F} the only admissible similarity solutions are constant states
separated by propagating contact discontinuities, cf. for
instance~\cite{rozhdestvenskii_systems_1983}. Discontinuous, weak,
solutions are physically acceptable here since the kinetic equation
describes the conservation of the number of solitons within any given
spectral interval, and the Rankine-Hugoniot type conditions can be imposed
to ensure the conservation of the number of solitons across the
discontinuities. As a result, the solution of the Riemann problem \eqref{cont}, \eqref{eq:init} for each component $w_i(x,t)$ is composed of
$n+1$ constant states, or plateaus, separated by $M$ discontinuities
(see e.g.~\cite{lax_hyperbolic_1973}):
\begin{equation}
\label{eq:sol}
w_i(x,t) =
\begin{cases}
w_i^{1} = w_i^{\rm L}, &x/t<z_1,\\
\dots\\
w_i^j, & z_{j-1} \leq x/t < z_j\\
\dots\\
w_i^{n+1} = w_i^{\rm R}, &z_n \leq x/t,
\end{cases}\quad
\end{equation}
where the lower index $i$ indicates the $i$-th component of the vector
$\bs{w}$, and the upper index $j$ is the
index of the plateau. For clarity we labeled the superscripts $j=1$ as
``L'' (left boundary condition) and $j=n+1$ as ``R'' (right boundary
condition).    
The contact discontinuities propagate at the characteristic
velocities~\cite{lax_hyperbolic_1973}:
\begin{equation}
\label{eq:z}
z_j =s_{j}(w_1^j,\dots,w_n^j) = s_{j}(w_1^{j+1},\dots,w_n^{j+1}),
\end{equation}
where the plateaus' values $w_i^j$ are found
from the Rankine-Hugoniot jump conditions \cite{whitham_linear_1999}:
\begin{align}
\label{eq:RH}
-z_j \Big[ w_i^{j+1} -w_i^{j}\Big] + \Big[
s_{i}(w_1^{j+1},\dots,w_n^{j+1})
w_i^{j+1} -  s_{i}(w_1^j,\dots,w_n^j)
w_i^{j} \Big] = 0,
\end{align}
where $i,j=1 \dots n$. 
The Rankine-Hugoniot conditions with $i=j$ are trivially satisfied by
the definition of contact discontinuity~\eqref{eq:z}. We also note that conditions \eqref{eq:z} are compatible with  \eqref{eq:RH}  for linearly degenerate systems of conservation laws.

Note that, if the solitons were not interacting, the initial step
distribution $w_i(x,0)$ for the component $\lambda = \zeta_i$ would
have propagated at the free soliton velocity $s_{0i}$:
\begin{equation}
\label{eq:sol_free}
w^{\rm free}_i(x,t) =
\begin{cases}
w_i^{\rm L}, &x/t< s_{0i},\\
w_i^{\rm R}, &s_{0i} \leq x/t,
\end{cases}
\quad i=1 \dots n,
\end{equation}
which is very different  compared to the solution~\eqref{eq:sol}.

\subsection{Riemann problem: Examples}

{\it  KdV  soliton gas}

\medskip
\begin{figure}[h]
\begin{center}
  \includegraphics[scale=0.4]{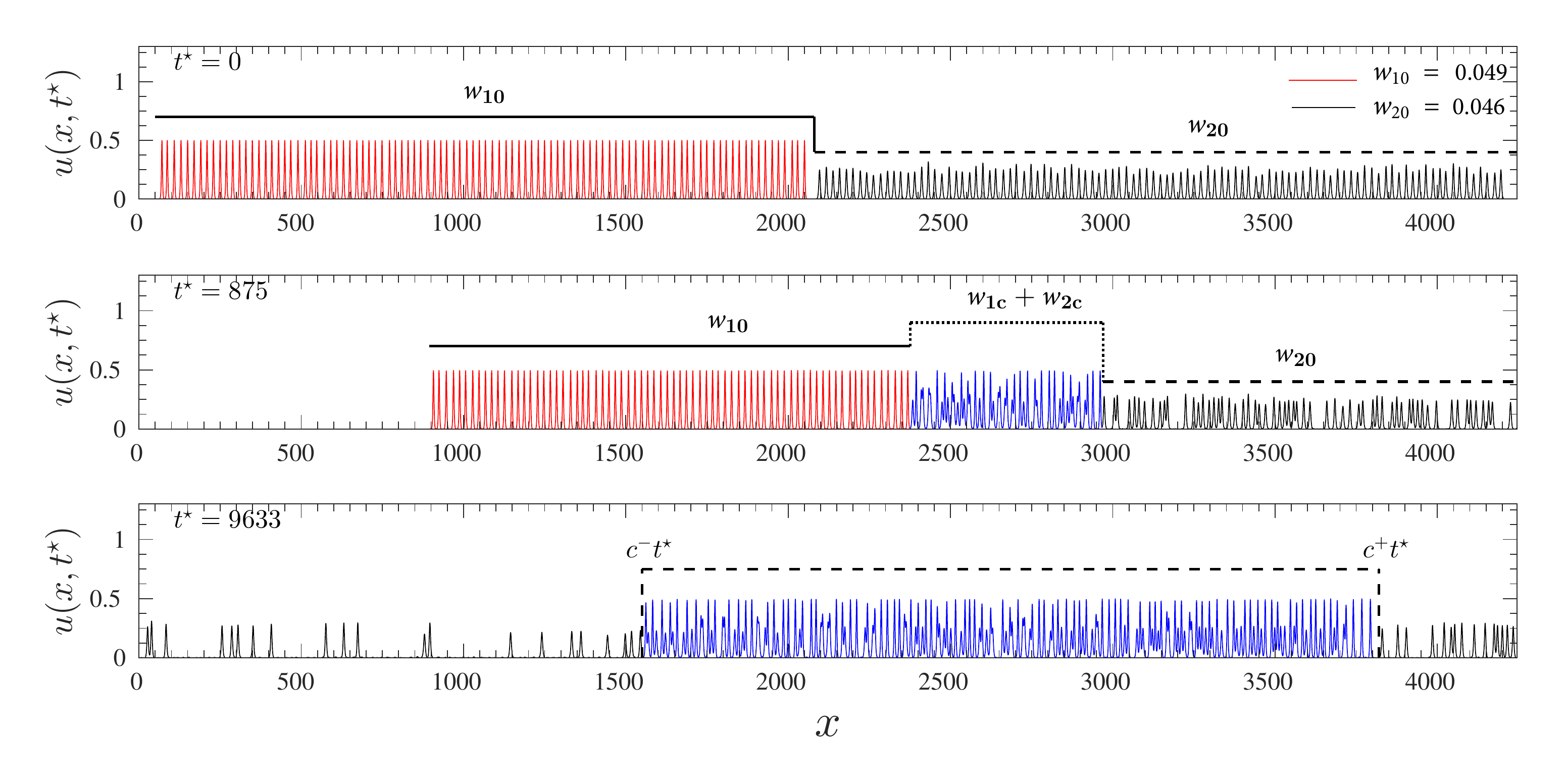}    
  \end{center}
    \caption{(adapted from \cite{carbone_macroscopic_2016}). Soliton gas shock tube problem: numerical solution of the KdV equation for one typical realisation of the collision of two cold soliton gases. The expanding interaction region  is shown in blue. The piece-wise constant  solution for the total density of solitons $\beta$ \eqref{RPs} is shown above the KdV solution plot.}
    \label{fig:trial_sol}  
  \end{figure} 
  We  now  consider the simplest concrete Riemann problem for soliton gas. Namely, we shall study the collision of two single-component KdV soliton gases  \cite{carbone_macroscopic_2016}.  To be consistent with the notations of Sections \ref{sec:unidir_sol_gas} and \ref{sec:spectral_theory_kdv}  we shall be using the  spectral variable $\eta=\sqrt{-\la}$. Also we simplify the general notation of the previous section to make the specific results more transparent.

  We consider the two-component ansatz \eqref{u_delta1} for the DOS, 
\begin{equation}\label{cold2}
  f(\eta; x,t)\ =\ w_1(x,t)\delta(\eta - \eta_1)\ +\ w_2(x,t)\delta(\eta - \eta_2)\, ,
\end{equation}
where 
\begin{equation}\label{pm}
0< \eta_2< \eta_1 < 1 \, , \quad w_{1,2}\ \geqslant\ 0\, .
\end{equation}
Substitution of (\ref{cold2}) into the kinetic equation (\ref{kin_eq0}), (\ref{eq_state_kdv}) for KdV soliton gas yields the system of two conservation laws
\begin{equation}\label{red1}
  \partial_t w_1\ +\ \partial_x (w_1 s_1)\ =\ 0\, , \qquad \partial_t w_2\ +\ \partial_x (w_2 s_2)\ =\ 0,
\end{equation}
where the transport velocities $s_{1,2}  \equiv s(\eta_{1,2}, x,t)$  are given by (cf. \eqref{s_12})
\begin{equation}\label{s120}
  s_1 = 4\eta_1^2+\frac{4(\eta_1^2-\eta_2^2)\alpha_2 w_2}{1-\alpha_1 w_1 - \alpha_2 w_2},\ \ 
  s_2 = 4\eta_2^2-\frac{4(\eta_1^2-\eta_2^2)\alpha_1 w_1}{1-\alpha_1 w_1-\alpha_2 w_2},
\end{equation}
where 
\begin{equation}\label{epsilon}
  \alpha_{1,2}\ =\ \frac{1}{\eta_{1,2}}\;\ln \left|\frac{\eta_1\ +\ \eta_2}{\eta_1\ -\ \eta_2}\right|\ >\ 0 \,.
\end{equation}
It is assumed that  $w_1 \alpha_1 + w_2 \alpha_2 < 1$ (see the remark after eq. \eqref{s_12}).

We now consider the  system \eqref{red1}, \eqref{s120} subject to the Riemann data \eqref{eq:init}  with ${\bs w}^L=(w_{10},0)$, ${\bs w}^R=(0, w_{20})$, i.e. 
\be\label{RPf}
\left\{\begin{array}{ll}
w_1(x,0) = w_{10}, \ \ w_2(x,0) = 0\, ,     \quad & x<0, \\ [6pt]
w_2 (x,0)= w_{20}, \ \ w_1(x,0) = 0  , &x \ge 0,
\end{array}
\right.
\ee
where $w_{10}, w_{20} > 0$ are some  constants satisfying $w_{i0} \leqslant \eta_i/3$ (see \eqref{crit_kdv}). Note that the initial velocities of the colliding soliton gases  are fully determined, via Eq.~(\ref{s120}), by the density distribution (\ref{RPf}).  An example of one realisation of direct numerical simulation of the collision of two single-component KdV soliton gases modelled by the spectral Riemann problem \eqref{red1}, \eqref{RPf} is shown in Fig.~\ref{fig:trial_sol}. Details of the numerical implementation of the KdV soliton gas  in the simulations  can be found in \cite{carbone_macroscopic_2016}.

The required solution for each component $w_i$ represents a combination of three constant states (plateaus) separated by two contact discontinuities given by equations \eqref{eq:sol} - \eqref{eq:RH} for $n=2$.  In the notations of this section, 
\begin{equation}
\label{eq:sol21}
w_i(x,t) =
\begin{cases}
 w_{i0} \,  \delta_{i,1}, &x/t<c^-,\\
w_{ic}, & c^- \leq x/t < c^+,\\
w_{i0} \, \delta_{i,2}, &c^+ \leq x/t,
\end{cases}\quad
\end{equation}
where $i \in \{1,2\}$ and $\delta_{ij}$ is the Kronecker delta.
The values $w_{1c}$ and $w_{2c}$ of the component densities in the middle plateau region, where the interaction of soliton gases occurs, together with the velocities $c^{\pm}$ of the contact discontinuities, are found from  the Rankine--Hugoniot jump conditions, 
\begin{equation}
\begin{array}{l}
  -c^-(w_{10}-w_{1c})\ +\ (w_{10} s_{10}-w_{1c} s_{1c})\ =\ 0 \, , \\
  -c^-(0-w_{2c})\ +\ (0-w_{2c} s_{2c})\ =\ 0\, ;
\end{array}
\label{j1}
\end{equation}
\begin{equation}
\begin{array}{l}
  -c^+(w_{1c} - 0)\ +\ (w_{1c} s_{1c} - 0)\ =\ 0 \, , \\
  -c^+(w_{2c}  - w_{20} )\ +\ (w_{2c} s_{2c} -  w_{20} s_{20})\ =\ 0 \, ,
\end{array}
\label{j2}
\end{equation}
where the velocities $s_{1c}=s_1(w_{1c}, w_{2c})$ and  $s_{2c}=s_2(w_{1c}, w_{2c})$ are determined by relations (\ref{s120}).
\begin{figure}[h]
\begin{center}
  \includegraphics[scale=0.35]{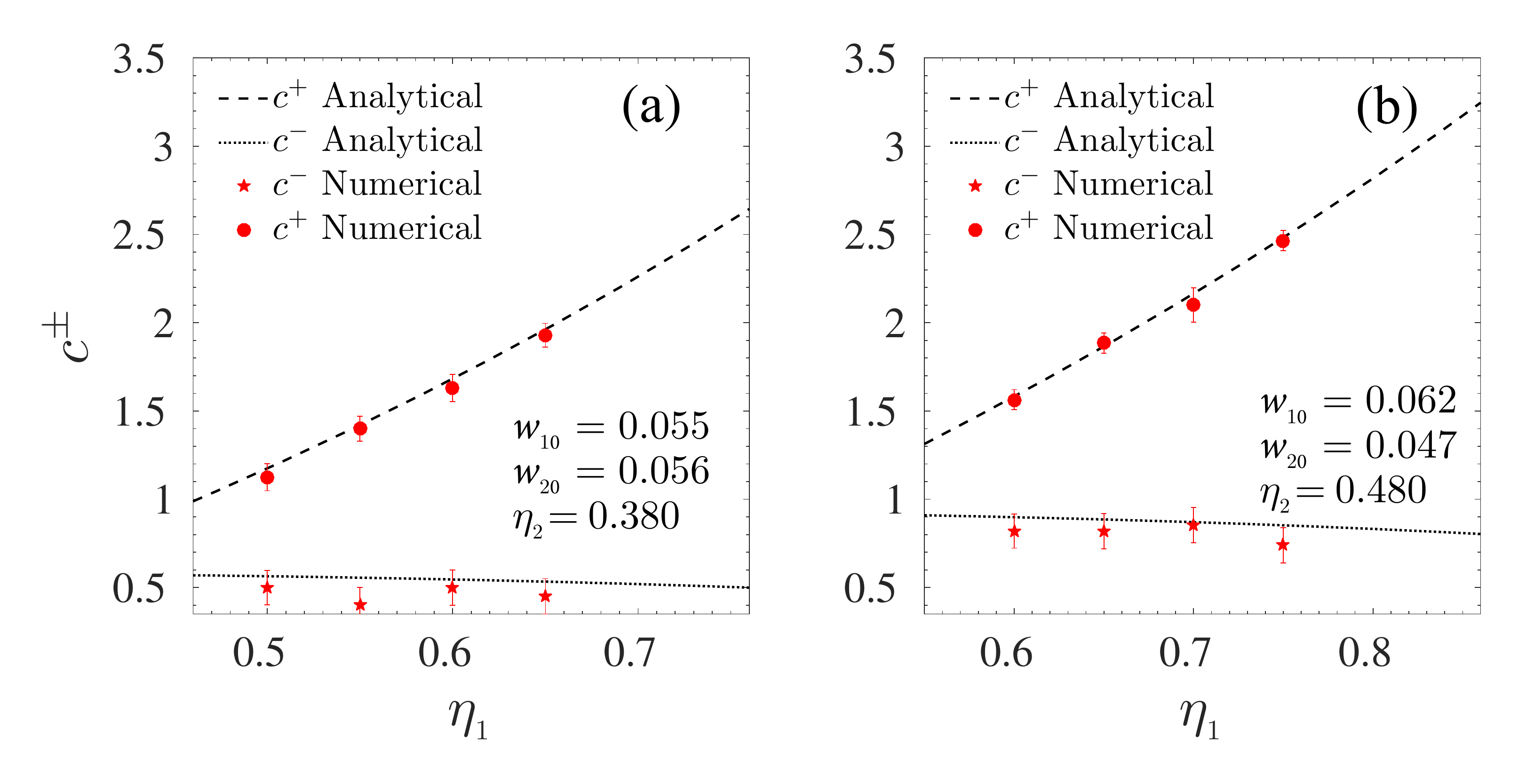}    
  \end{center}
    \caption{(adapted from \cite{carbone_macroscopic_2016}). Comparisons for the shock tube problem: the speeds $c^\pm$ of the edges of the interaction region for two different sets of parameters.}
    \label{fig:kdv_compar_cpm}
 \end{figure}
 \begin{figure}[h]
\begin{center}
  \includegraphics[scale=0.35]{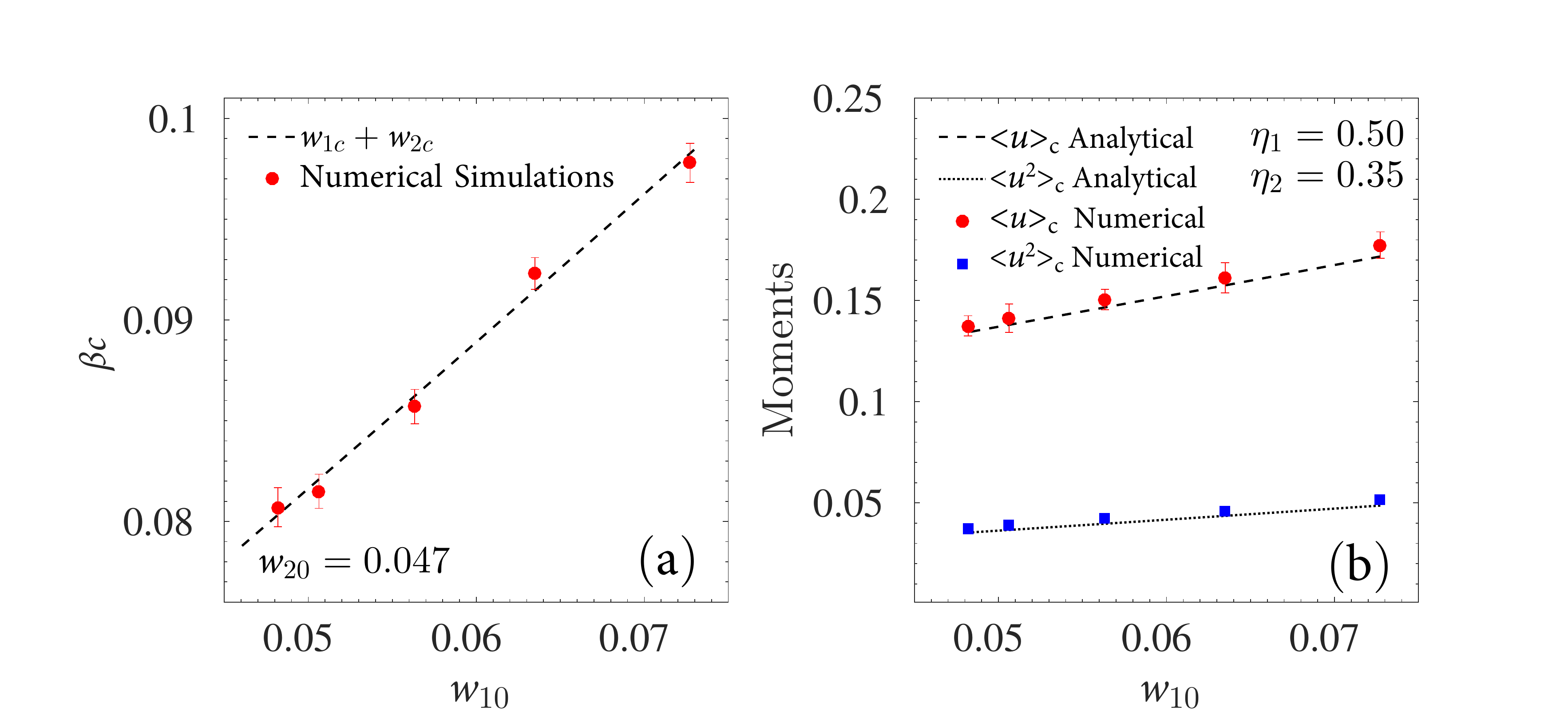}    
  \end{center}
    \caption{(adapted from \cite{carbone_macroscopic_2016}). (a) The equilibrium total density 
    $\beta_c=w_{1c}+ w_{2c}$ in the interaction region as a function of the density $w_{10}$ of the `left' gas. The density of the `right' gas $w_{20} = 0.047$; (b) the moments $\langle u \rangle_c, \langle u^2 \rangle_c$ of the random wave field in the interaction region.}
\label{fig_kdv_mom_comp}
    \end{figure} 

Solving (\ref{j1}), (\ref{j2}) we obtain:
\begin{equation}\label{fc}
  w_{1c}\ =\ \frac{w_{10}(1-\alpha_2 w_{20})}{1 - \alpha_1 \alpha_2 w_{10} w_{20}} \, , \quad 
  w_{2c}\ =\ \frac{w_{20}(1-\alpha_1 w_{10})}{1- \alpha_1 \alpha_2 w_{10} w_{20}} \, .
\end{equation}
\begin{equation}\label{cpm}
  c^-\ =\ 4\eta_2^2\ -\ \frac{4(\eta_1^2 - \eta_2^2)\alpha_1 w_{1c}}{1- \alpha_1 w_{1c} - \alpha_2 w_{2c}}, \quad
  c^+\ =\ 4 \eta_1^2\ +\ \frac{4(\eta_1^2 - \eta_2^2) \alpha_2 w_{2c}}{1-\alpha_1 w_{1c} - \alpha_2 w_{2c}}\, .
\end{equation} 
One can see that velocities $c^\pm$ of the edges of the expansing interaction region coincide with the characteristic velocities \eqref{s120} which is the basic property of contact discontinuities, cf. \eqref{eq:z}.

The  total density of solitons 
$\beta = \int_0^1 f(\eta) \rmd \eta = w_1 + w_2$ satisfies
\be\label{RPs}
  \beta(x,t)=\left\{\begin{array}{ll}
  w_{10},      \quad & x\ <\ c^-t, \\ [6pt]
  w_{1c} + w_{2c}, \quad & c^-t\ \le \ x\  < \ c^+ t ,\\ [6pt]
  w_{20}, \quad   & x\  \ge \ c^+t.
\end{array}
\right.
\ee
The piece-wise constant solution for $\beta(x,t)$ is shown in Fig.~\ref{fig:trial_sol} above the soliton gas plot.
It is not difficult to show that the  density of the two-component soliton gas in the interaction region,  $\beta_c = w_{1c}+w_{2c} >w_{10}, w_{20}$ but $\beta_c  < w_{10} + w_{20}$. 
 We emphasize
that, although the soliton gas is initially prepared in a rarefied
regime where solitons are spatially well-separated, the total density of solitons increases
in the interaction region, and a more dense soliton gas is formed
 for which solitons exhibit significant
overlap.

The expanding interaction region in the soliton gas Riemann problem  can be viewed as a stochastic counterpart of the traditional, coherent dispersive shock wave  forming due to a dispersive regularisation of the Riemann initial data in the KdV equation \cite{gurevich_nonstationary_1974}, \cite{el_dispersive_2016}. 

Upon using the ansatz (\ref{cold2}) the  values of the  two first moments (\ref{mean}) in the interaction region assume the form,  
\begin{equation}\label{mom_2}
  \langle u \rangle_c\ =\ 4(\eta_1 w_{1c}\ +\ \eta_2 w_{2c}),  \quad \langle{u^2}\rangle_c\ =\ \frac{16}{3} (\eta_1^3 w_{1c}\ +\ \eta_2^3 w_{2c}) \, ,
\end{equation}
where $w_{1c}$ and $w_{2c}$ are determined in terms of the initial data by formulae (\ref{fc}). 

Comparisons of the theoretical results \eqref{fc}, \eqref{cpm},  \eqref{mom_2} for the soliton gas physical `observables'  $c^\pm, \beta, \langle u\rangle, \langle u^2 \rangle$ with direct numerical simulations of the shock tube problem for the KdV soliton gas are presented in Figs.~\ref{fig:kdv_compar_cpm}, \ref{fig_kdv_mom_comp}.

\bigskip
{\it Solition gases for the defocusing and resonant NLS equations}

\medskip
The spectral Riemann problem \eqref{eq:F} \eqref{eq:init} for  bidirectional soliton gases for the defocusing and resonant NLS  equations (see Section~\ref{sec:bidirectional}) was considered in \cite{congy_soliton_2021}, where the  spectral Riemann problem solution \eqref{eq:sol} -- \eqref{eq:RH} 
was constructed for the collisions of two- and three-component gas  and comparisons of some analytically obtained `observables' (the speeds of contact discontinuities and the spatial profiles of the mean density $\langle \rho \rangle$)
 with direct numerical simulations of the soliton gas collision  have been made. 
 \begin{figure}[h]
\centering
\includegraphics{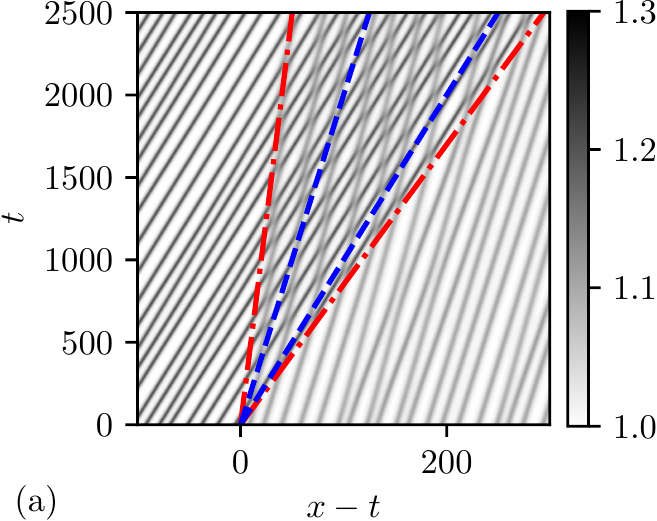}
\includegraphics{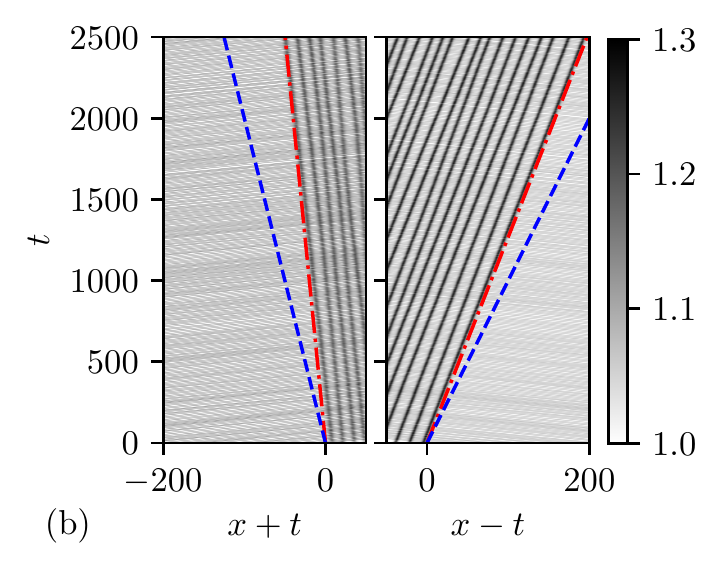}
\caption{(Adapted from \cite{congy_soliton_2021}). Spatiotemporal plots of the field $\rho(x,t)$ for one
realization of the anisotropic soliton gas collision. Trajectories of the solitons appear in
solid lines. Red dash-dotted lines correspond to the trajectories of the
contact discontinuities: $x = z_1 t$, $x = z_2 t$,
cf.~\eqref{eq:z}, and blue dashed lines to the free soliton
trajectories: $x = s_{01} t$, $x = s_{02} t$. (a)~Overtaking collisions
$(\zeta_1,\zeta_2)=(1.05,1.10)$, cf. initial condition (i) in
Table~\ref{tab:num}. (b)~Head-on collisions
$(\zeta_1,\zeta_2)=(-1.05,1.10)$, cf. initial condition (ii) in
Table~\ref{tab:num}.}
\label{fig:xt}
\end{figure}
\begin{figure}[h]
\includegraphics{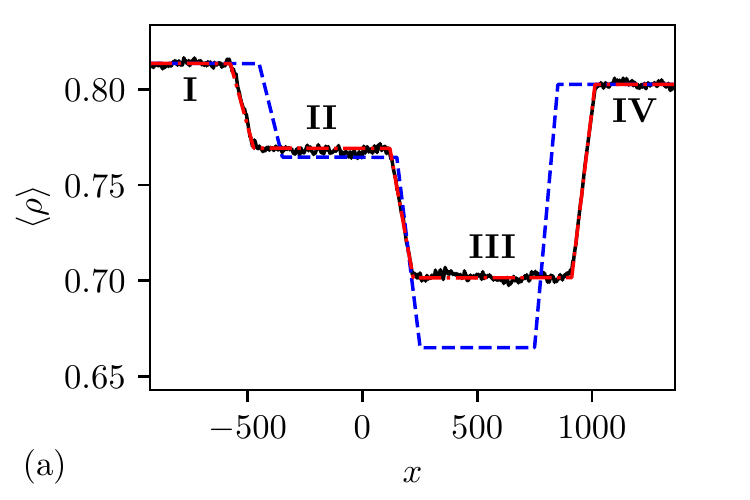}
\includegraphics{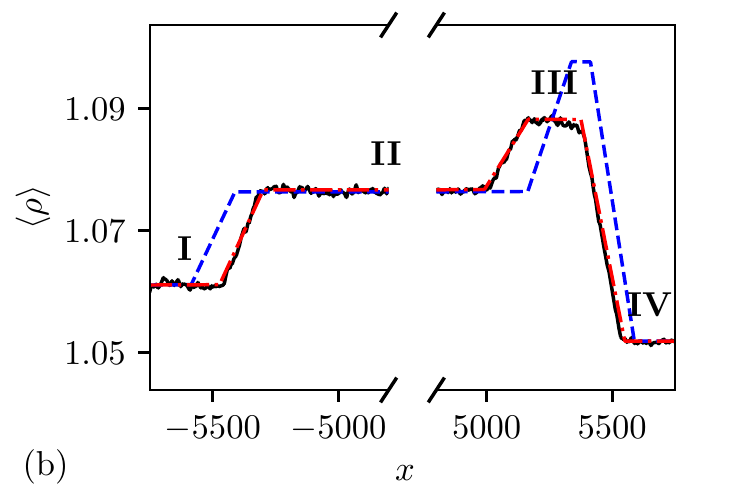}
\caption{(Adapted from \cite{congy_soliton_2021}). Comparison between the ensemble average
$\langle \rho(x,t) \rangle$ obtained by averaging over 100 realisations  of the numerical solution of the three-component soliton gas shock tube problem (solid line) for bidirectional dispersive hydrodynamics and the analytical
solution obtained via the corresponding spectral Riemann problem (dash-dotted line). The dashed
line corresponds to the average field density in the artificial soliton gas composed of
non-interacting solitons with the spectral distribution given by
\eqref{eq:sol_free}. (a)~Defocusing NLS soliton gas at
$t=2000$, case (iii) Table~\ref{tab:num}. (b)~Resonant NLS soliton gas at
$t=5000$, case (iv) in Table~\ref{tab:num}.}
\label{fig:psi2b}
\end{figure}

Fig.~\ref{fig:xt} displays the numerically obtained $x-t$ diagrams extracted from a single realisation of the soliton gas collision and illustrating the difference between overtaking and head-on interactions  in the anisotropic soliton gas of the resonant NLS equation.
 
In Fig.~\ref{fig:psi2b} the comparisons are shown for the profiles of ensemble average $\langle \rho \rangle$  of the wave field $\rho(x,t)$ associated with the spectral Riemann problem solution for the collision of three-component soliton gases for the defocusing and  resonant NLS  equations. The spectral data used for the numerical realisation of the soliton gases are collected in Table~\ref{tab:num}. The numerical evaluation of the ensemble average $\langle \rho \rangle$ involves averaging over 100 realisations. 
As expected, the solution is composed of 4 plateaus, where
regions II and III contain at least two distinct soliton components
and are region of interactions. 
The comparison shows an excellent agreement confirming the efficacy of the developed spectral kinetic theory.  Note that in Fig.~\ref{fig:psi2b} the discontinuities in
$\langle \rho(x,t)\rangle$ have a finite slope 
which is an artefact of the statistical averaging procedure detailed in \cite{congy_soliton_2021}.
 \begin{table*}
\centering
\begin{tabular}{p{1cm} p{5cm} p{4cm} p{4cm}}
\hline\hline
& spectral parameters & left boundary condition & right boundary
condition\\
\hline\hline
(i) & $(\zeta_1 = 1.05 ,\zeta_2 \in [1.06,1.10])$ &
$\bs w^{\rm L}=(0,6.6) \times 10^{-2}$ & $\bs w^{\rm R}=(6.6,0) \times
10^{-2}$ \\ 
(ii) & $(\zeta_1 = -1.05,\zeta_2 \in [1.06,1.1])$ &
$\bs w^{\rm L}=(0,6.6) \times 10^{-2}$ & $\bs w^{\rm R}=(6.6,0) \times
10^{-2}$ \\ 
(iii) & $(\zeta_1,\zeta_2,\zeta_3)=(-0.2,0.1,0.4)$ &
$\bs w^{\rm L}=(2.5,0,7.5) \times 10^{-2}$ & $\bs w^{\rm R}=(5,5,0)
\times 10^{-2}$ \\ 
(iv) & $(\zeta_1,\zeta_2,\zeta_3)=(-1.1,1.05,1.1)$ &
$\bs w^{\rm L}=(1.6,0,5) \times 10^{-2}$ & $\bs w^{\rm R}=(3.3,3.3,0)
\times 10^{-2}$ \\ 
\hline
\end{tabular}
\caption{Spectral  Riemann problem data
used in the numerical simulations for Figs.~\ref{fig:xt}, \ref{fig:psi2b}.}
\label{tab:num}
\end{table*}

\section{Summary and Outlook}

In this article we have reviewed the results on the spectral theory of soliton gases obtained by the author and his collaborators over the last two decades, although many  of the developments are relatively recent. The theory is inspired by Zakharov's 1971 paper \cite{zakharov_kinetic_1971} where the ideas of the spectral kinetic description of a weakly interacting, rarefied soliton gas for the KdV equation were introduced for the first time. 
It has transpired later that the perspective of dispersive hydrodynamics, particularly the Whitham theory of modulated spectral  finite-gap solutions provides an appealing mathematical  framework for the generalisation of Zakharov's kinetic equation to the case of strongly interacting, dense soliton gas \cite{el_thermodynamic_2003}, \cite{el_kinetic_2005}, \cite{el_spectral_2020}.  It has also been realised  that soliton gas represents a ubiquitous physical phenomenon, a kind of strongly nonlinear turbulent wave motion, that can observed in the environmental conditions \cite{costa_soliton_2014}, \cite{osborne_highly_2019} and realised in laboratory   experiments \cite{redor_experimental_2019}, \cite{suret_nonlinear_2020}. The recently discovered  connections of the soliton gas dynamics and statistics with the topical  areas of modulational instability and rogue wave formation \cite{gelash_bound_2019}, \cite{gelash_strongly_2018}, \cite{kraych_statistical_2019} have provided the soliton gas research with further relevance.  

One can envisage many important and interesting mathematical and physical problems arising in connection with the developed spectral theory of soliton gas. Probably the most pertinent one is the determination of the statistical parameters of the integrable turbulence associated with a given spectral DOS of a soliton gas, in particular, obtaining the probability density function,  the power spectrum and the autocorrelation function of the nonlinear wave field. This would  provide the theoretical explanation of the intriguing statistical properties of integrable turbulence  observed in the numerical simulations and physical experiments \cite{Agafontsev:15}, \cite{kraych_statistical_2019}, \cite{gelash_bound_2019}, \cite{agafontsev_rogue_2021}, \cite{agafontsev_extreme_2021}.
Further, the spectral hydrodynamics/kinetics of non-equilibrium soliton gas could  provide the means for the manipulation of the integrable turbulence statistics, and therefore, control of the rogue wave emergence in physical systems described at leading order by integrable equations.

Another pertinent question is the generation of soliton gas out of `non-gas' initial or boundary conditions. This problem is important  from both the theoretical and experimental points of view. One possibility is to use the spontaneous `soliton fission' mechanism \cite{trillo_experimental_2016}, \cite{maiden_solitary_2020} inspired by the original Zabusky-Kruskal work \cite{zabusky_interaction_1965} and employed in the experiments on the creation of a shallow water bidirectional anisotropic soliton gas in \cite{redor_experimental_2019}. In modulationally unstable media  a semiclassical  scenario of the transition to integrable soliton turbulence via a chain of topological bifurcations of the finite-gap spectra  was theoretically proposed in \cite{el_dam_2016} and to some extent realised in an optics experiment reported in \cite{marcucci_topological_2019-1}. 
Of particular relevance is the IST based approach to the controlled (as opposed to the spontaneous) experimental realisation of soliton gas developed in \cite{suret_nonlinear_2020}. This approach enables the generation of a soliton gas with a prescribed DOS, which is crucial for the verification and utilisation of the spectral kinetic theory.

Yet another prospective area of the further development of the soliton gas theory is  the interaction of soliton gas with external potentials, either `static' as in the soliton gas in a trapped BECs \cite{schmidt_non-thermal_2012}, \cite{hamner_phase_2013} or dynamic, as in the recently proposed hydrodynamic soliton tunnelling framework \cite{maiden_solitonic_2018}, \cite{sprenger_hydrodynamic_2018}, \cite{sande_dynamic_2021}. 
Related to this, the development of the theory of soliton gas in perturbed integrable systems is of particular importance for applications where the higher order, non-integrable corrections to the core integrable dynamics are always present and generally result in the slow evolution of the DOS, that is distinct from the spectral kinetic transport by the continuity equation, see  \cite{suret_nonlinear_2020}.

Finally, the recently emerged intriguing parallels between the spectral theory of soliton gas and generalised hydrodynamics \cite{doyon_lecture_2020}, \cite{spohn_hydrodynamic_2021} provide yet another avenue for the novel cross-disciplinary research that could be of significant benefit for both fields of research.

\section*{Acknowledgments}
This  work  was partially supported by EPSRC grant EP/R00515X/2. 

\noindent I would like to express my gratitude to the late A.L. Krylov, who had suggested this topic to me many years ago.
I am grateful to   M. Hoefer,  A. Kamchatnov,  M. Pavlov, S. Randoux, P. Suret and  A. Tovbis  for the collaboration  and many  stimulating discussions over the years.  I  would like to thank T. Congy and G. Roberti for their more recent contributions and for the help with preparing the figures. Special thanks to T. Bonnemain for reading the manuscript and providing  a number of helpful comments.

\begin{appendix}
\section{Nonlinear dispersion relations for the finite-gap solutions of the focusing NLS equation}
 We introduce the fundamental wavenumber and frequiency vectors: ${\bs k} = (k_1, \dots, k_N, \tilde k_1, \dots, \tilde k_N)$ and  ${\bs \o} = (\o_1, \dots, \o_N, \tilde \o_1, \dots, \tilde \o_N)$ respectively, whose components are defined as follows 
\begin{eqnarray}
k_j &=& -\oint_{\alpha_j}\rmd p, \quad \o_j = \oint_{\alpha_j}\rmd q, \quad j=1, \dots, N, \label{kdp}\\
\tilde k_j &=& \oint_{\beta_j}\rmd p, \quad \tilde \o_j = -\oint_{\beta_j}\rmd q, \quad j = 1, \dots, N. \label{omdq}
\end{eqnarray}
Here  
$\rmd p(z)$ and $\rmd q(z)$ are the meromorphic quasimomentum and quasienergy differentials  with the only poles at $z=\infty$ on both sheets of the Riemann surface \eqref{rsurf}  defined by (see e.g. \cite{dafermos_geometry_1986}, \cite{tovbis_semiclassical_2016})
\begin{equation}\label{pq}
\rmd p=1 + \mathcal{O}(z^{-2}) , \qquad \rmd q = 4z+\mathcal{O}(z^{-2}) 
\end{equation}
near $z=\infty$ on the main sheet respectively, and the normalisation conditions requiring that all the periods of $\rmd p, \rmd q$ are real (real normalised
differentials). The homology basis  $\alpha_j$, $\beta_j$ is defined in Section~\ref{sec:FNLS_sol}, see Fig.~\ref{Fig:Cont}. The signs of integrals in \eqref{kdp}, \eqref{omdq} will be opposite if we replace $j$ by $-j$.  The signs of the integrals in \eqref{kdp}, \eqref{omdq} will be opposite if we replace $j$ by $-j$.

{The wavenumbers and frequencies are symmetrically extended to negative indices by $k_{-j}=k_j,\o_{-j}=\o_j$, $j=1, \dots, N, $ 
and similar equations for the `tilded' quantities. They also satisfy the corresponding equations \eqref{kdp}, \eqref{omdq}, but 
the signs of integrals in \eqref{kdp}, \eqref{omdq} will be opposite when we replace $j$ by $-j$.}

The proof that $k_j$, $\tilde k_j$, $\o_j$, $\tilde \o_j$ defined by \eqref{kdp}, \eqref{omdq} are indeed  the wavenumbers and  frequencies of the finite-gap focusing NLS solution associated with the spectral surface  $\Rscr$ of \eqref{rsurf} can be found in \cite{el_spectral_2020}.  
It was then shown in   \cite{el_spectral_2020} that  for the wavenumbers and frequencies \eqref{kdp}, \eqref{omdq} associated with the finite-gap focusing NLS solution the  nonlinear dispersion relations are given by
\bea
&& \tilde k_j+\sum_{|m|=1}^{N} k_m\oint_{\beta_m}{{P_j(\zeta)\rmd \zeta}\over{R(\zeta)}}=-\frac12 \oint_{\tilde \gamma}{{\zeta P_j(\zeta) \rmd \zeta}\over{R(\zeta)}}, \nonumber\\
&&\tilde \o_j+\sum_{|m|=1}^{N} \o_m\oint_{\beta_m}{{P_j(\zeta)\rmd \zeta}\over{R(\zeta)}}=- \oint_{\tilde \gamma}{{\zeta^2 P_j(\zeta) \rmd \zeta}\over{R(\zeta)}}, \nonumber \\
&& |j|= 1,\dots, N,  \label{WUPjM}
\eea
where $\tilde \gamma$ is a large clockwise oriented contour containing $\G$, and $P_j(z)$ and $w_j$ are defined in \eqref{Pj0}, \eqref{D-P0}
Taking real and imaginary parts of \eqref{WUPjM} and using the residues in the right hand side,  we obtain separate systems 
for the  solitonic components $k_m$, $\o_m$ and the carrier components $\tilde k_m$, $\tilde \o_m$.

\end{appendix}

\end{document}